\documentclass[
preprint,
 amsmath,amssymb,
 aps,
prd,
floatfix,
nofootinbib]{revtex4-1}

\usepackage[utf8]{inputenc}
\usepackage[]{mathrsfs}
\usepackage[]{amsbsy}
\usepackage[]{slashed}
\usepackage[inline]{enumitem}
\usepackage{graphicx}% Include figure files
\usepackage{dcolumn}% Align table columns on decimal point
\usepackage{bm}% bold math
\usepackage[bookmarks=false]{hyperref}% add hypertext capabilities

\usepackage[]{comment}
\usepackage[caption=false]{subfig}
\usepackage[]{xcolor}
\usepackage[]{multirow}

\DeclareMathOperator{\Li}{Li}

\allowdisplaybreaks

\newcommand{\be}{\begin{equation}}
\newcommand{\ee}{\end{equation}}
\newcommand{\bea}{\begin{eqnarray}}
\newcommand{\eea}{\end{eqnarray}}

\def\tev{\, {\rm TeV}}
\def\gev{\, {\rm GeV}}

\newcommand{\gsim}{\lower.7ex\hbox{$\;\stackrel{\textstyle>}{\sim}\;$}}
\newcommand{\lsim}{\lower.7ex\hbox{$\;\stackrel{\textstyle<}{\sim}\;$}}

\newcommand{\pb}{{\rm pb}}

\begin{document}

\preprint{UH-511-1263}

\title{Gamma-ray signals from dark matter annihilation via charged mediators}% Force line breaks with \\
\author{Jason Kumar}
\email{jkumar@hawaii.edu}
\affiliation{Department of Physics and Astronomy, University of Hawaii, Honolulu, Hawaii 96822}
\author{Pearl Sandick}
 \email{sandick@physics.utah.edu}

\author{Fei Teng}%
 \email{Fei.Teng@utah.edu}

\author{Takahiro Yamamoto}
 \email{takahiro.yamamoto@uath.edu}
\affiliation{%
 Department of Physics and Astronomy, University of Utah, Salt Lake City, Utah 84112.
}
\date{\today}% It is always \today, today,

\begin{abstract}
We consider a simplified model in which Majorana fermion dark matter annihilates to
charged fermions through the exchange of charged mediators.  We consider the gamma-ray signals
arising from the processes $XX \rightarrow \bar f f \gamma$, $\gamma \gamma$, and $\gamma Z$
in the most general case, including nontrivial fermion mass and nontrivial
left-right mixing and the $CP$-violating phase for the charged mediators.  In particular,
we find the most general spectrum for internal bremsstrahlung, which interpolates
between the regimes dominated by virtual internal bremsstrahlung and by final state radiation.
We also examine the variation in the ratio $\sigma(\gamma \gamma) / \sigma (\gamma Z)$ and the
helicity asymmetry in the $XX \rightarrow \gamma \gamma$ process, each as a function of the mixing
angle and $CP$-violating phase.  As an application, we apply these results to searches for
a class of minimal supersymmetric Standard Model.
%\begin{description}
%\end{description}
\end{abstract}

%Use showkeys class option if keyword
\maketitle

%\tableofcontents

\section{\label{sec:Intro}Introduction}
The nature of dark matter has long been a puzzle in modern physics. It is suspected to be a very long-lived massive particle, while currently no evidence shows that it carries electrical or color charge. Such a particle cannot be described by the Standard Model (SM) of particle physics. The most recent measurement of the dark matter abundance from the Planck satellite is $\Omega h^{2}=0.1199\pm0.0027$ \cite{planck}. If we assume that dark matter consists of weakly interacting massive particles (WIMPs) with mass ranging from $\sim 10\gev$ to $\sim 10\tev$, the standard thermal freeze-out mechanism yields the qualitatively correct relic density~\cite{Zeldovich,Chiu:1966kg,Lee:1977ua,Steigman:1979kw,Scherrer:1985zt,Jungman96}.
As a result, the WIMP hypothesis is very attractive, but by no
means required.  Searches for the interaction of dark matter with SM matter are ongoing, utilizing a variety of strategies, including direct, indirect, and
collider-based searches.

The main purpose of this paper is to analyze the associated gamma-ray signals that may be observable in indirect dark matter searches if Majorana fermion dark matter
couples to light SM fermions via charged mediators.  Such couplings arise in a variety of dark matter scenarios, including the minimal supersymmetric
Standard Model (MSSM), in which the lightest supersymmetric particle (LSP) is a dark matter candidate, and its bino component can couple to SM fermions through the
$t$- or $u$-channel exchange of sfermions.  Moreover, the gamma-ray signals from dark matter annihilation in this scenario are often crucial to observational
strategies, because gamma-ray signals are relatively clean, and because the direct annihilation process $XX \rightarrow \bar f f$ is often suppressed.

The processes which we will consider are $XX \rightarrow \gamma \gamma$, $XX \rightarrow \gamma Z$, and $XX \rightarrow \bar f f \gamma$. All of these processes
have been considered in the past \cite{Bergstrom97,Bern97,Ullio98,Bringmann08}, but either for special cases or different purposes.
Our goal here will be to consider the most general spectra that can arise for these processes in a simplified model in which a Majorana fermion dark matter
particle couples to a Dirac fermion (which may or may not be a SM fermion) through the exchange of two charged scalars, with an arbitrary left-right mixing angle and $CP$-violating phase.  Examples of this simplified model exist within the parameter space of the MSSM, including the ``Incredible Bulk" models
described in Ref.~\cite{Fukushima14},
but the applicability is much broader.

The main new features which  we will find are:
\begin{itemize}
\item{The complete spectrum for the process $XX \rightarrow \bar f f \gamma$ as a function of mixing angle, which interpolates between the
hard regime, dominated by virtual internal bremsstrahlung, and the soft regime, dominated by soft and
collinear final state radiation }
\item{The ratio of the cross sections for $XX \rightarrow \gamma \gamma$ and $XX \rightarrow \gamma Z$ as a function of
mixing angle and $CP$-violating phase}
\item{The difference in rates for the production of left-circularly and right-circularly polarized photons via the process
$XX \rightarrow \gamma \gamma$}
\end{itemize}

This article is organized as follows: In Sec.~\eqref{sec:Theory}, we briefly describe the effective model
and the parameter space in which we are interested.
We then discuss the general features of the relevant gamma-ray signals in Sec.~\eqref{sec:Gamma}.
In Sec.~\eqref{sec:Linecal}, we describe the monochromatic line signals, and their observational
impact.  In Sec.~\eqref{sec:IBcal}, we similarly describe the general internal bremsstrahlung gamma-ray signature.
Finally our chief results are summarized in Sec.~\eqref{sec:Con}.

\section{\label{sec:Theory}Model and Its General Features}
We consider a simplified model in which the dark matter candidate is a SM gauge singlet Majorana fermion and the only relevant interaction is
\begin{equation}
\label{eq:Lag}
\mathcal{L}_{\text{int}}=\lambda_{L}\widetilde{f}^{\ast}_{L}\overline{{X}}P^{}_{L}f+\lambda_{R}\widetilde{f}^{\ast}_{R}\overline{{X}}P^{}_{R}f
+\text{c.c}\,,
\end{equation}
where $P^{}_{L(R)}$ are the chiral projectors.
Here, $f$ is a fermion charged\footnote{For simplicity, we assume the charge be $Q=-1$.}
under $U(1)_{\text{em}}$, and $\widetilde{f}_{L(R)}$ are the charged scalar mediators.
We also assume that the dark matter is absolutely stable because it is the lightest
particle charged under an unbroken hidden symmetry, and the $\widetilde{f}_{L(R)}$ are also charged under the same symmetry.
But the fermion $f$ is uncharged under the symmetry that stabilizes the dark matter.

The mass eigenstates and chiral eigenstates of the scalar mediators are related by a mixing angle $\alpha$,
\begin{equation}
\label{eq:slepton-mixing}
\left(\begin{array}{c}
\widetilde{f}_{1} \\ \widetilde{f}_{2}
\end{array}\right)=\left(\begin{array}{cc}
 \cos\alpha & -\sin\alpha \\
 \sin\alpha & \cos\alpha
\end{array}\right)\left(\begin{array}{c}
\widetilde{f}_{L} \\ \widetilde{f}_{R}
\end{array}\right)\,.
\end{equation}
We denote the two mass eigenvalues as $m_{\widetilde{f}_{1}}$ and $m_{\widetilde{f}_{2}}$ in the following.  To ensure
that the dark matter is stable, we assume $m_{\widetilde{f}_{1,2}} > m_X$.  However, $m_f$ can be either larger or smaller
than $m_X$.

We also allow a nonzero $CP$-violating phase, $\varphi$, such that the coupling constants may be expressed as
\begin{align}
&\lambda_{L}=|\lambda_{L}|\,e^{i\varphi/2}\,, &\lambda_{R}=|\lambda_{R}|\,e^{-i\varphi/2}\,.
\end{align}
We are thus left with seven free parameters for this simplified model: $$(m_{{X}},m_{{\widetilde{f}}_{1}},m_{\widetilde{f}_{2}},\lambda_L, \lambda_R, \alpha,\varphi)\,.$$
In the MSSM framework, if $X$ is a purely binolike LSP and there is a single generation of light sfermions,
then we have $|\lambda_{L}|=\sqrt{2}\,g |Y_{L}|$ and
$|\lambda_{R}|=\sqrt{2}\,g|Y_{R}|$,
where $g$ is the $U(1)_{Y}$ gauge coupling and $Y_{L(R)}$ are the scalar hypercharges.  This scenario has been considered
recently in Ref.~\cite{Fukushima14,Buckley:2013sca,Pierce:2013rda}.
We also briefly consider the possibility of a new heavy fermion, in which case there is an additional parameter necessary to specify its mass, $m_f$.

Note that only the relative phase between $\lambda_L$ and $\lambda_R$ is physically significant, since any
overall phase can be removed by a vectorlike phase rotation of $\widetilde f_{L,R}$.  Similarly, although the
most general matrix relating the scalar mass and chiral eigenstates contains three complex phases, they can
be absorbed by a phase rotation of the chiral eigenstates, ${\widetilde{f}}_{L,R}$, and the mass
eigenstates, ${\widetilde{f}}_{1,2}$.  Having chosen to make the mixing matrix real, one cannot then
use a chiral rotation of the ${\widetilde{f}}_{L,R}$ to rotate away the phase $\varphi$.  However, if $\sin 2\alpha=0$,
the requirement that the mixing matrix be real only fixes two phases; in this case, the phase $\varphi$ can then be
absorbed into a chiral rotation of the ${\widetilde{f}}_{L,R}$.
Similarly, if $m_f=0$, then the phase $\varphi$ can be removed by a chiral rotation of $f$.
As a result, $CP$-violating effects must scale as $(m_f /m_X) \sin 2\alpha $.

\subsection{General features}
\begin{figure*}[t]
	\includegraphics[width=\textwidth]{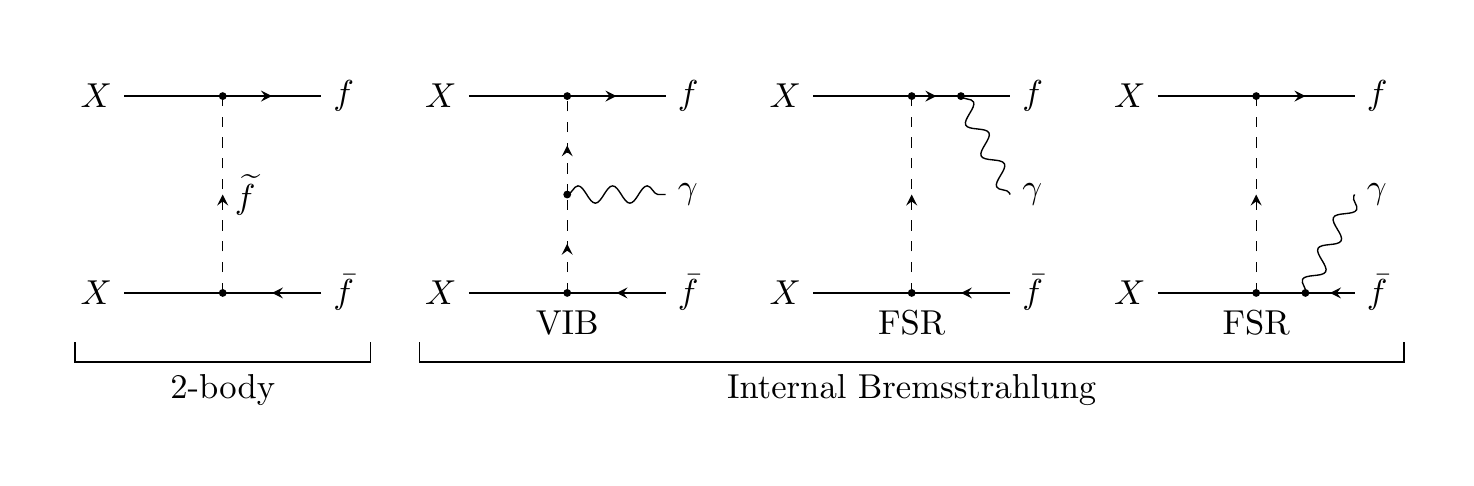}
	\vspace{-4.5em}
	\caption{\label{fig:2bIB}The Feynman diagrams for two-body annihilation and IB.}
\end{figure*}

If the cross section for the process $XX \rightarrow \bar f f$ is not suppressed, it will dominate over processes
such as $XX \rightarrow \bar f f \gamma$ and $XX \rightarrow \gamma \gamma, \gamma Z$, which will be suppressed by
factors of at least  $\alpha_{\text{em}}$ and $\alpha_{\text{em}}^2$, respectively.  In this case, prompt gamma-ray signals are sometimes considered to be less promising from an observational standpoint, because models that would produce prompt gamma-ray signals large enough to
be observed with current or near future experiments can already be probed by searches for cosmic rays produced
by the process $XX \rightarrow \bar f f$.  However, current exclusion limits based on searches for positrons or
antiprotons are subject to large systematic uncertainties related to assumptions about the astrophysical background and propagation of charged particles in our Galaxy; if these
assumptions are weakened, then the exclusion limits from cosmic-ray searches can be similarly weakened, permitting
even the suppressed $XX \rightarrow \bar f f \gamma$, $\gamma \gamma$, and $\gamma Z$ signals to be relevant. {The Feynman diagrams for the two-body annihilation $XX\rightarrow \bar{f}f$ and internal bremsstrahlung $XX\rightarrow\bar{f}f\gamma$ are shown in Fig.~\ref{fig:2bIB}, while those for the one-loop process $XX\rightarrow\gamma\gamma$ are shown later in Fig.~\ref{fig:feyndiag} of Appendix~\ref{sec:photoncs}, where a detailed discussion on this process is presented.}

But there are two scenarios in which the $XX \rightarrow \bar f f$ annihilation cross section is suppressed:
\begin{itemize}
\item{If $m_f > m_X$, then $XX \rightarrow \bar f f$ is not kinematically allowed.}
\item{If there is minimal flavor violation (MFV), then the cross section for the process
$XX \rightarrow \bar f f$ is suppressed by a factor $(m_f / m_X)^2$.}
\end{itemize}
In the case where $m_f / m_X >1$, the processes $XX \rightarrow \bar f f, \bar f f \gamma$ are
forbidden, allowing the processes $XX \rightarrow \gamma \gamma$ and $XX \rightarrow \gamma Z$
to be the most important (other processes, such as $XX \rightarrow ZZ, W^+ W^-$ may have similar
cross sections but are likely to be less observationally important compared to a clean gamma-ray signal).  This scenario is relevant in the case where dark matter couples to a new, heavy charged
fermion.

In the case where $m_f / m_X \rightarrow 0$, the cross section for the
process $XX \rightarrow \bar f f$ must scale with the remaining parameters, which breaks flavor symmetry.
The reason is that, because the dark matter is Majorana and the initial state wave function must be
antisymmetric, the $s$-wave initial state must have $J=0$.  The final state $\bar f$ and $f$ must
then have the same helicity, implying that the $f$ and $\bar f$ arise from different Weyl spinors.
The final state is thus not invariant under chiral flavor symmetries and must vanish in the
$m_f / m_X \rightarrow 0$ limit in the case of MFV .

In the simplified model that we consider here, the only deviation from MFV arises from
the presence of nontrivial mixing of the scalar chiral eigenstates.  This requires both a
nontrivial left-right mixing angle $\alpha$ and nondegeneracy of the mass eigenstates (if the mass
eigenstates are degenerate, then a redefinition of the eigenstates is sufficient to absorb the mixing angle).
In the massless fermion limit, left-right mixing
 gives ${X}{X}\rightarrow \bar f f$ an  $s$-wave two-body annihilation amplitude of
\begin{widetext}
\begin{align}
    \mathcal{A}_{\text{2-b}}=i m_{X} \frac{|\lambda_{L}\lambda_{R}| }{ 2}\sin(2\alpha)\left(\frac{\overline{u} (k_1) \gamma^{5}v (k_2)}{2m_{X}}\right)&\left[\cos\varphi\,\overline{u} (k_3) \gamma^{5}v(k_4)-i\sin\varphi\,\overline{u}(k_3)v(k_4)\right]\nonumber\\*
    &\times\left(\frac{1}{m_{X}^{2}+m_{\widetilde{f_1}}^{2}}-\frac{1}{m_{X}^{2}+m_{\widetilde{f_2}}^{2}}\right),
\end{align}
\end{widetext}
where we denote the initial state dark matter momenta as $k_{1}=k_{2}=k$ and the final state fermion momenta as $k_{3}$ and $k_{4}$; $u(k_i)$ and $v(k_i)$ are spinor wave functions, following the definition of Ref.~\cite{Peskin}. This amplitude leads to the cross section
\begin{align}
\label{eq:AnniCS}
\left(\sigma v\right)_{f\bar{f}}&\xrightarrow{m_{f}= 0}\frac{m^{2}_{X}}{32\pi} |\lambda_{L}\lambda_{R}|^{2}\sin^{2}(2\alpha)
%\nonumber\\*
%&\quad
\left(\frac{1}{m^{2}_{\widetilde{f}_{1}}+m^{2}_{{X}}}-\frac{1}{m^{2}_{\widetilde{f}_{2}}+m^{2}_{{X}}}\right)^{2}.
\end{align}
In the $m_f / m_X \rightarrow 0$ scenario, the charged fermion $f$ must necessarily be a SM fermion.

\subsection{\label{sec:Constraint}Constraints from colliders and lepton dipole moments}
The discovery of the $125\gev$ SM-like Higgs boson \cite{CMShiggs,*ATLAShiggs} at the LHC is a triumph of the SM. Meanwhile, null searches for supersymmetric particles imply a lower limit of $780\gev$ for light degenerate first and second generation squarks \cite{ATLASsquark}.
Constraints on squark masses in the simplest predictive supersymmetric model, the constrained MSSM (CMSSM), are even more stringent,
and exclude squarks below $1.7\tev$ for certain benchmark models \cite{ATLASsquark,*CMSsquark}.
If we relax some unification constraints imposed in the CMSSM at the grand unified theory scale, then it has been shown that the MSSM-9 model \cite{cabrera1,*cabrera2}
can contain a $\sim 1\tev$ Higgsino  LSP or a $\sim 3\tev$ wino LSP, which are viable thermal dark matter candidates satisfying the relic density.

On the other hand, the current limit on the mass of any slepton is much weaker. We still have the possibility that heavy squarks provide the necessary
loop corrections to the mass of the SM-like Higgs while light sleptons provide the main dark matter annihilation channel. Large Electron-Positron Collider (LEP) experiments only put a lower limit at $\sim 100\gev$ \cite{ALEPH,*L3,*DELPHI,*OPAL}, while the LHC $8\tev$ run has excluded left-handed sleptons below $310\gev$ and right-handed sleptons
below $235\gev$, assuming a massless bino LSP \cite{ATLASslepton,*CMSslepton}.  For a massive bino LSP with mass $m_{X}$, a new allowed region opens up for sleptons lighter than
approximately $m_{X}+80\gev$ \cite{ATLASslepton,*CMSslepton}. The LHC $14\tev$ run has the potential to push the upper exclusion limit to as high as $900\gev$ (for a Higgsino LSP) but cannot move the lower exclusion limit \cite{Eckel1}.

Although these constraints are phrased as bounds
on scalar superpartners, the lesson is more general: LHC constraints place tight bounds on colored scalars but weaker bounds on
QCD-neutral scalars.  Since
the $\widetilde{f}_{L,R}$ must necessarily be QCD charged if $f$ is a quark, we will assume that, if $f$ is a SM fermion, it is a lepton.

A new correction to the electric and magnetic dipole moments of the SM fermions arises from diagrams
with $X$ and the new charged mediator running in the loop; if $f$ is a SM fermion, then the $XX \rightarrow \bar f f$
can be constrained by bounds on fermion dipole moments~\cite{Fukushima:2013efa,Fukushima14} (in the absence of fine-tuned cancellations against
other contributions to the dipole moments from independent new physics).  In the case where $f$ is a
SM charged lepton, the constraints can be summarized as follows:
\begin{itemize}
\item The $XX \rightarrow e^{+}e^{-}$ cross section is constrained to be $\ll 1~\pb$, absent fine-tuning.
\item The $XX \rightarrow \mu^{+}\mu^{-}$ cross section is constrained to be $\ll 1~\pb$, absent fine-tuning, unless
$CP$ violation is close to maximal ($\varphi \sim \pi/2$).  This constraint arises because the muon magnetic dipole moment
is much more tightly constrained than its electric dipole moment.  For near-maximal $CP$ violation, the annihilation
cross section must be less than ${\cal O}(100)~\pb$, absent fine-tuning.
\item The $XX \rightarrow \tau^{+}\tau^{-}$ cross section can easily be ${\cal O}(1)~\pb$, or larger.  For our purposes,
it is unconstrained by dipole moment bounds.
\end{itemize}

We close this section by noting that although our model fits within the MSSM, it can also serve as a simplified model for other scenarios in which a gauge singlet Majorana dark matter couples only to a fermion  and two scalar particles.

\section{\label{sec:Gamma}Gamma-Ray Signals }

In the current era, it is believed that there are potentially observable excesses of dark matter particles near our Galactic center and in nearby dwarf galaxies. The ongoing annihilations of these dark matter particles may result in observable cosmic-ray signals, such as in the cosmic gamma-ray spectrum and/or in the cosmic-ray positron and/or antiproton fraction. Typically, various reactions involving the final state lepton pairs result in an almost featureless secondary photon spectrum. In the simplified model we consider, however, a distinctive feature may be contained in the internal bremsstrahlung (IB) spectrum and the associated line signals.

The search for line signals of dark matter annihilation has been one of the primary goals of various ground-based and satellite-based experiments. In general, the ground-based atmospheric Cherenkov telescopes \cite{HESS,*CTA} are most effective for dark matter that is somewhat heavier than 100 GeV.  For example, with 500 h of observing time, Cherenkov Telescope Array (CTA) will be sensitive to cross sections $\sim 10^{-27}\text{cm}^{3}/\text{s}$ for dark matter with mass of $\sim 300~\gev$ annihilating to $\tau^+\tau^-$ in the Galactic center region \cite{Wood:2013taa}. Due to its much lower energy threshold, the Fermi Gamma-Ray Space Telescope is better suited to study dark matter masses in the range $0.1$ to a few hundred GeV, which is the range we are interested in here. For $m_{{X}}\lesssim 100$ GeV, the Fermi Large Area Telescope (LAT) has set a limit on the thermally averaged annihilation cross section to $\gamma\gamma$ of $\langle\sigma v\rangle_{\gamma\gamma}\approx10^{-28}\sim 10^{-29}\text{cm}^{3}/\text{s}$ with the $95\%$ C.L. containment spanning approximately one order of magnitude using the PASS 8 analysis of $5.8$ yr of data~\cite{Ackermann:2015lka}. However, this limit, as well as any projected sensitivities, is sensitive to the dark matter profile of the Milky Way halo and may move up or down by about one order of magnitude for different profiles. We hope that the sensitivity to $\langle\sigma v\rangle_{\gamma\gamma}$ will be improved with additional data and/or new technology.
Future satellite-based experiments GAMMA-400~\cite{G400} and HERD~\cite{HERD} are expected to reach $\langle\sigma v\rangle_{\gamma\gamma}\lesssim 10^{-28}\text{cm}^{3}/\text{s}$ for $m_X=100\gev$.
In addition, each of these experiments is expected to have energy resolution of $\sim 1\%$, which is much better than Fermi-LAT's ($\sim 10\%$ at $100\gev$), making it possible to distinguish between a sharply peaked IB spectrum and a true line signal.

Here, we present our results for the bremsstrahlung and other prompt photon emissions arising from dark matter annihilation to fermions
and monochromatic emissions from annihilation to $\gamma\gamma$ and $\gamma Z$. By prompt emission, we mean the photons produced  directly at the dark matter annihilation, including, for example, the hadronic decay of the final state $\tau^{\pm}$. On the other hand, the photon emission due to inverse Compton scattering and bremsstrahlung in the Galactic electromagnetic fields (so-called secondary emission), which depends on modeling of the dark matter distribution and cosmic-ray propagation, is not included.

The photon spectrum is defined as the photon number per annihilation per energy bin, and can be broken into a continuum spectrum and a contribution from monochromatic photons:
\begin{equation}
    \frac{dN}{dx}=\left(\frac{dN}{dx}\right)_{\text{cont.}}+\left(\frac{dN}{dx}\right)_{\text{line}}\,.
\end{equation}
In our case, the continuum spectrum comes mainly from IB and other prompt emission from the final state particles,
\begin{equation}
    \left(\frac{dN}{dx}\right)_{\text{cont.}}=\frac{1}{(\sigma v)_{\text{ann.}}}\left[\frac{d(\sigma v)_{\text{IB}}}{dx}+\sum_{i}N_{i}\frac{d(\sigma v)_{i}}{dx}\right],
\end{equation}
where $N_i$ is the number of photons produced in a single annihilation process  and the sum over $i$ includes all higher order prompt emissions, and $(\sigma v)_{\text{IB}}$ is the IB cross section, which will be described in Sec.~\ref{sec:IBcal}.  Note that the spectrum is normalized by $(\sigma v)_{\text{ann.}}$, the total annihilation cross section. When there is no chiral mixing, its dominant component is the total IB cross section.
As discussed below, the line spectrum consists of the $\gamma\gamma$ and $\gamma Z$ peaks,
\begin{align}
    \left(\frac{dN}{dx}\right)_{\text{line}}=\frac{1}{(\sigma v)_{\text{ann.}}}&\Big[2(\sigma v)_{\gamma\gamma}\,\delta(x-1)
%    \nonumber\\
%    &
    +(\sigma v)_{\gamma Z}\,\delta(x-x_{Z})\Big],
\end{align}
where $x_{Z}=E_{\gamma Z}/m_{{X}}$ as given in Eq.~\eqref{eq:lineenergy}. These definitions follow Ref.~\cite{Bringmann14}.

\section{\label{sec:Linecal}Monochromatic Gamma-Ray Line Signals}
\label{sec:monochromatic}
Monochromatic lines in the gamma-ray spectrum arise due to the one-loop annihilation process ${X}{X}\rightarrow\gamma Y$, where $Y=\gamma,\,Z,$ or $h_{0}$. The photon(s) in the final state has (have) energy
\begin{equation}
\label{eq:lineenergy}
E_{\gamma Y}=m_{{X}}-\frac{m^{2}_{Y}}{4m_{{X}}}\,,
\end{equation}
where $m_{Y}$ is the mass of the particle $Y$. $1\%$ energy resolution is sufficient to differentiate $\gamma\gamma$ and $\gamma Z$ lines for $m_{{X}}\lesssim 450\gev$. In the Galactic center, the relative velocity between dark matter particles is $v \sim 10^{-3}$, so the $p$-wave component of the dark matter annihilation cross section is suppressed.

The $s$-wave ($L=0$) component must arise from a spin-singlet initial state ($S=0$), since the dark matter particles are Majorana fermions
and must be in a totally antisymmetric initial state.  This state thus necessarily has vanishing total angular momentum ($J=0$), implying that the final state particles must have the same helicity.
As a result, only the $\gamma\gamma$ and $\gamma Z$ cross sections can develop nonvanishing $s$-wave components, while the leading $\gamma h_{0}$ cross section must be $p$-wave suppressed and is thus too small to be observed.

To ensure the accuracy of the results presented here, we perform a scan over
the parameter space, conducted as follows.  We first generate the analytic amplitudes for both $XX \rightarrow \gamma \gamma$ and $XX\rightarrow\gamma Z$ using {\tt FeynArts} \cite{Hahn01}, including left-right scalar mixing and a $CP$-violating phase (the $\gamma \gamma$ amplitude is presented in Appendix \ref{sec:photoncs}).
The numerical calculation is performed using {\tt FormCalc} \cite{Hahn98}. The package {\tt LoopTools} is internally invoked by {\tt FormCalc} to calculate the loop integrals involved in the amplitudes. However, since our initial state particles are at rest, we have $k_{1}=k_{2}$. It is well known that if two external momenta are collinear, the Gram matrix becomes singular and the tensor loop integrals fail to be linearly independent. For analytic calculation, this is a virtue and essentially the reason why all the four-point loop integrals that appear in the $\gamma\gamma$ amplitude can be reduced to three-point scalar loop integrals (see Appendix \ref{sec:photoncs})\footnote{For a comprehensive review on the calculation techniques of general tensor loop integrals, see Ref.~\cite{Ellis12}.}. Numerically, {\tt FormCalc} breaks down for collinear external momenta, since {\tt LoopTools} uses precisely the Gram matrix to derive higher-point and higher-rank integrals.

To circumvent this issue and arrive at a reliable result, we introduce a small relative velocity, so that the results of {\tt LoopTools} remain stable. For example, we use a center-of-mass energy of $\sqrt{s}=200.01\gev$ for $m_{{X}}=100\gev$ in our numerical calculations. We have checked that the numerical error in the cross section is $\lesssim 1\%$ for annihilations to both $\gamma\gamma$ and $\bar f f$ (with scalar mixing), independent of the model parameters.
For $p$-wave dominant cross sections, we perform a linear fit with respect to $v^{2}$ to find the coefficients $a$ and $b$ in the expansion $\sigma v=a+bv^{2}$. We have also checked that the error is $\lesssim 1\%$ in this scenario.

Analytic MSSM calculations of the annihilation cross section to $\gamma\gamma$ and $\gamma Z$ have been presented in Ref.~\cite{Bergstrom97,Bern97,Ullio98} in the limit of no $CP$ violation,
and those expressions are consistent with the ones presented here. However, if $\varphi\neq 0$, then the amplitude of the $(++)$ photon helicity state will be different from that of the $(--)$ state, unlike the $\varphi=0$ case. As we argued previously, the difference in the scattering amplitude is chirally suppressed by the fermion mass $m_{f}$,
\begin{equation}
\delta\pmb{\mathcal{A}}\sim\alpha_{\text{em}}|\lambda_{L}\lambda_{R}|\sin(2\alpha)\sin\varphi\left(\frac{m_{f}}{m_{{X}}}\right),
\end{equation}
a term which does not appear previously in the literature.
We defer the full analytic expressions for the $\gamma\gamma$ cross section, including chiral mixing and $CP$ violation, to Appendix \ref{sec:photoncs}.

In order for $CP$ violation to yield differing cross sections for the $(++)$ and  $(--)$ final photon states, it
is necessary for $m_f < m_X$.  If $m_f > m_X$, then the $CP$-violating part of the amplitude is purely
imaginary (as a result of the optical theorem), and $CP$ conjugation of the matrix element is equivalent to
complex conjugation.  But if $m_f < m_X$, then the intermediate states of the one-loop diagram can go on shell,
providing an imaginary component to the $CP$-conserving matrix element, which is necessary
for a nontrivial asymmetry. On the other hand, if $m_f > m_X$, then the only final states that are kinematically allowed are
$\gamma \gamma$  and $\gamma Z$.  In this kinematic regime, if the couplings $\lambda_{L,R}$ are large,
these final states will be most easily observable.

We would like to make a few general comments regarding the sensitivity of the dark matter annihilation cross sections into $\gamma \gamma$ and $\gamma Z$ to scalar chiral mixing and $CP$ violation. First, both cross sections decrease as the scalar masses increase. Thus to make a sizable line signal, we need to have at least one scalar mass not too much heavier than the dark matter. If they are very degenerate, of course, coannihilations, not considered here, would also play a role in determining the relic density. 
Second, the ratio $2(\sigma v)_{\gamma\gamma}/(\sigma v)_{\gamma Z}$ increases as the difference between the two scalar masses increases. As a very crude estimate, this ratio is approximately $2\tan^{-2}\theta_{W}\sim 7$, which works well at no mixing. However, by varying the mixing angle and $CP$-violating phase, we can make it as large as $40$ within the MSSM.

In the following discussion, we will focus on several benchmark models, displayed in Table~\ref{tab:bm}.  Models $A$, $B$, and $C$ are consistent with a supersymmetric implementation of the Lagrangian in Eq.~\eqref{eq:Lag}, while models $D$ and $E$ are explicitly nonsupersymmetric due to the couplings $\lambda_{L,R}$ and, in the case of model $E$, an additional heavy fermion.
Note that, for all of these benchmark points, the new charged particles are fully consistent with constraints from the LHC and LEP.

Benchmark $E$ presents
an interesting case, as it contains a new charged fermion with $m_f = 105\gev$.  Such a particle is within the energy reach of the LHC, and one must
worry if such a particle would already be excluded by current data.  But LHC sensitivity to new charged particles depends greatly on the particle decay
chains; it is easy to choose a decay scenario for which $f$ would escape current LHC limits.  For example, if the new $105\gev$ fermion decayed to a
SM charged lepton and a new $\sim 100\gev$ invisible scalar, then this charged fermion would escape detection for the same reason that light sfermions do in the
compressed spectrum scenario.  The new invisible scalar need not contribute to dark matter, or even be long lived, provided its lifetime was
long enough to decay outside the detector.  We will not focus further on this particular decay
scenario, which we describe only to demonstrate that benchmark $E$ can
be completely consistent with LHC constraints.  For models where $m_f$ is even larger, LHC constraints may be more easily satisfied, without
qualitatively changing the analytic results we obtain.

\begin{table*}[ht]
    \centering
    \renewcommand{\arraystretch}{1.2}
    \begin{tabular}{*{8}{|c}|}
         \hline
         & Channel & $\lambda_L$ & $\lambda_R$  & $\alpha$ & $\varphi$ & Marker \\ \hline
         $A$ & $\mu^{+}\mu^{-}$  &  \multirow{3}{*}{$\sqrt{2}Y_L g$} &  \multirow{3}{*}{$\sqrt{2}Y_R g$} & \multirow{3}{*}{$\displaystyle\frac{\pi}{4}$} & $\pi/2$ & Star \\ \cline{1-2} \cline{6-7}
         $B$ & \multirow{2}{*}{$\tau^{+}\tau^{-}$}                 &                      &                      &
                                                       & $0$     & Circle \\ \cline{1-1} \cline{6-7}
         $C$ &                                                       &                &                      &
                                                       & {$\pi/2$} &  Cross                            \\ \hline
         $D$ &        \multirow{2}{*}{$\mu^{+}\mu^{-}$}    &       $0.8$                &     $0.8$                    & \multirow{2}{*}{$\displaystyle\frac{\pi}{6}$}
                                                       &  \multirow{2}{*}{$\displaystyle\frac{\pi}{2}$} &       Square                       \\ \cline{1-1} \cline{3-4} \cline{7-7}
         $D'$ &                           & $0.75$              & $0.75$
                                &                       &       & Diamond \\            \hline
         $E$ &     $\bar f f$, $m_f = 105$ GeV      &   2            &
               2                 & $\pi/4$
                                                       &  $3\pi/4$ &    Triangle                          \\ \hline
    \end{tabular}
    \caption{We take $m_X = 100\gev$, $m_{\widetilde{f}_1} = 120\gev$, and $m_{\widetilde{f}_2}=450\gev$ for the SUSY ($A$, $B$, and $C$) and non-SUSY ($D$ and $E$) benchmarks, but $m_{\widetilde{f}_{1}}=102.5\gev$ for the non-SUSY benchmark $D'$.  We take $|Y_L| =1/2$, $|Y_R|=1$ for the case of a bino
    coupling to leptons.}
    \label{tab:bm}
\end{table*}

In the following subsections, we examine the line signal strengths in the context of different SUSY and non-SUSY models.

\begin{figure*}
\centering
\includegraphics[width=0.47\textwidth]{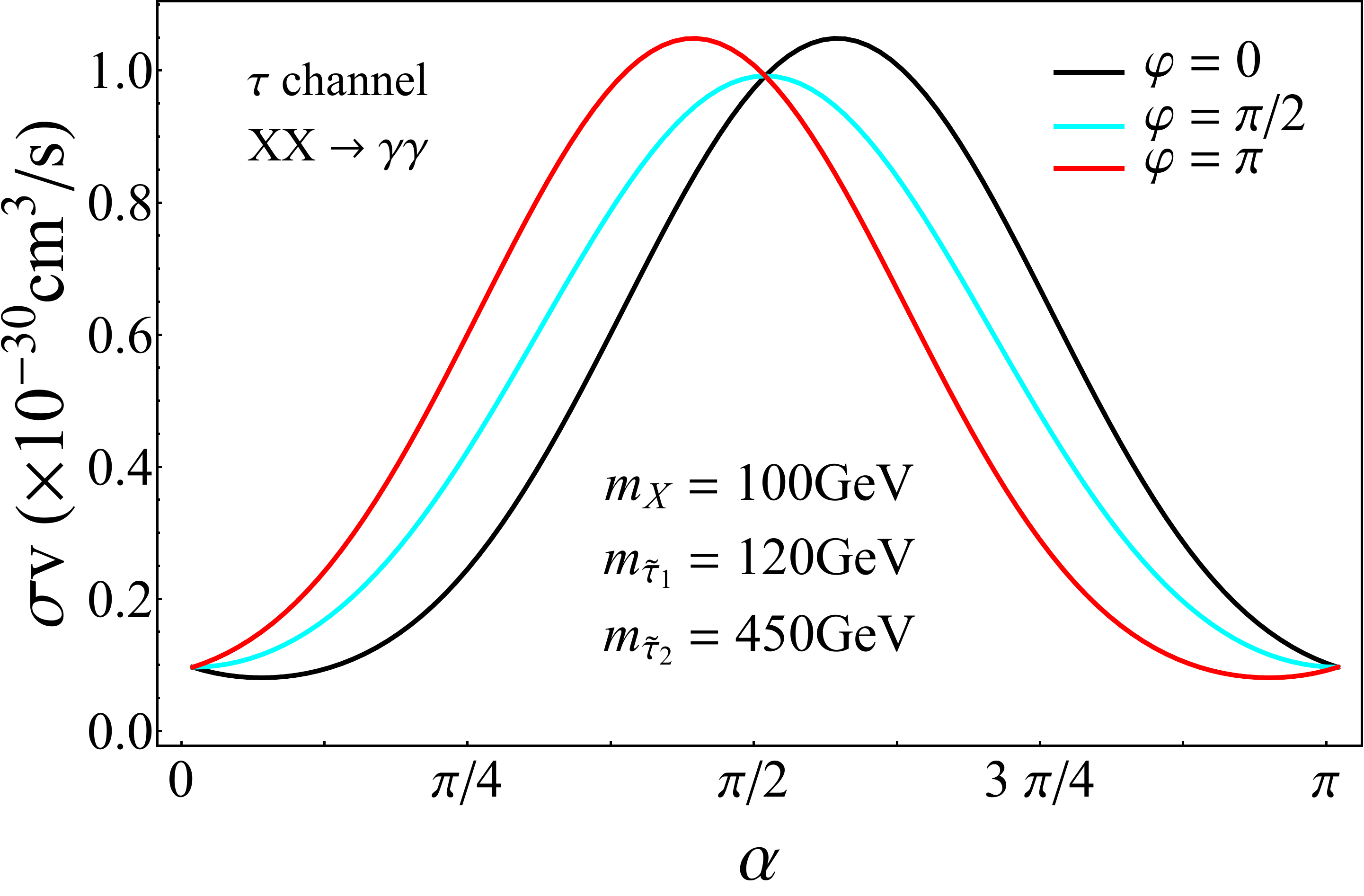}\quad
\includegraphics[width=0.47\textwidth]{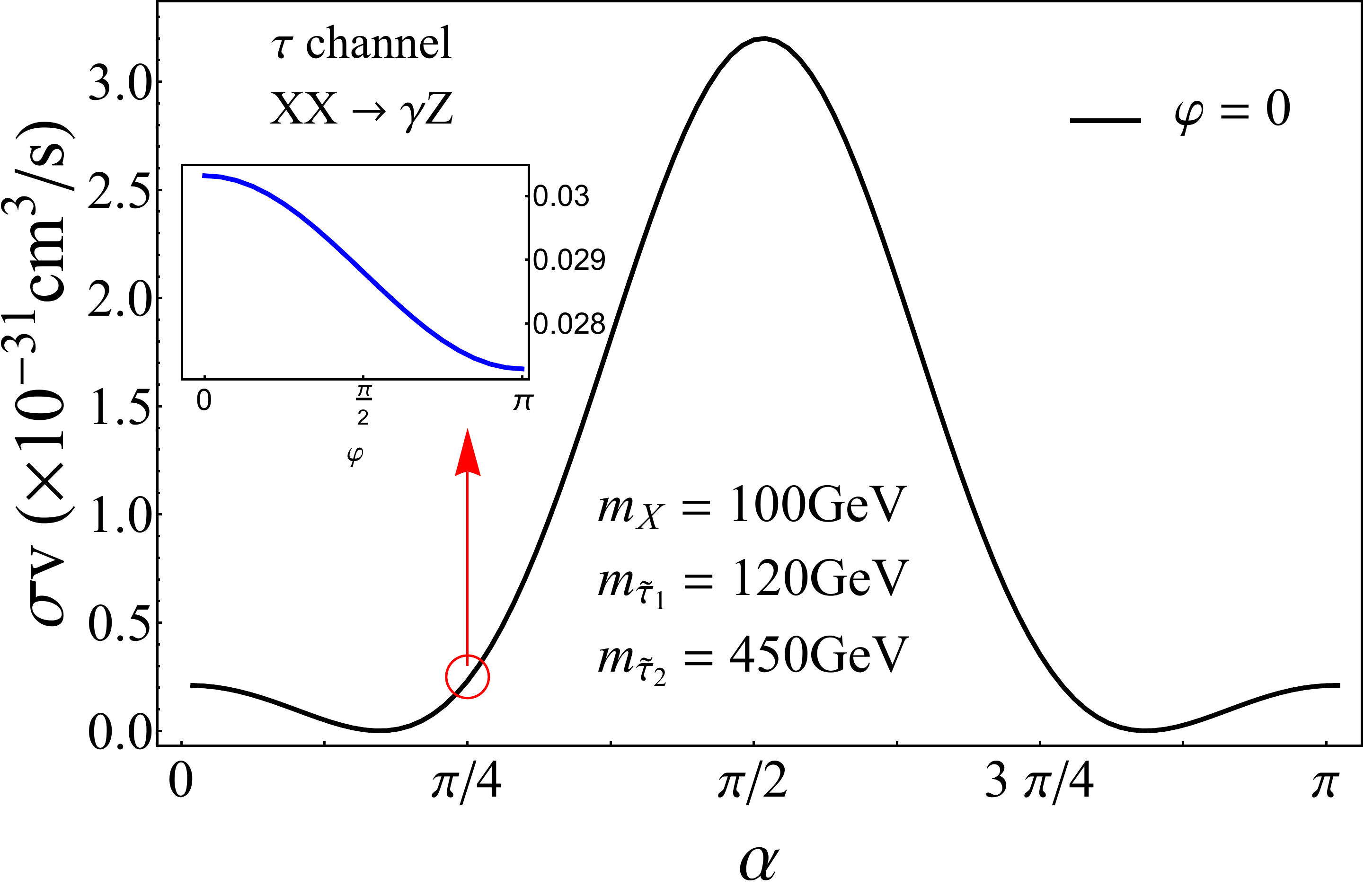}
\caption{$\left( \sigma v \right)_{\chi\chi \to \gamma \gamma}$ (left) and $\left( \sigma v \right)_{\chi \chi \to \gamma Z}$ (right) dependence on $\alpha$ and $\varphi$ for SUSY bino dark matter with coupling only to $\tau$ and $\widetilde{\tau}$. 
}
\label{fig:XXAZtau}
\end{figure*}

\subsection{SUSY case}
We first consider the SUSY case, as in Ref.~\cite{Fukushima14}, where $f$ and $\widetilde{f}$ are SM leptons and MSSM sleptons, therefore denoted as $\ell$ and $\widetilde{\ell}$ in this subsection. In Fig.~\ref{fig:XXAZtau}, we display the cross sections $(\sigma v)_{\gamma \gamma}$ (left) and $(\sigma v)_{\gamma Z}$ (right) as functions of $\alpha$ for $\varphi=0$, $\pi/2$, and $\pi$ .  Since $(\sigma v)_{\gamma Z}$ is only mildly sensitive to the $CP$-violating phase, $\varphi$, we show the cross section as a function of $\varphi$ for $\alpha = \pi/4$ in the inset of the right panel. Turning first to the left panel, we see an increase of $(\sigma v)_{\gamma \gamma}$ by a factor of $6$ as $\alpha$ ranges from zero to $\pi/4$ at $\varphi=\pi/2$ (and as much as a factor of $\gtrsim 10$ over the full range of $\alpha$ shown).
For the $\tau$ channel, displayed in Fig.~\ref{fig:XXAZtau}, the dependence on $\varphi$ is significant (in contrast to the $\mu$ channel): At $\alpha \approx \pi/4$, $(\sigma v)_{\gamma\gamma}$ varies by a factor of 2 as $\varphi$ ranges from zero to $\pi/2$. Turning to the right panel, we see that there is an increase in $(\sigma v)_{\gamma Z}$ by about a factor of $16$ for $\alpha=\pi/2$ relative to $\alpha=0$. This arises from the fact that $Y_R =2 Y_L$: for $\alpha = \pi/2$ ($\alpha=0$), the lighter scalar mass eigenstate consists entirely of the right-handed (left-handed) component, the contribution to the cross section of which is proportional to $Y_{R}^{4}$ ($Y_{L}^{4}$). This enhancement is possible only when the two-body annihilation cross section is suppressed ($\alpha = n\pi/2$ for $n$ odd), and thus the relic abundance of binos is too large. If another mechanism, such as coannihilation, helped to lower the relic abundance, or if dark matter were nonthermal, it may be possible for the line signal to be much larger than that suggested by the benchmark points. As we see in the right panel of Fig.~\ref{fig:XXAZtau}, the dependence of $(\sigma v)_{\gamma Z}$ on the $CP$-violating phase is not as significant as it is for annihilation to $\gamma \gamma$, even for the $\tau$ channel.

\begin{figure*}
\centering
\subfloat[$\mu$ channel]{\includegraphics[width=0.47\textwidth]{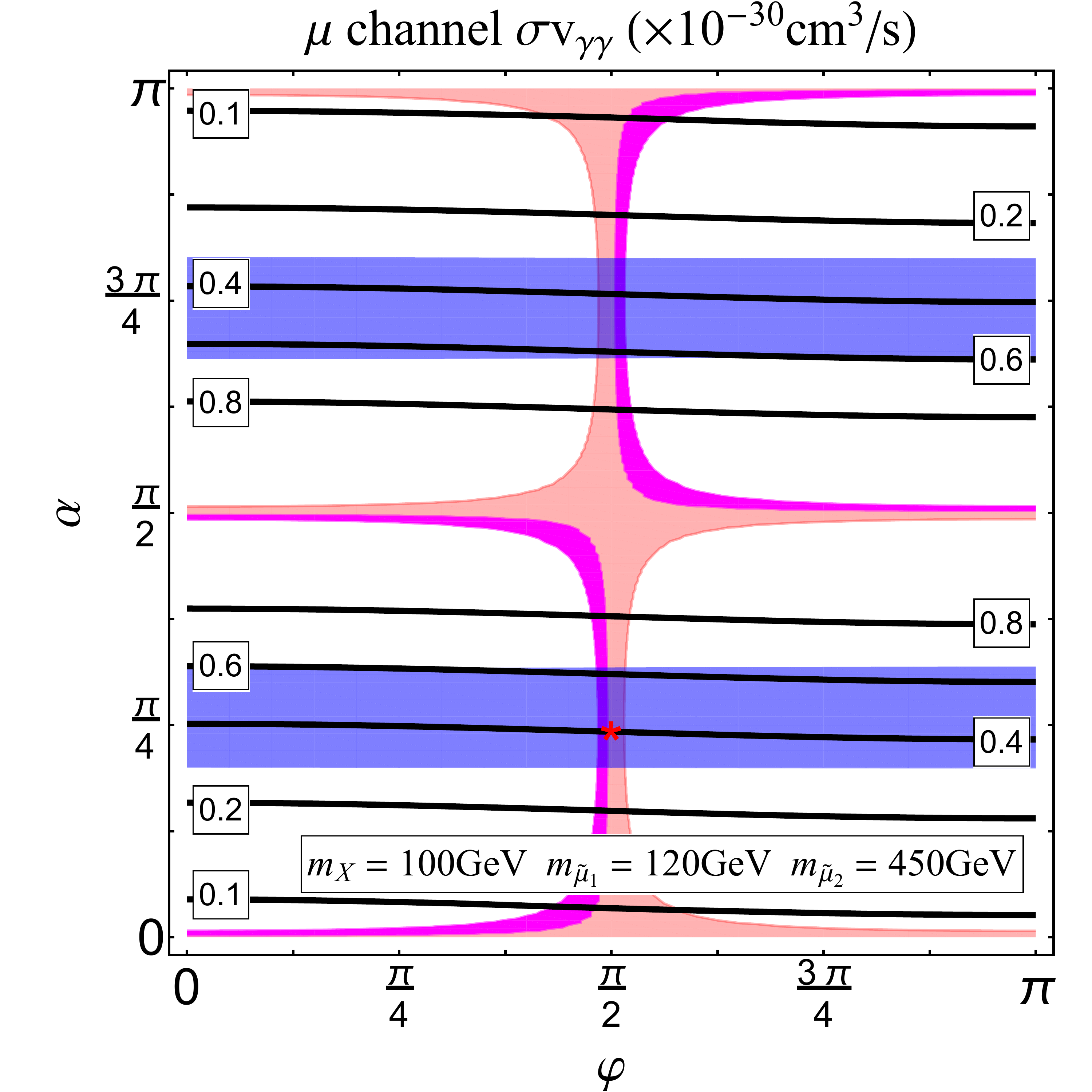}}\quad
\subfloat[$\tau$ channel]{\includegraphics[width=0.47\textwidth]{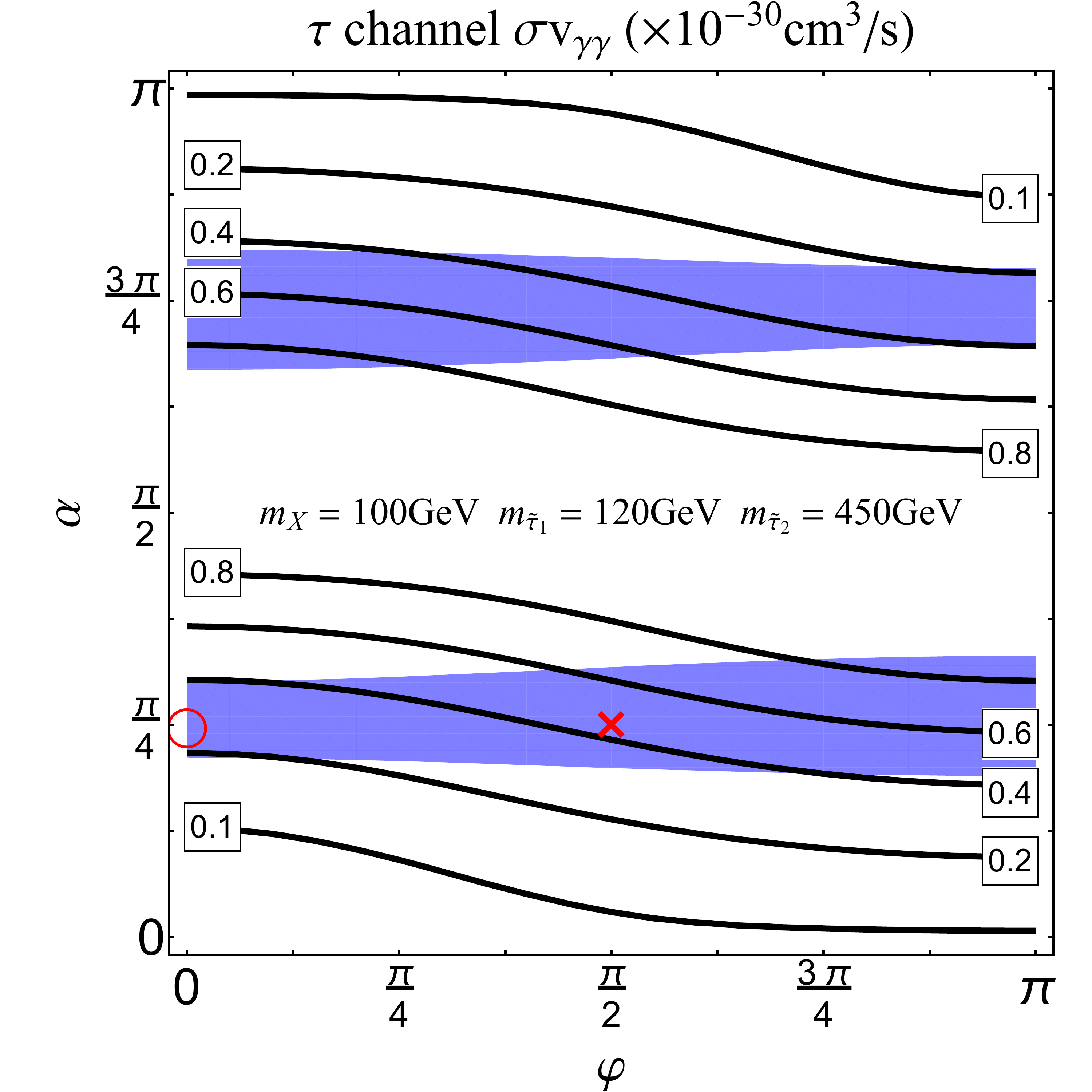}}
\caption{\label{fig:XXAA}
The dependence of $(\sigma v)_{\gamma\gamma}$ on the slepton mixing angle, $\alpha$, and the $CP$-violating phase, $\varphi$, for the $\mu$ channel (left panel) and $\tau$ channel (right panel) for the SUSY case $\lambda_L = 2\lambda_R$. In each plot, the blue stripe indicates the region that satisfies $0.11<\Omega h^{2}<0.13$. In the $\mu$ channel plot (left), the light magenta region of our parameter space leads to $128\times 10^{-11}<a_{\mu}<448\times 10^{-11}$, which resolves the issue of the muon anomalous dipole moment. In the light red region, we have instead $-448\times 10^{-11}<a_{\mu}<128\times 10^{-11}$, which neither solves nor exacerbates the discrepancy between the observed muon anomalous magnetic moment and the SM expectation. For the $\tau$ channel, the dipole moment measurements do not constrain the parameter space. The red markers (star, circle, and cross) indicate the positions of our benchmark models (A, B, and C, respectively).}
\end{figure*}

In Fig.~\ref{fig:XXAA}, we display a contour plot of $(\sigma v)_{\gamma\gamma}$ with respect to the chiral mixing, $\alpha$, and $CP$-violating phase, $\varphi$. The regions of parameter space in which the dark matter is a thermal relic are shaded blue, and, for the $\mu$ channel, the regions compatible with the measurement of the muon anomalous magnetic moment are shaded red/magenta (for the $\tau$ channel, the dipole moment measurements do not constrain the parameter space). Benchmarks A, B, and C are also marked.

Unfortunately for the SUSY case, these monochromatic photon signals lie well below the current experimental sensitivity.  Nonetheless, it is worth considering the possibility of an eventual detection.  As discussed in Sec.~\ref{sec:monochromatic}, once a statistical excess of these line signals is observed, and if the dark matter mass lies in the range $m_Z < m_X \lesssim  140\gev$ (for Fermi-LAT) or $\lesssim 450\gev$ (for GAMMA-400 or HERD), the ratio of the dark matter annihilation cross section into $\gamma \gamma$ and $\gamma Z$ will be of significant interest for determining the nature of the dark matter particle and the theory of physics beyond the SM in which it resides.  Indeed, this ratio does not suffer from astrophysical uncertainties in the dark matter distribution in our Galaxy~\cite{Bergstrom97}.  In Ref.~\cite{Yaguna09}, a wide range of MSSM parameter space is examined, and an attempt is made to use the ratio of $2(\sigma v)_{\gamma\gamma}/(\sigma v)_{\gamma Z}$ to distinguish among coannihilation, funnel, and focus point scenarios in mSUGRA,  as well as within more general MSSM scenarios.

\begin{figure*}
\centering
\captionsetup[subfigure]{labelformat=empty}
\subfloat[]{%
    \includegraphics[width=0.46\textwidth]{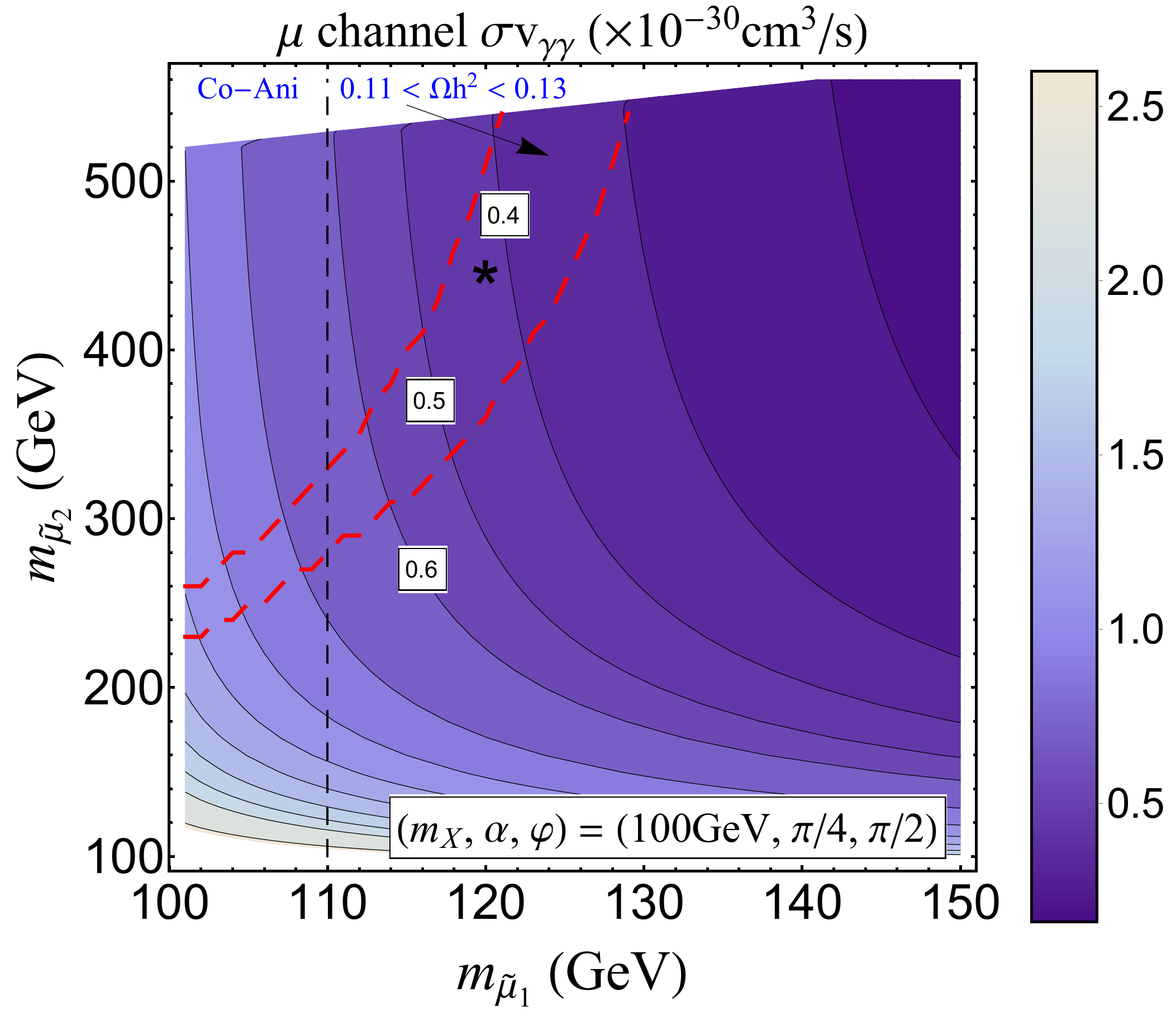}%
}
\subfloat[]{%
    \includegraphics[width=0.46\textwidth]{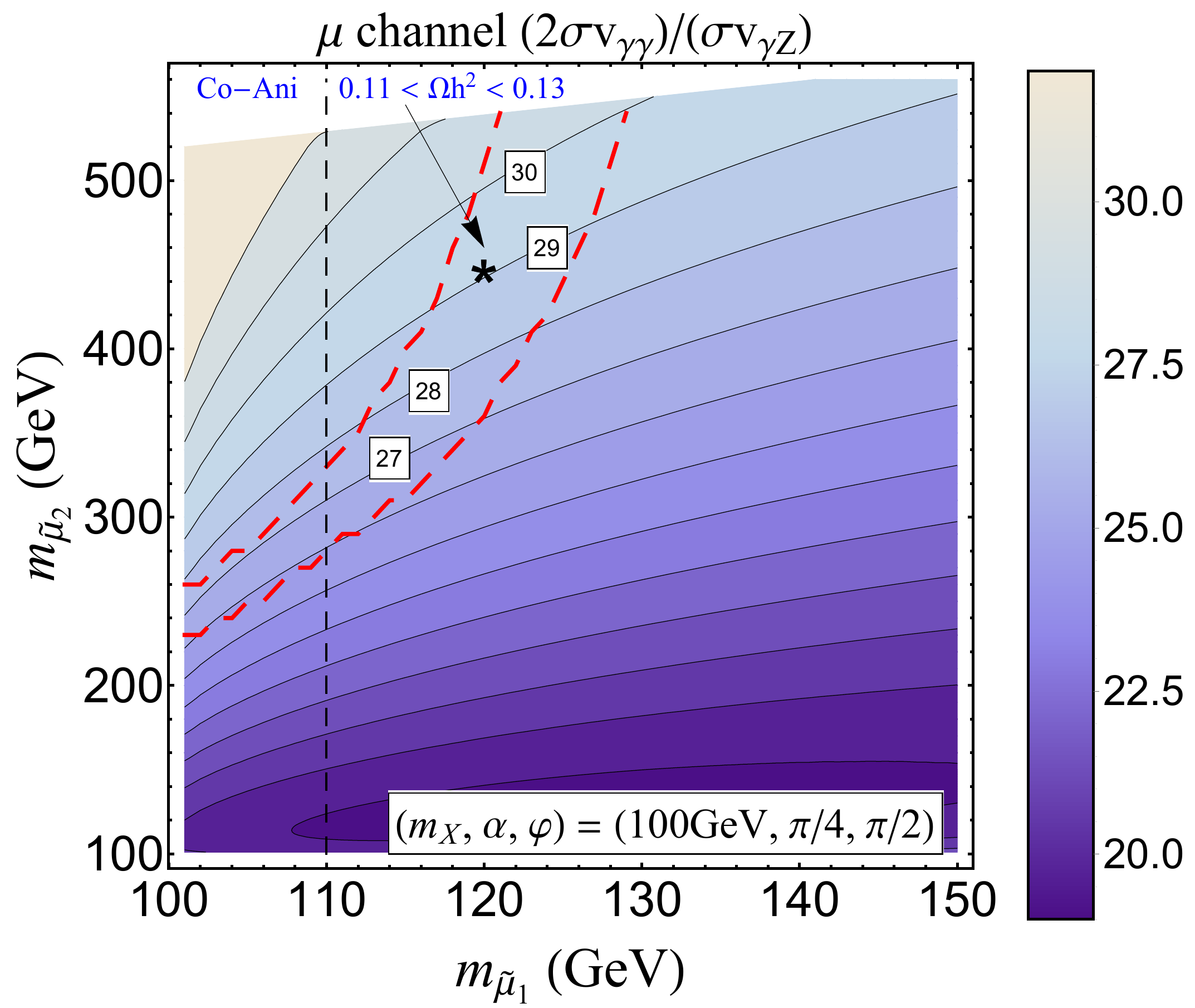}%
}
\vspace{-2.55em}
\subfloat[]{%
    \includegraphics[width=0.46\textwidth]{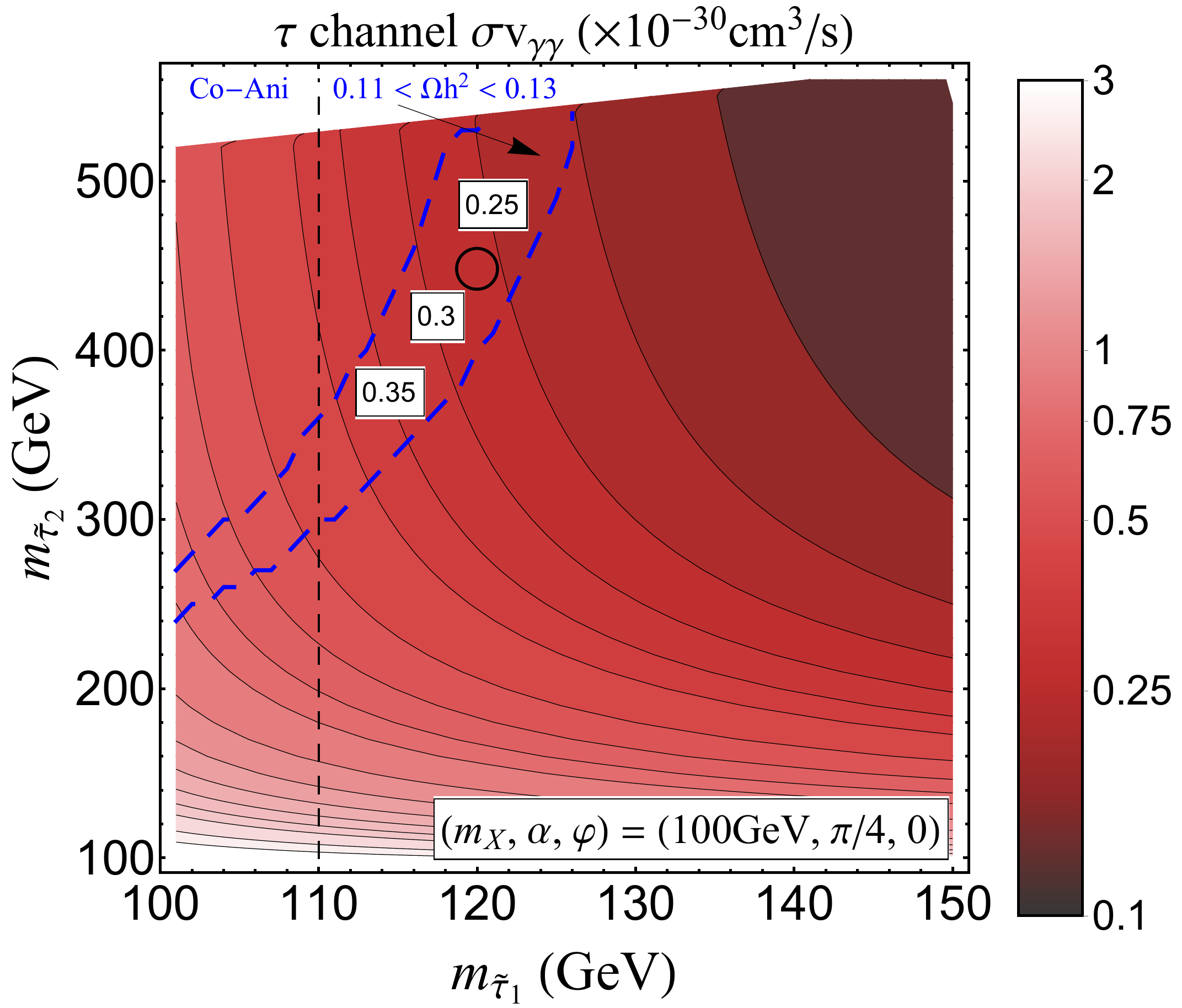}%
}
\subfloat[]{%
    \includegraphics[width=0.46\textwidth]{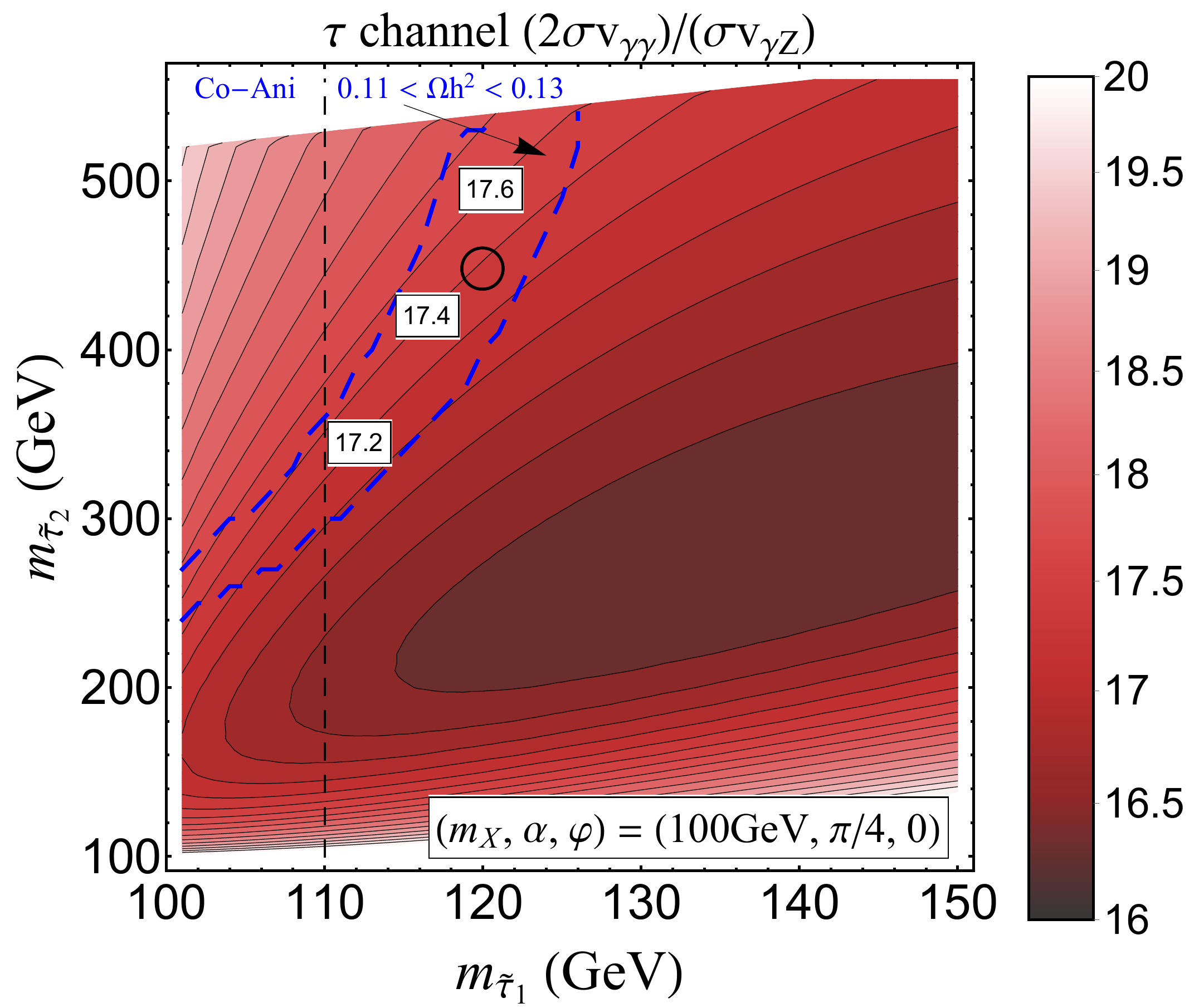}%
}
\vspace{-2.55em}
\subfloat[]{%
    \includegraphics[width=0.46\textwidth]{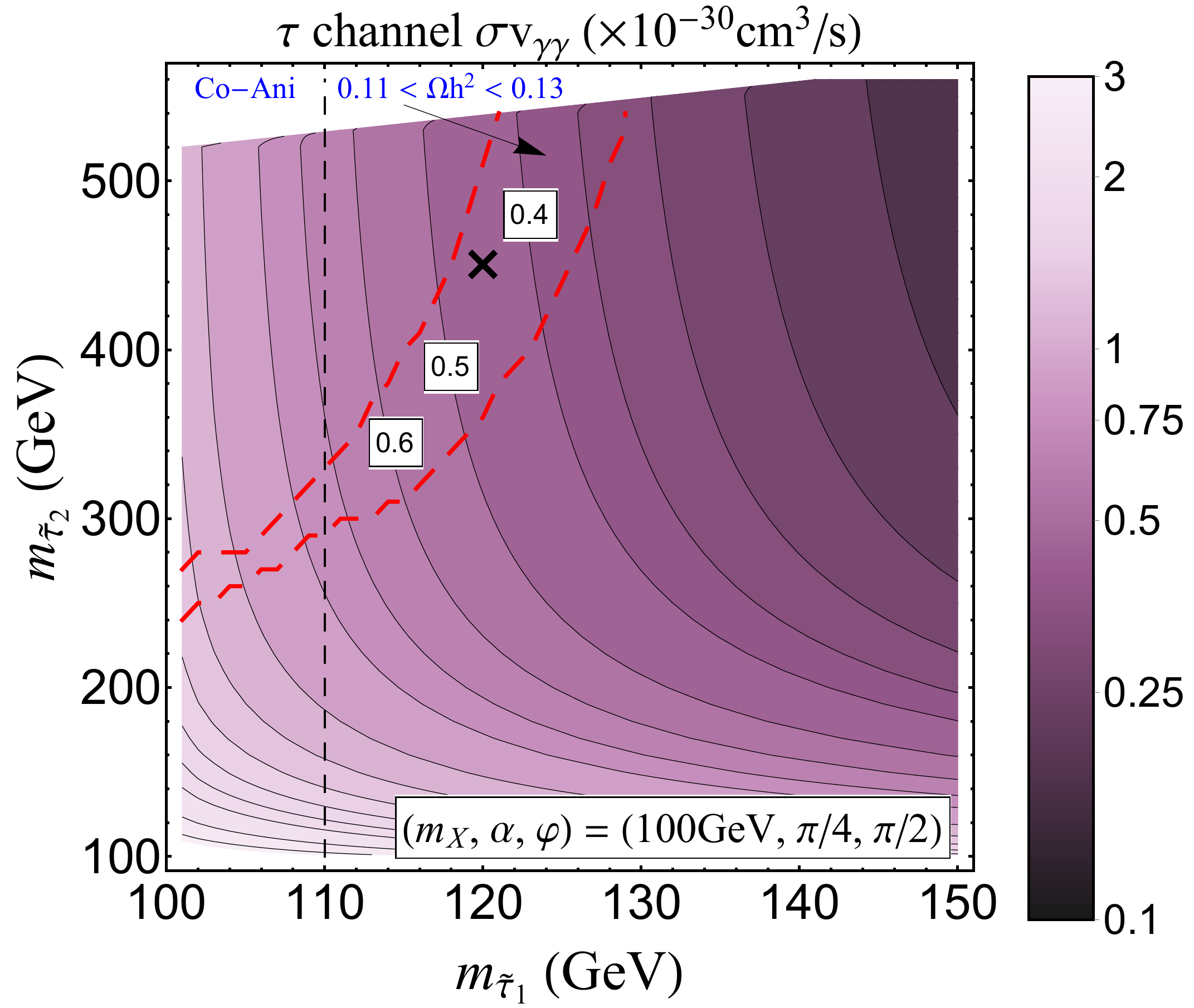}%
}
\subfloat[]{%
\includegraphics[width=0.46\textwidth]{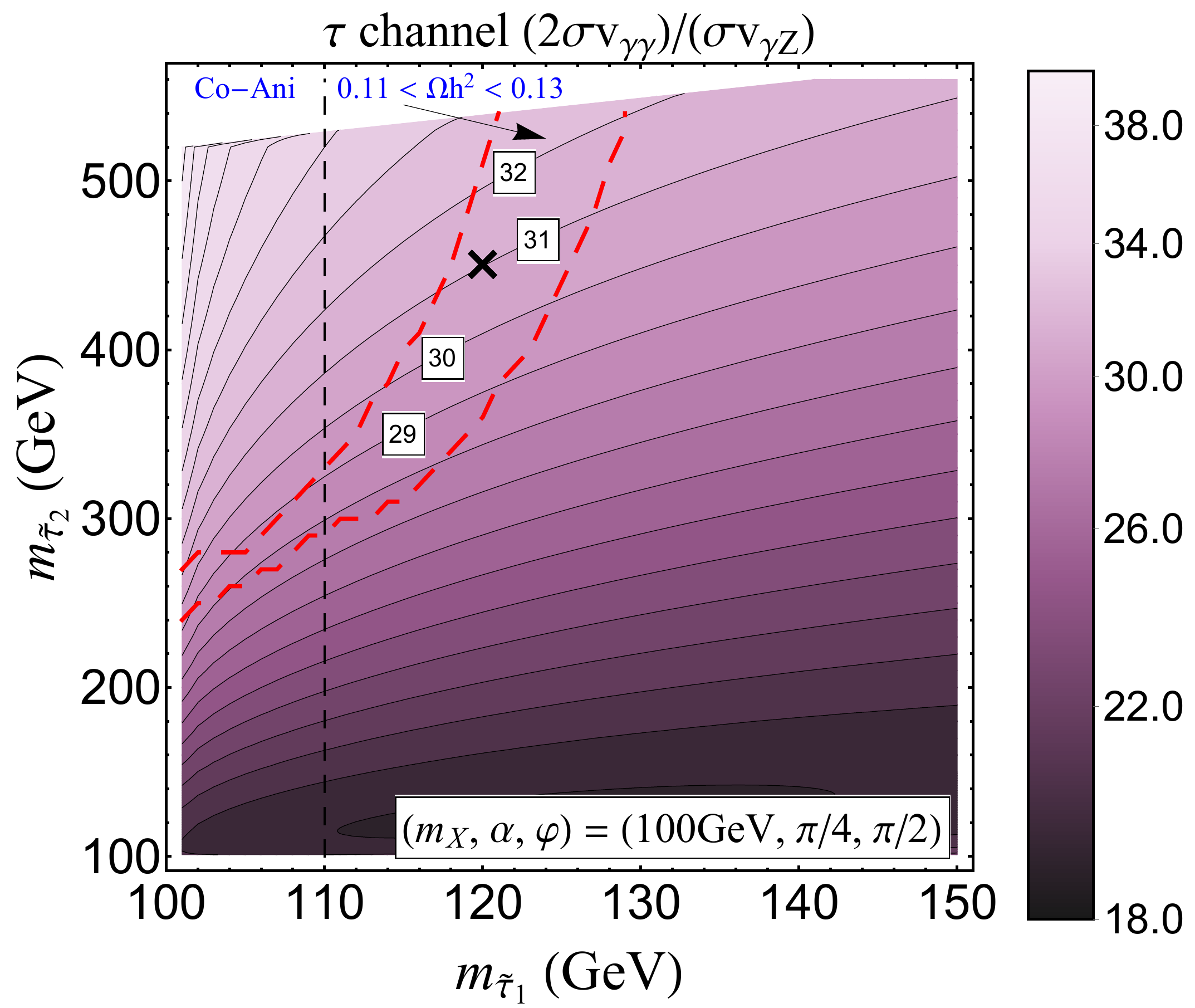}%
}
\vspace{-2em}
\caption{\label{fig:ratio}The dependence of the $XX\rightarrow\gamma\gamma$ cross section (left) and the ratio $2(\sigma v)_{\gamma\gamma}/(\sigma v)_{\gamma Z}$ (right) on the slepton masses for the SUSY case, $\lambda_L = 2\lambda_R$. The black markers in each plane indicate the positions of our benchmark points. Note that in three of the plots we have used a log-scaled color function.}
\end{figure*}

Following the same line of analysis, we plot the annihilation cross section to $\gamma \gamma$ (left) and the ratio $2(\sigma v)_{\gamma\gamma}/(\sigma v)_{\gamma Z}$ (right) as functions of the slepton masses in Fig.~\ref{fig:ratio} for the $\mu$ (top) and $\tau$ (middle and bottom) channels for $m_{X} = 100\gev$ and $\alpha=\pi/4$.  In the top panels, we display the $\mu$ channel
with $\varphi = \pi/2$ and in the middle and lower panels, we display the tau channel with $\varphi = 0$ and $\pi / 2$, respectively. In the left panels, we display the $XX\rightarrow\gamma\gamma$ cross sections in units of $10^{-30} ~\textrm{cm}^3~ \textrm{s}^{-1}$, while in the right panels we show the ratio $2(\sigma v)_{\gamma\gamma}/(\sigma v)_{\gamma Z}$.
The parameter space that accommodates thermal relic dark matter lies between the thick dashed contours that cut diagonally across each plane.
As expected, $(\sigma v)_{\gamma\gamma}$ decreases as the slepton masses increase. Similarly, the ratio $2(\sigma v)_{\gamma\gamma}/(\sigma v)_{\gamma Z}$ increases as the difference between $m_{\tilde \ell_1}$ and $m_{\tilde \ell_2}$ increases.
$2(\sigma v)_{\gamma\gamma}/(\sigma v)_{\gamma Z}$
is larger than 17.0 for $\varphi = 0$ in the $\tau$ channel. Although not presented in Fig.~\ref{fig:ratio}, this approximately holds true for the $\mu$ channel as well.
In the case of $\varphi = \pi/2$, $2(\sigma v)_{\gamma\gamma}/(\sigma v)_{\gamma Z}$ is greater than $26$ for the $\mu$ channel and greater than $28$ for the $\tau$ channel. By contrast, in the coannihilation region in minimal supergravity (mSUGRA), $2(\sigma v)_{\gamma\gamma}/(\sigma v)_{\gamma Z}$ ranges from $7-12$ (see Fig.~5 in Ref.~\cite{Yaguna09}).  It is therefore possible that if both the $\gamma \gamma$ and  $\gamma Z$ lines are observable, the ratio of the signal strengths could be used to distinguish between, for example, the coannihilation region and a model similar to the Incredible Bulk.
Though these scenarios could, in principle, also be distinguished by the cosmic-ray signal arising from $XX \rightarrow \ell^+ \ell^-$, such a signal would be subject to astrophysical uncertainties and would therefore leave much room for doubt.

In summary, the $XX\rightarrow\gamma\gamma$ cross sections increase by a factor of $\sim2$ as $\varphi$ varies from 0 to $\pi/2$, as does the the ratio $2(\sigma v)_{\gamma\gamma}/(\sigma v)_{\gamma Z}$ (since the $XX\rightarrow\gamma Z$ cross section is insensitive to the value of $\varphi$). As the slepton masses increase, $(\sigma v)_{\gamma\gamma}$ becomes smaller, while $2(\sigma v)_{\gamma\gamma}/(\sigma v)_{\gamma Z}$ increases as the difference between $m_{\tilde \ell_1}$ and $m_{\tilde \ell_2}$ increases.

\subsection{Beyond the MSSM}

There are two other scenarios we consider, beyond the Incredible Bulk scenario of binolike dark matter in the MSSM,
in which the monoenergetic gamma-ray line signals are particularly interesting: $f=\mu$ (benchmark $D$), and $m_f > m_X$ (benchmark $E$), each with arbitrary but perturbative couplings.  If
$m_f > m_X$, as in benchmark $E$, the processes $XX \rightarrow \bar f f (\gamma)$ are kinematically forbidden, and the processes
$XX \rightarrow \gamma \gamma, \gamma Z$ will be the most important for indirect detection.

If $f=\mu$, as in benchmark $D$, then
the process $XX \rightarrow \mu^+ \mu^-$ produces few photons or antiprotons through final state decay.  As a result, the
$2 \rightarrow 2$ cross section is constrained only by positron searches and dipole moment constraints.
Tight constraints on the $XX \rightarrow \mu^+ \mu^-$ cross section have been presented in the literature based on AMS-02 positron
searches~\cite{Bergstrom:2013jra}, which would require $(\sigma v)_{\mu^+ \mu-} \lesssim 1~\pb$.
But these analyses have relatively large systematic uncertainties arising from assumptions about the sources of astrophysical backgrounds
and propagation effects~\cite{DiMauro:2015jxa}.
A full discussion of these issues is beyond the scope of this work, but it suffices to note that gamma-ray signal arising
from the process $XX \rightarrow \gamma \gamma, \gamma Z$ is much cleaner than the positron signal arising from the process $XX \rightarrow \mu^+ \mu^-$,
particularly since the gamma-ray signal can point back to sources which are well understood, such as dwarf spheroidals.
As a result, these gamma-ray signals are of interest even for large $\alpha$ and $\lambda_{L,R}$, where the $XX \rightarrow \mu^+ \mu^-$ cross section
would be in tension with analyses of AMS-02 electron flux data.  Note, however, that this rationale would be less compelling in the case
where $f=\tau$, as in this case, the process $XX \rightarrow \tau^+ \tau^-$ can produce gamma-ray signals from dwarf spheroidals, arising
from hadronic $\tau^\pm$ decay.

\begin{figure*}
\centering
{\includegraphics[width=0.47\textwidth]{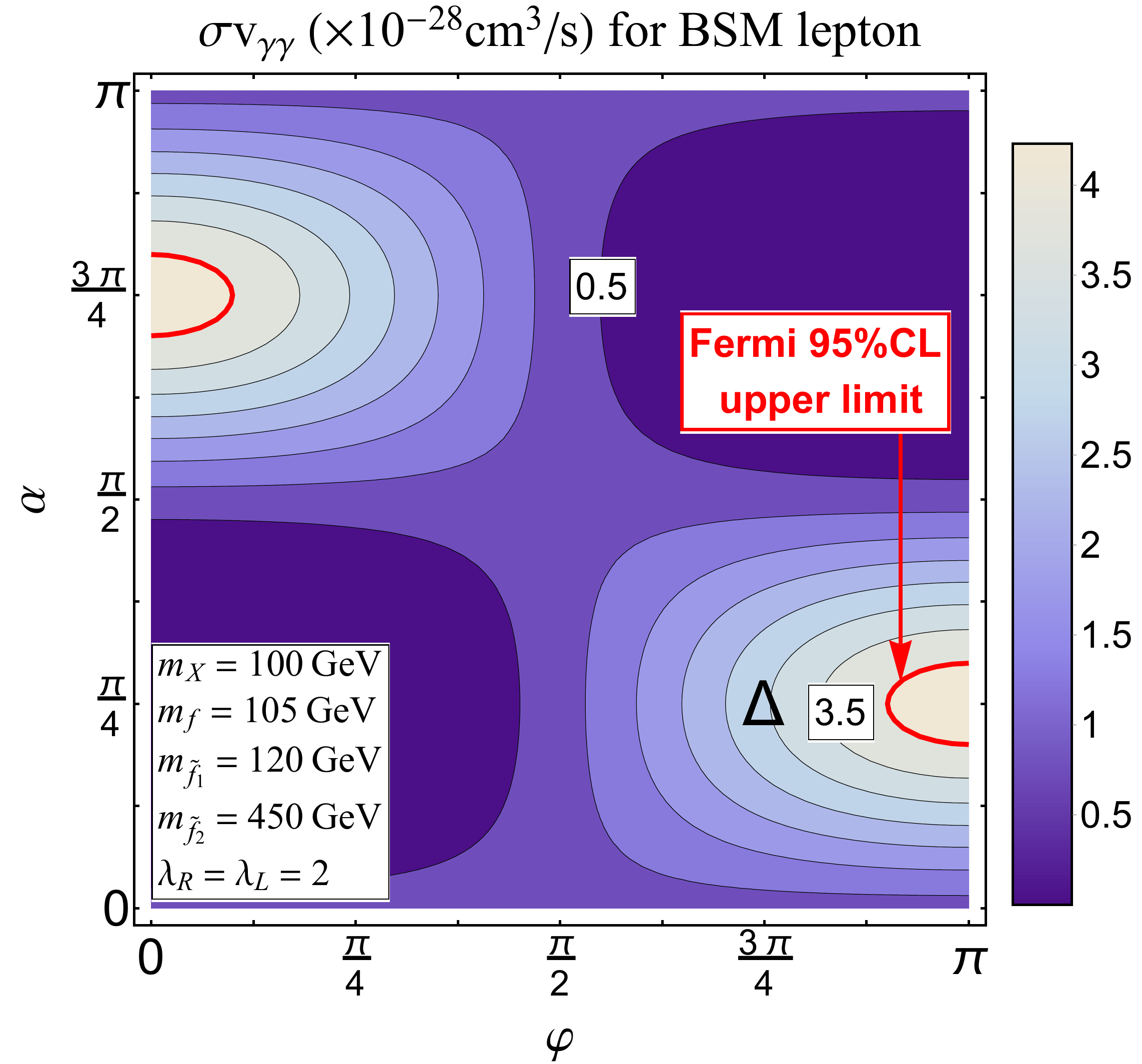}}\hspace{3mm}
\caption{\label{fig:GGlargeMF}
The cross section for the process $XX \rightarrow \gamma \gamma$ with $m_{f}>m_{X}$. Benchmark $E$ is labeled by the triangle. Here and in the following figures, the Fermi line constraint is taken as $4\times 10^{-28}\text{cm}^{3}/\text{s}$ for $m_X=100$ GeV~\cite{Ackermann:2015lka}.}
\end{figure*}

{In Fig.~\ref{fig:GGlargeMF}, we show the cross section for the process $XX \rightarrow \gamma \gamma$ as a function of
$\alpha$ and $\varphi$ for benchmark $E$. %
We see that Fermi line searches \cite{Ackermann:2015lka}  tend to
constrain models with large left-right mixing and small $CP$ violation.}  Note that this is in contrast to the
case of $m_f < m_X$, where larger $CP$ violation tends to lead to a larger $\sigma (\gamma \gamma)$ cross section.
Future experiments with a larger effective area and/or energy resolution could improve on these sensitivities.

\begin{figure*}
\centering
\subfloat[$\lambda_{R}=2\lambda_{L}$]{\includegraphics[width=0.47\textwidth]{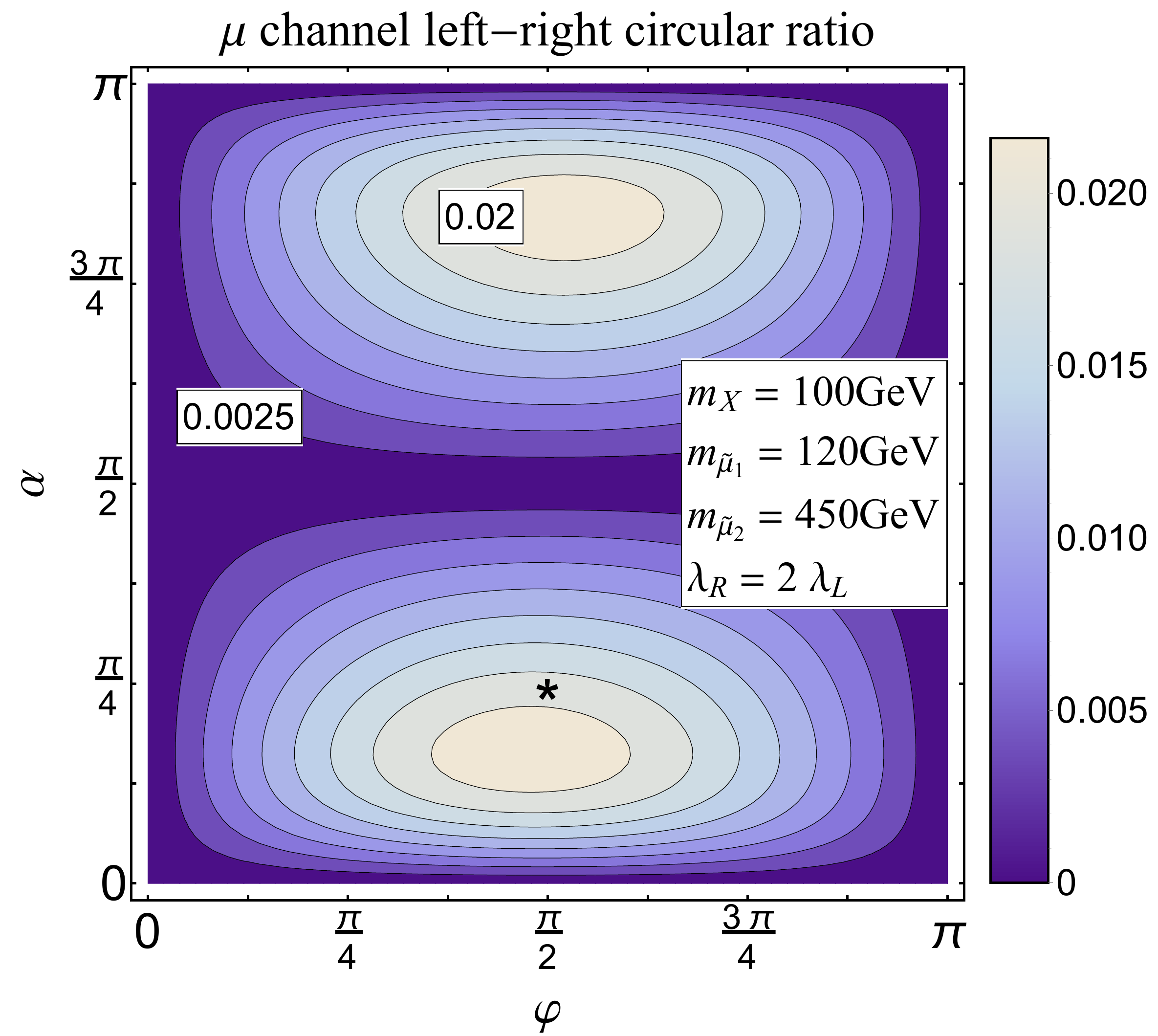}}\quad
\subfloat[$\lambda_{R}=\lambda_{L}$]{\includegraphics[width=0.47\textwidth]{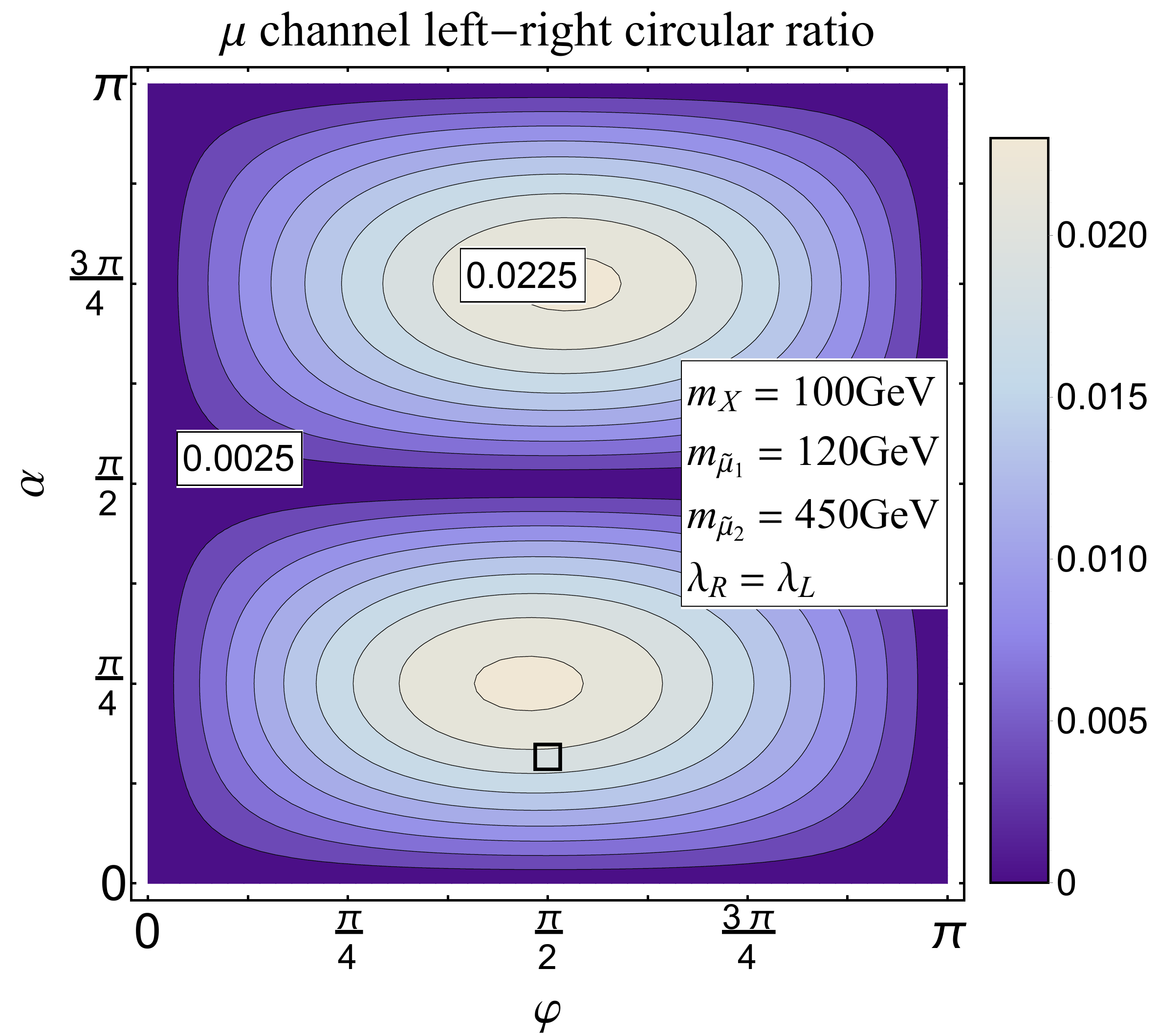}}
\caption{\label{fig:muasym}
The asymmetry ratio $R$ for the $\mu$ channel $XX\rightarrow\gamma\gamma$ process. The left panel shows $\lambda_{R}=2\lambda_{L}$, while the right panel shows $\lambda_{L}=\lambda_{R}$. Benchmarks $A$ and $D$ are labeled by the star and square, respectively.}
\end{figure*}

{In Figure~\ref{fig:muasym} we plot, for the $\mu$ channel, the asymmetry ratio \begin{equation}
    R = \frac{\sigma(++) - \sigma(--)}{\sigma(++)+\sigma(--)}\,,
\end{equation}
where $\sigma(\pm\pm)$ is the annihilation cross section with two positive or negative helicity final state photons.} 
Note that this ratio is independent of the common scaling of $\lambda_{L}$ and $\lambda_{R}$. As expected, this asymmetry
is maximized at large left-right mixing and the maximal $CP$-violating phase.  At its maximum, the asymmetry is
$\sim 2\%$, which is larger than one might naively expect from the $m_f / m_X$ suppression of the $CP$-violating
term in the matrix element.  This arises because the loop integral relevant for the $CP$-violating term happens to be
about an order of magnitude larger than the integral which is relevant for the $CP$-conserving term in the $m_f \ll m_X$
limit.
{If we have $m_{\widetilde{f}_{i}}>m_{f}>m_{X}$, as in benchmark $E$, then $R$ is identically zero, as expected from the optical theorem.  The detailed reason is that beyond
the branching point $m_{f}=m_{X}$, all the loop integrals are real, and the amplitudes of the $(++)$ and $(--)$ final states are conjugate with each other and lead to the
same cross section. See the Appendix for details.}

In Fig.~\ref{fig:tau_LRratio} we plot the asymmetry ratio $R$ for Benchmark $D$, except that we instead take
$f=\tau$.  In this case, as expected, the asymmetry is about an order of magnitude larger, because the
$m_f / m_X$ suppression factor is about a factor of 10 larger.  Note that the cross section asymmetry is linear
in this suppression factor, since it arises from the interference of the $CP$-conserving and $CP$-violating pieces.
Although the asymmetry is more pronounced in this case than in the case where the fermion is a muon, the couplings
$\lambda_{L,R}$ are also more tightly constrained in this case due to tight bounds on the process $XX \rightarrow
\tau^+ \tau^-$ arising from Fermi searches for the continuum photons from $\tau$-decay via a neutral pion \cite{Ackermann:2015zua,*Rico:2015nya}.

\begin{figure*}
\centering
{\includegraphics[width=0.47\textwidth]{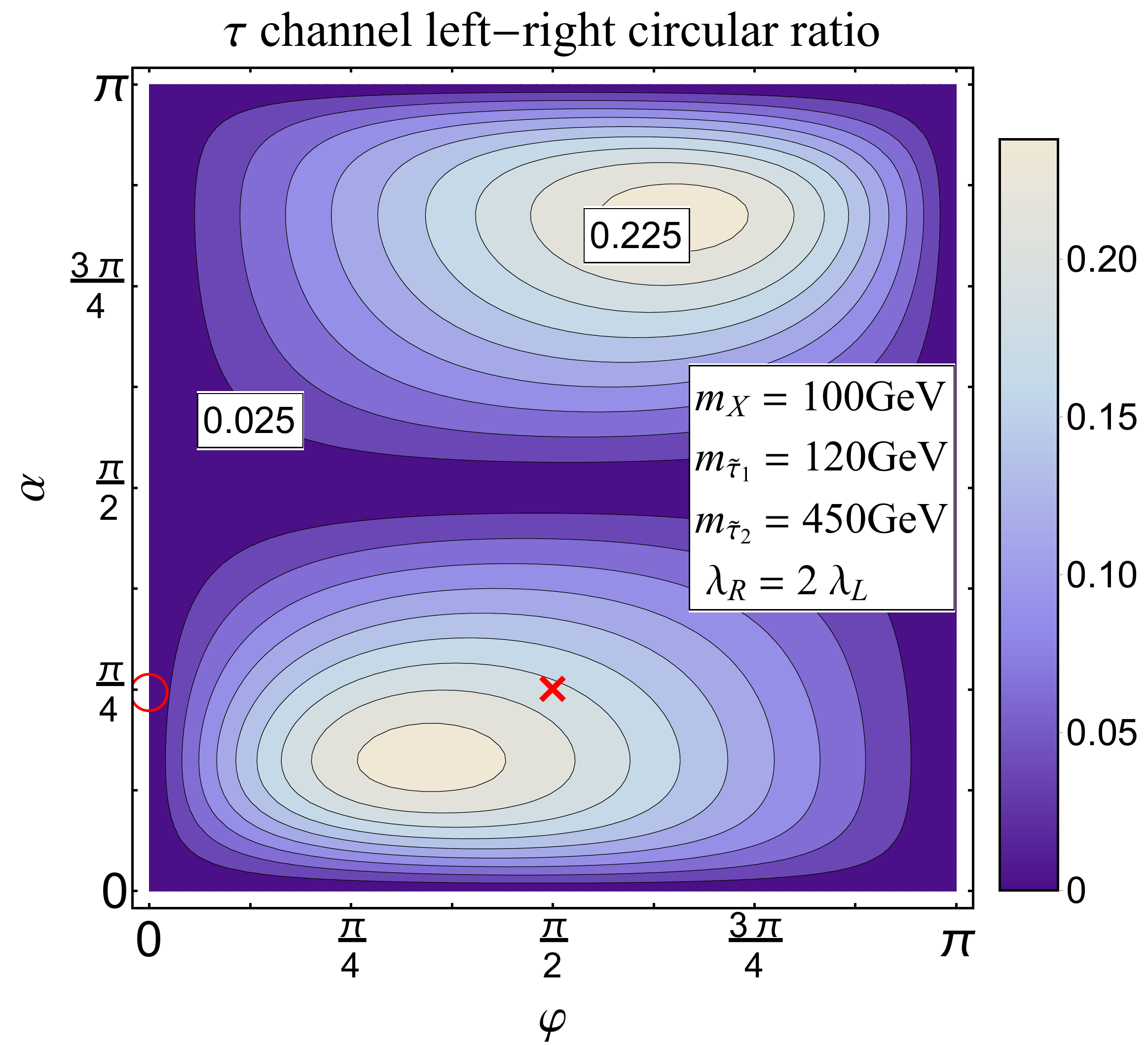}}
\caption{\label{fig:tau_LRratio}
{The asymmetry ratio $R$
as in Fig.~\ref{fig:muasym}, but for the $\tau$ channel, with $\lambda_{R}=2\lambda_{L}$. Benchmarks $B$ and $C$ are labeled by the circle and cross, respectively.}}
\end{figure*}

\section{\label{sec:IBcal} Internal Bremsstrahlung}

The IB process ${X}{X}\rightarrow \bar f f \gamma$ can in general produce a steplike feature in the photon spectrum at $E_{\gamma}\approx m_{{X}}$. 
Further study shows that IB can even lead to a significant peak at the end of the spectrum when the dark matter and $\widetilde{f_1}$ are nearly degenerate~\cite{Bringmann14}. We can separate the IB amplitude into two (not gauge invariant) parts as in Ref.~\cite{Bringmann08}: virtual internal bremsstrahlung (VIB) for the photon attached to the internal scalar propagator and final state radiation (FSR) for the photon attached to the external fermion lines.  The full IB amplitude can then be written
\begin{equation}
\label{eq:VIBFSR}
\mathcal{A}_{\text{IB}}=\mathcal{A}_{\text{VIB}}+\mathcal{A}_{\text{FSR}}\,.
\end{equation}
From $\mathcal{A}_{\text{IB}}$, the total differential cross section can be calculated by
\begin{equation}
    \frac{d(\sigma v)_{\text{IB}}}{dx}=\frac{x}{512\pi^{4}}\sqrt{1-\frac{m_{f}^{2}}{m_{X}^{2}(1-x)}}\int{d\Omega_{34}}\overline{\left|\mathcal{A}_{\text{IB}}\right|^{2}}\,,
\end{equation}
where $d\Omega_{34}$ is the integration over the fermions' direction in the fermion pair center-of-mass frame, and $\overline{\left|\mathcal{A}_{\text{IB}}\right|^{2}}$ is the squared amplitude with initial spin averaged and final spin summed.

If there is no mixing between the scalars associated with the left- and right-handed fermions, there may be a hard feature at the end of the IB spectrum that comes from the VIB. However, in the presence of scalar mixing, FSR will introduce another $s$-wave amplitude that dominates over VIB at high energies. In particular, it comes mainly from the collinear limit of the FSR photon, and the total IB spectrum is fairly flat relative to the case with no mixing. To illustrate this point, we rewrite the total IB amplitude $\mathcal{A}_{\text{IB}}$ in
terms of three gauge invariant subamplitudes,
\begin{equation}
\mathcal{A}_{\text{IB}}=\frac{ie}{2}\left(\frac{\overline{u}(k_1)\gamma^{5}v(k_2)}{2m_{{X}}}\right)\left(\mathcal{A}_{\text{vb}}+\mathcal{A}_{\text{mix}}+\mathcal{A}_{m_f}\right)\,,
\end{equation}
where again $k_{1,2}$ are the momenta of the two dark matter particles.
The first term, $\mathcal{A}_{\text{vb}}$, is the intrinsic $s$-wave amplitude, which survives in both the massless fermion limit ($m_{f}\rightarrow 0$) and the no-mixing limit ($\alpha=0$).  This term is the amplitude for the production of a fermion and antifermion with opposite helicities, arising from the same Weyl spinor, with the
remaining angular momentum carried by the vector boson.
If we denote the photon momentum and polarization as $k_{5}$ and $\epsilon_{5}$, and the outgoing fermion (antifermion)
momentum by $k_3$ ($k_4$), this amplitude can be written as
\begin{align}
\mathcal{A}_{\text{vb}}&=\overline{u}(k_3)\,\mathcal{O}_{1}(|\lambda_{L}|^{2}\cos^{2}\alpha P_{L}-|\lambda_{R}|^{2}\sin^{2}\alpha P_{R})v(k_4)\nonumber\\*
&\quad+\overline{u}(k_3)\,\mathcal{O}_{2}(|\lambda_{L}|^{2}\sin^{2}\alpha P_{L}-|\lambda_{R}|^{2}\cos^{2}\alpha P_{R})v(k_4)\,.
\end{align}
The matrices $\mathcal{O}_{i}$ are given by
\begin{equation}
\mathcal{O}_{i} \equiv \gamma_\mu \left[ \frac{k_{5}^\mu (k_{3}-k_{4})\cdot\epsilon_{5}- \epsilon_{5}^\mu(k_{3}-k_{4})\cdot k_{5}}{(s_{3}-m^{2}_{\widetilde{{f}_{i}}})(s_{4}-m^{2}_{\widetilde{{f}_{i}}})} \right],
\end{equation}
with $s_{3} \equiv (k-k_{3})^{2}$ and $s_{4} \equiv (k-k_{4})^{2}$. When $m_{f}=0$, the cross section due solely to $\mathcal{A}_{\text{vb}}$ is
\begin{align}
\frac{d(\sigma v)_{\text{vb}}}{dx}&=\sum_{i=1,2}\frac{\alpha_{\text{em}}\lambda_{i}^{4}(1-x)}{64\pi^{2}m^{2}_{X}}\nonumber\\*
&\times\left[\frac{4x}{(1+\mu_{i})(1+\mu_{i}-2x)}-\frac{2x}{(1+\mu_{i}-x)^{2}}\right.\nonumber\\*
&\quad\left.-\frac{(1+\mu_{i})(1+\mu_{i}-2x)}{(1+\mu_{i}-x)^{3}}\log\frac{1+\mu_{i}}{1+\mu_{i}-2x}\right],
\end{align}
where
\begin{align*}
    &\lambda_{1}^{2}\equiv|\lambda_{L}|^{2}\cos^{2}\alpha-|\lambda_{R}|^{2}\sin^{2}\alpha\,,\\
    &\lambda_{2}^{2}\equiv|\lambda_{L}|^{2}\sin^{2}\alpha-|\lambda_{R}|^{2}\cos^{2}\alpha\,,
\end{align*}
$\mu_{i} \equiv m^{2}_{\widetilde{f}_{i}}/m^{2}_{X}$, and $x \equiv E_{\gamma}/m_{X}$ is the photon energy fraction. 
In the limit $\alpha=0$, we have $\mathcal{A}_{\text{IB}}\sim\mathcal{A}_{\text{vb}}$, and we recover the well-known result given, for example, in Ref.~\cite{Bringmann08,Bergstrom89}. Note that if $\mu_i \sim 1$, the photon spectrum becomes very hard, due the enhancement as
$x \rightarrow 1$.  This enhancement arises in the limit where an outgoing fermion becomes soft; if the dark matter
and the scalar are nearly degenerate, then one intermediate scalar propagator goes on shell. The total cross section in the
$\alpha, m_f/m_X \rightarrow 0$ limit is finite:
\begin{widetext}
\begin{align}
\label{eq:sigmavb}
(\sigma v)_{\text{vb}}=\sum_{i=1,2}\frac{\alpha_{\text{em}}\lambda_{i}^{4}}{64\pi^{2}m_{X}^{2}}&\left\{(\mu_{i}+1)\left[\frac{\pi^{2}}{6}-\log^{2}\left(\frac{\mu_{i}+1}{2\mu_{i}}\right)-2\text{Li}_{2}\left(\frac{\mu_{i}+1}{2\mu_{i}}\right)\right]\right.\nonumber\\*
&\quad\left.+\frac{4\mu_{i}+3}{\mu_{i}+1}+\frac{4\mu_{i}^{2}-3\mu_{i}-1}{2\mu_{i}}\log\left(\frac{\mu_{i}-1}{\mu_{i}+1}\right)\right\}\,.
\end{align}
\end{widetext}
If $\mu_{i}-1\ll 1$, the combination in the curly brackets approaches a constant, $(7/2)-(\pi^{2}/3)$; if $\mu_{i}\gg 1$, it behaves as $4/(15\mu^{4}_{i})$.

When $\alpha\neq 0$, there is another contribution to the $s$-wave amplitude arising from mixing,
\begin{align}
\mathcal{A}_{\text{mix}}&=m_{X}|\lambda_{L} \lambda_{R}|\sin(2\alpha)\nonumber\\*
&\quad\times\left[\cos\varphi\,\overline{u}(k_3)\gamma^{5}(\mathcal{V}_{1}+\mathcal{S}_{1}-\mathcal{V}_{2}-\mathcal{S}_{2})v(k_4)\right.\nonumber\\*
&\quad\quad\left.-i\sin\varphi\,\overline{u}(k_3)(\mathcal{V}_{1}+\mathcal{S}_{1}-\mathcal{V}_{2}-\mathcal{S}_{2})v(k_4)\right],
\end{align}
where the matrices $\mathcal{V}_{i}$ and $\mathcal{S}_{i}$ are given by
\begin{align}
\mathcal{V}_{i}
&\equiv
-\frac{i}{2} \sigma_{\mu \nu} k_5^\mu \epsilon^\nu
\left[ \frac{1}{(k_{3}\cdot k_{5})(s_{4}-m^{2}_{\widetilde{{f}_{i}}})}+\frac{1}{(k_{4}\cdot
k_{5})(s_{3}-m^{2}_{\widetilde{{f}_{i}}})} \right] \nonumber\\*
\mathcal{S}_{i}&\equiv \frac{(k_{3}-k_{4})\cdot\epsilon_{5}}{(s_{3}-m^{2}_{\widetilde{{f}_{i}}})(s_{4}-m^{2}_{\widetilde{{f}_{i}}})}\nonumber\\*
&\quad+\left[\frac{k_{3}\cdot\epsilon_{5}}{(k_{3}\cdot k_{5})(s_{4}-m^{2}_{\widetilde{{f}_{i}}})}-\frac{k_{4}\cdot\epsilon_{5}}{(k_{4}\cdot k_{5})(s_{3}-m^{2}_{\widetilde{{f}_{i}}})}\right]\,.
\end{align}

The last piece, $\mathcal{A}_{m_f}$, is proportional to the fermion mass, $m_{f}$,
\begin{align}
\mathcal{A}_{m_f}&=-m_{f}(|\lambda_{L}|^{2}\cos^{2}\alpha+|\lambda_{{R}}|^{2}\sin^{2}\alpha)\,\overline{u}(k_3)\gamma^{5}(\mathcal{V}_{1}+\mathcal{S}_{1})v(k_4)\nonumber\\*
&\quad-m_{f}(|\lambda_{L}|^{2}\sin^{2}\alpha+|\lambda_{{R}}|^{2}\cos^{2}\alpha)\,\overline{u}(k_3)\gamma^{5}(\mathcal{V}_{2}+\mathcal{S}_{2})v(k_4)\,.
\end{align}
Both $\mathcal{A}_{\text{mix}}$ and $\mathcal{A}_{m_f}$ are contributions to the amplitude for producing a fermion
and antifermion with the same helicity, where the mixing between Weyl spinors arises from either the nonvanishing mixing angle or
the fermion mass term.
Comparing with the separation \eqref{eq:VIBFSR}, we find that $\mathcal{A}_{\text{vb}}$ contains the entire $\mathcal{A}_{\text{VIB}}$ and part of $\mathcal{A}_{\text{FSR}}$, while $\mathcal{A}_{\text{mix}}$ and $\mathcal{A}_{m_f}$ receive contributions only from $\mathcal{A}_{\text{FSR}}$.

Each term in the matrix element can be written as the contraction of a
spinor product, with some Lorentz structure, and some function of the momenta.
$\mathcal{A}_{\text{vb}}$ contains spinor products with vector and axial vector Lorentz structure.
The $CP$-conserving parts of $\mathcal{A}_{\text{mix}}$ contain spinor products with scalar and tensor Lorentz structures,
while the $CP$-violating parts contain spinor products with pseudoscalar and tensor Lorentz structures.
We do not present the complete differential scattering cross section because it is quite lengthy.  However, the spinor
products can be found, for example, in Ref.~\cite{Kumar:2013iva}, allowing one to evaluate the entire expression.

We note also that each of these terms ($\mathcal{A}_{\text{vb}}$, $\mathcal{A}_{\text{mix}}$, and $\mathcal{A}_{m_f}$) is
suppressed at most by $\sin 2\alpha$ or $m_f / m_X$, but not by both.  Thus, we expect $CP$-violating contributions to
bremsstrahlung processes to be subleading, as they are doubly suppressed.  Indeed, our explicit calculation verifies that
this effect is small.  

There is a well-known enhancement in the cross section for emitting soft or collinear photons via final state radiation, arising from a nearly on-shell fermion propagator. In both the soft and collinear limits, we have $s_{3}\approx s_{4}\approx -m_{X}^{2}$, and the only divergent quantity is
\begin{equation}
    \mathcal{S}_{i}+\mathcal{V}_{i}\rightarrow-\left(\frac{\slashed{\epsilon}_{5}\slashed{k}_{5}+2k_{3}\cdot\epsilon_{5}}{2k_{3}\cdot k_{5}}-\frac{\slashed{k}_{5}\slashed{\epsilon}_{5}+2k_{4}\cdot\epsilon_{5}}{2k_{4}\cdot k_{5}}\right)\frac{1}{m_{X}^{2}+m_{\widetilde{f}_{i}}^{2}}\,.
\end{equation}
In particular, for the soft limit, we can further neglect the $\slashed{k}_{5}$ in the numerator and get the correct factorization behavior,
\begin{equation}
\label{eq:softcol}
\mathcal{A}_{\text{IB}}\xrightarrow{\text{soft}}-e\left(\frac{k_{3}\cdot\epsilon_{5}}{k_{3}\cdot k_{5}}-\frac{k_{4}\cdot\epsilon_{5}}{k_{4}\cdot k_{5}}\right)\mathcal{A}_{\text{2-b}}\,.
\end{equation}
This leads to the Sudakov log enhancement of the probability for photon emission from final state radiation,
\begin{align}
(\sigma v)_{\text{IB}} \sim \frac{\alpha_{\text{em}} }{ \pi} \log \left(\frac{s }{ E_{th}^2 } \right) \log \left(\frac{s }{ m_{f}^2 } \right)
\times (\sigma v)_{f\bar{f}}\,,
\end{align}
where $s=4m_X^2$ and we have kept only the leading logarithmic enhancement. The first logarithm is the soft photon enhancement, which is cut off
by $E_{th}$, the energy threshold of the photon detector. The second logarithm is the collinear photon enhancement and is cut off
by the mass of the fermion. More generally, if the photon is collinear but not necessarily soft, we obtain
\begin{equation}
    \frac{d(\sigma v)_{\text{IB}}}{dx}\sim\frac{\alpha_{\text{em}}}{\pi}\frac{(1-x)^{2}+1}{x}\log\frac{s(1-x)}{m_{f}^{2}}\times(\sigma v)_{f\bar{f}}
\end{equation}
from Eq.~\eqref{eq:softcol}, which agrees with the Weizs\"acker-Williams formula for FSR. The soft and collinear enhancements thus have little effect on the spectrum as $m_f / m_X \rightarrow 0$ and
$\alpha \rightarrow 0$, since $(\sigma v)_{f\bar{f}} \rightarrow 0$ in this limit.
But if $\alpha \neq 0$, then the collinear enhancement will have a large effect; one cannot strictly
take $m_{f} / m_X \rightarrow 0$ limit, as the nonzero fermion mass cuts off the collinear divergence.

There has been a variety of past work on the spectrum of the $XX \rightarrow \bar f f \gamma$ process, and this spectrum
is well known in two limits:
\begin{itemize}
\item{{\it $\alpha =0$, $m_f /m_X \rightarrow 0$}: This corresponds to the case where the process $XX \rightarrow \bar f f$
is suppressed, and the dominant process is $XX \rightarrow \bar f f \gamma$, yielding a hard spectrum which is dominated by
$\mathcal{A}_{\text{vb}}$.  The soft and collinear emission of photons via FSR has little effect on the spectrum.}
\item{{\it $\alpha =$} ${\cal O}(1)${\it , $m_f / m_X \rightarrow 0$}: In this case, the $XX \rightarrow \bar f f$ cross section is
unsuppressed, and the dominant contribution to $XX \rightarrow \bar f f \gamma$ arises from FSR in the soft and collinear
regimes.  Here, the details of the interaction are not very important; provided the $XX \rightarrow \bar f f$ cross section is
unsuppressed, the dominant contribution to $XX \rightarrow \bar f f \gamma$ arises from a simple rescaling of the $2 \rightarrow 2$
cross section by the Sudakov log factor.  }
\end{itemize}
We will now discuss the regime of intermediate $\alpha$, which interpolates between these two limits.

\begin{figure*}
\captionsetup[subfigure]{labelformat=empty}
\centering
\subfloat[]{\includegraphics[width=0.47\textwidth]{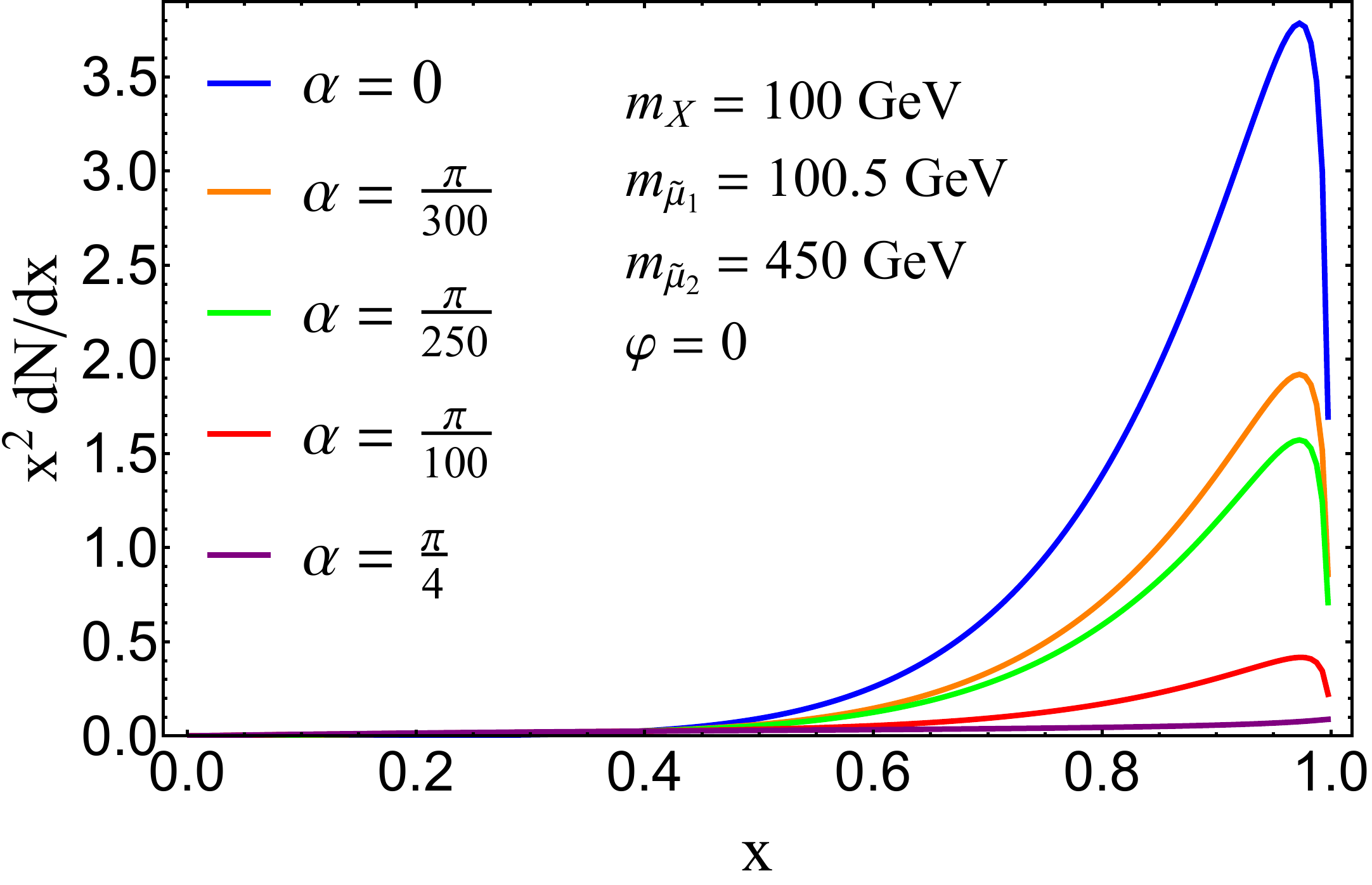}}
\quad
\subfloat[]{\includegraphics[width=0.47\textwidth]{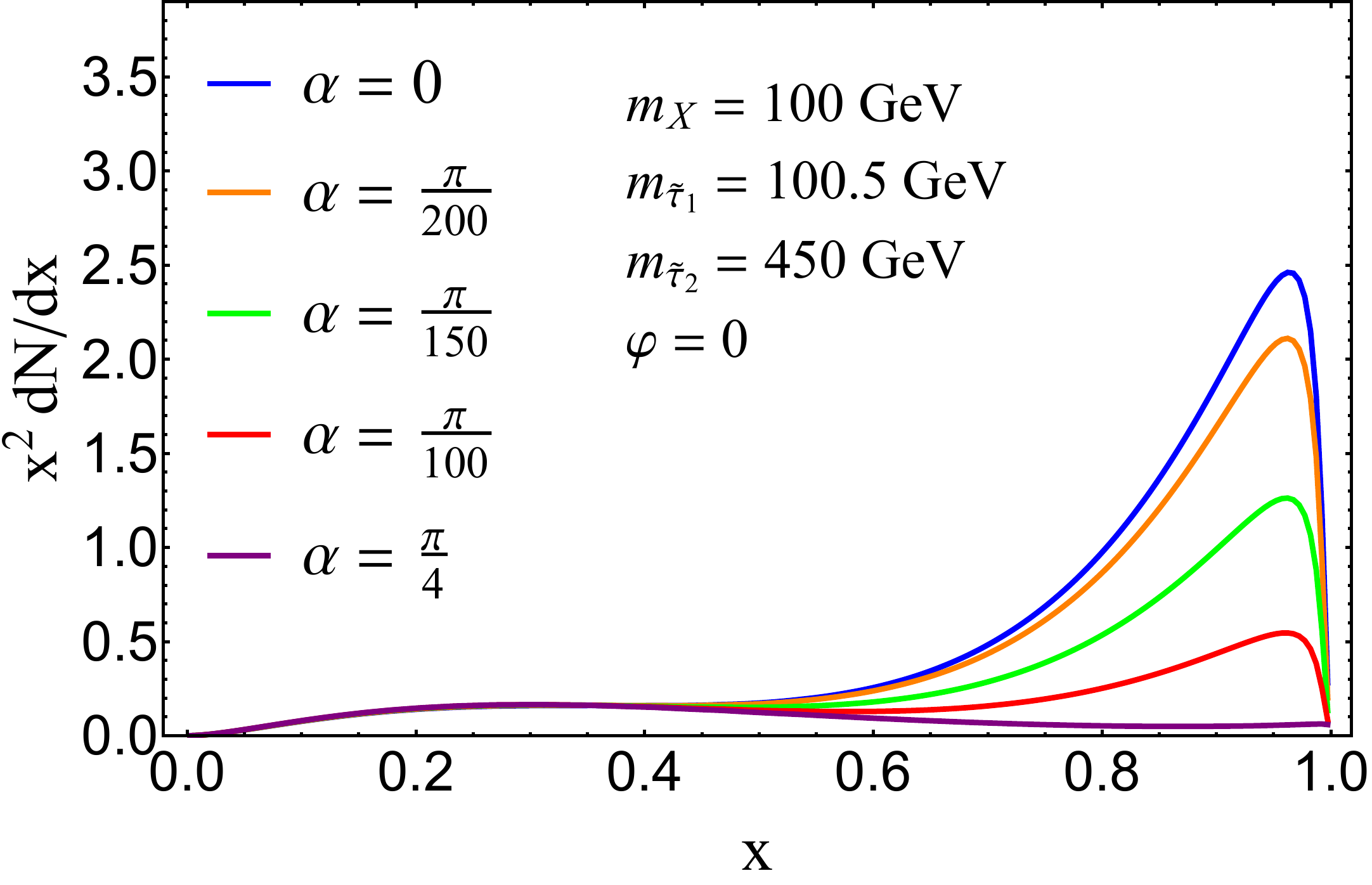}}\\
\subfloat[]{\includegraphics[width=0.47\textwidth]{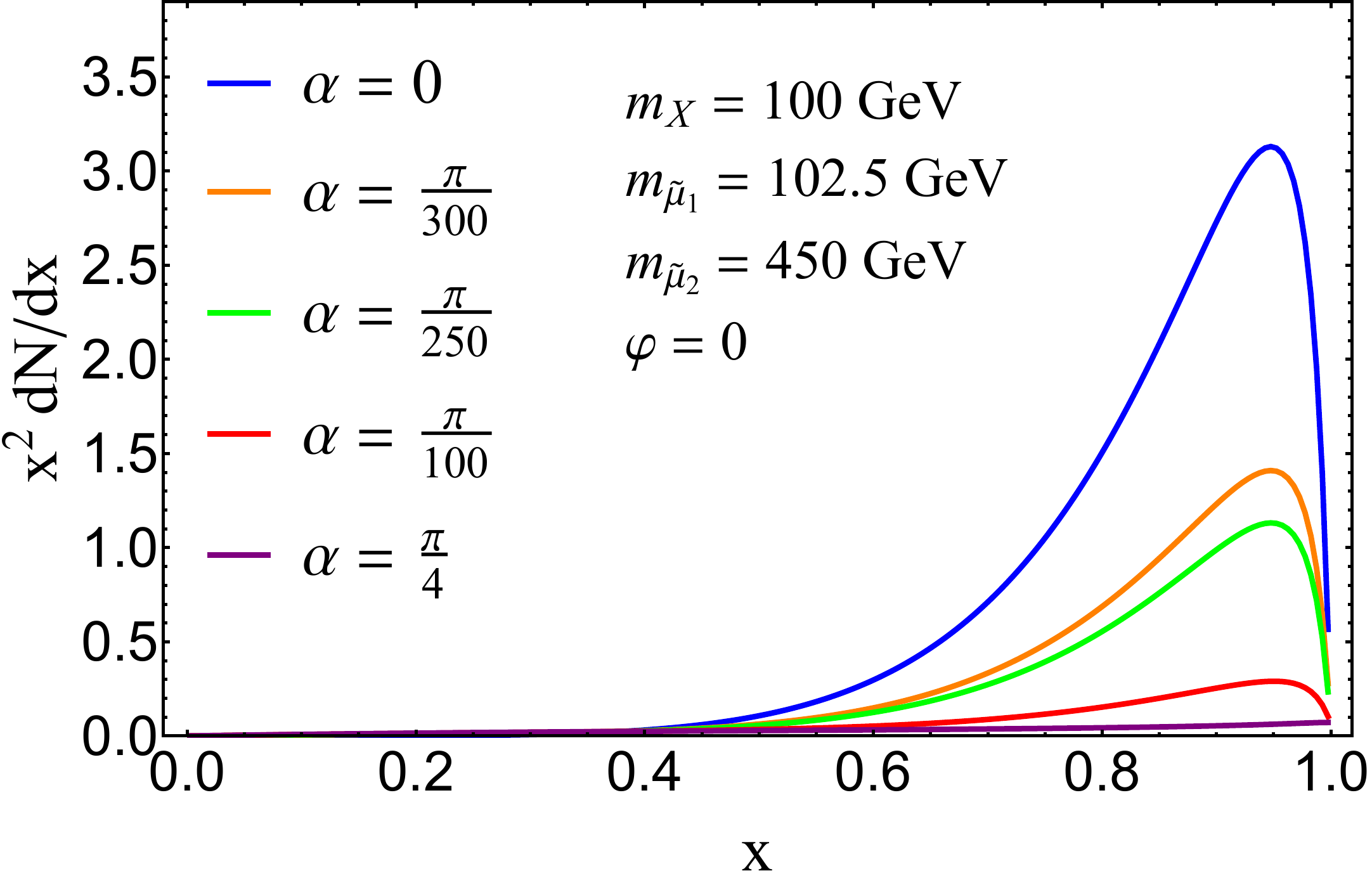}}
\quad
\subfloat[]{\includegraphics[width=0.47\textwidth]{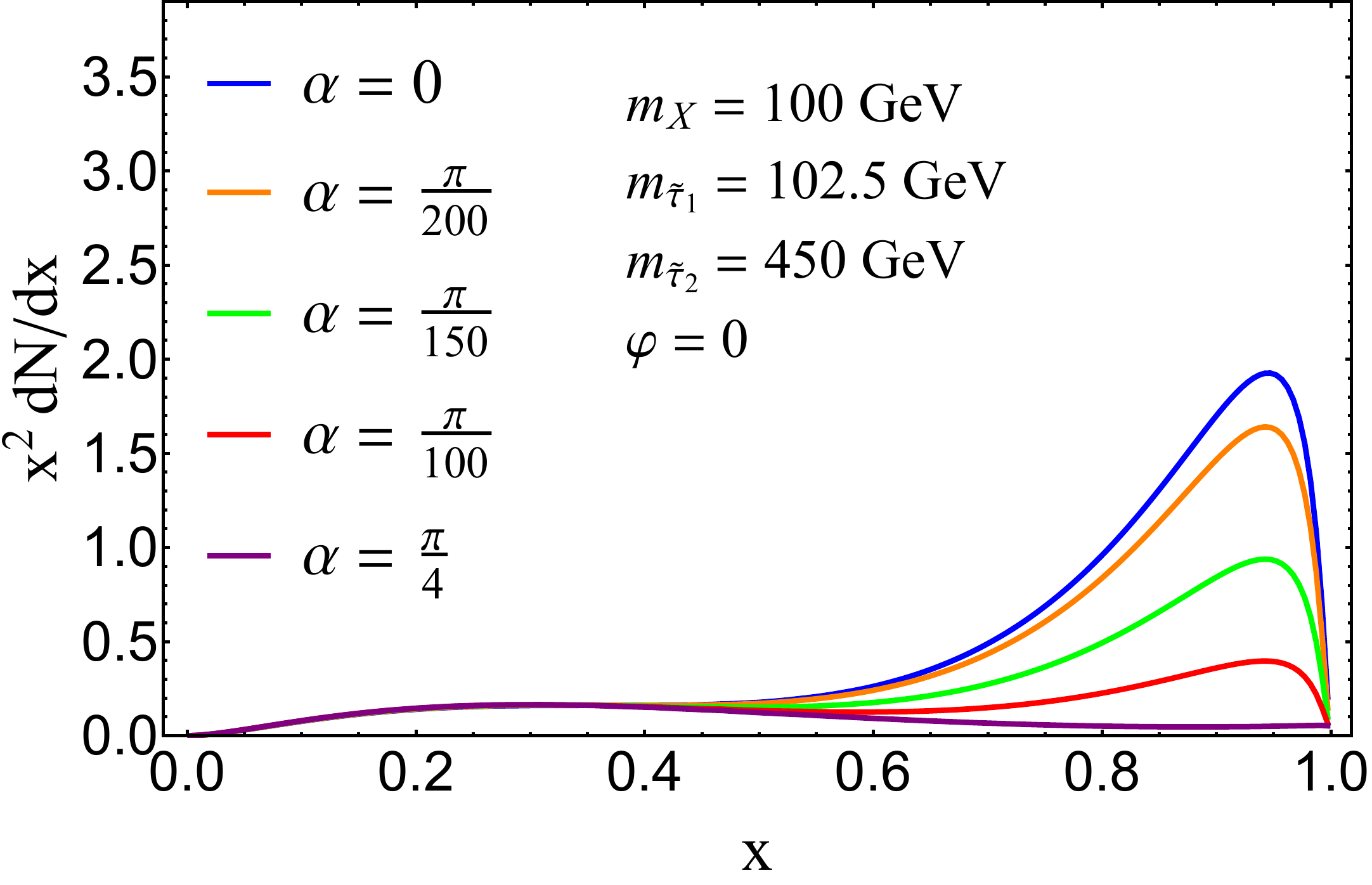}}\\
\subfloat[]{\includegraphics[width=0.47\textwidth]{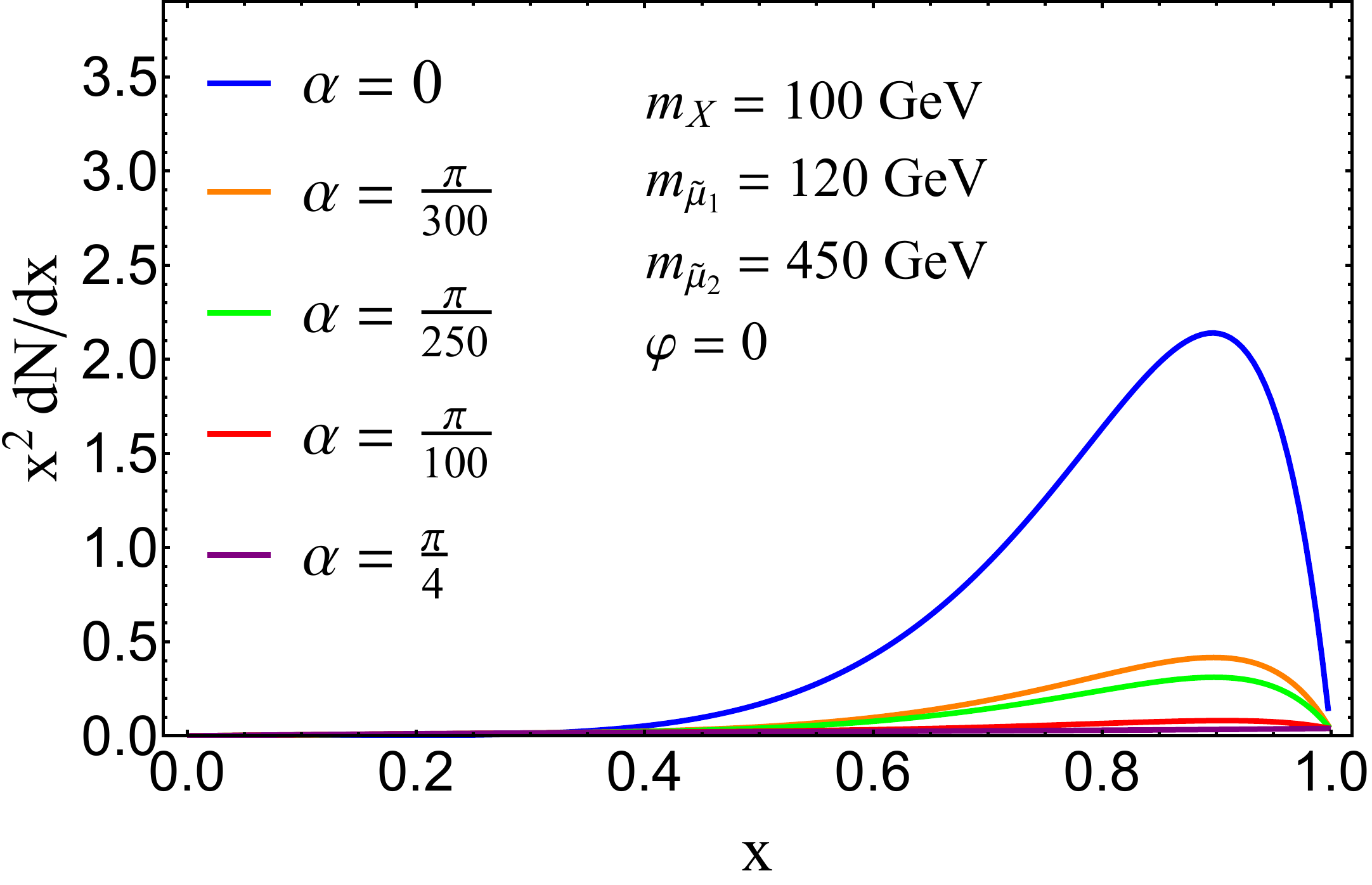}}
\quad
\subfloat[]{\includegraphics[width=0.47\textwidth]{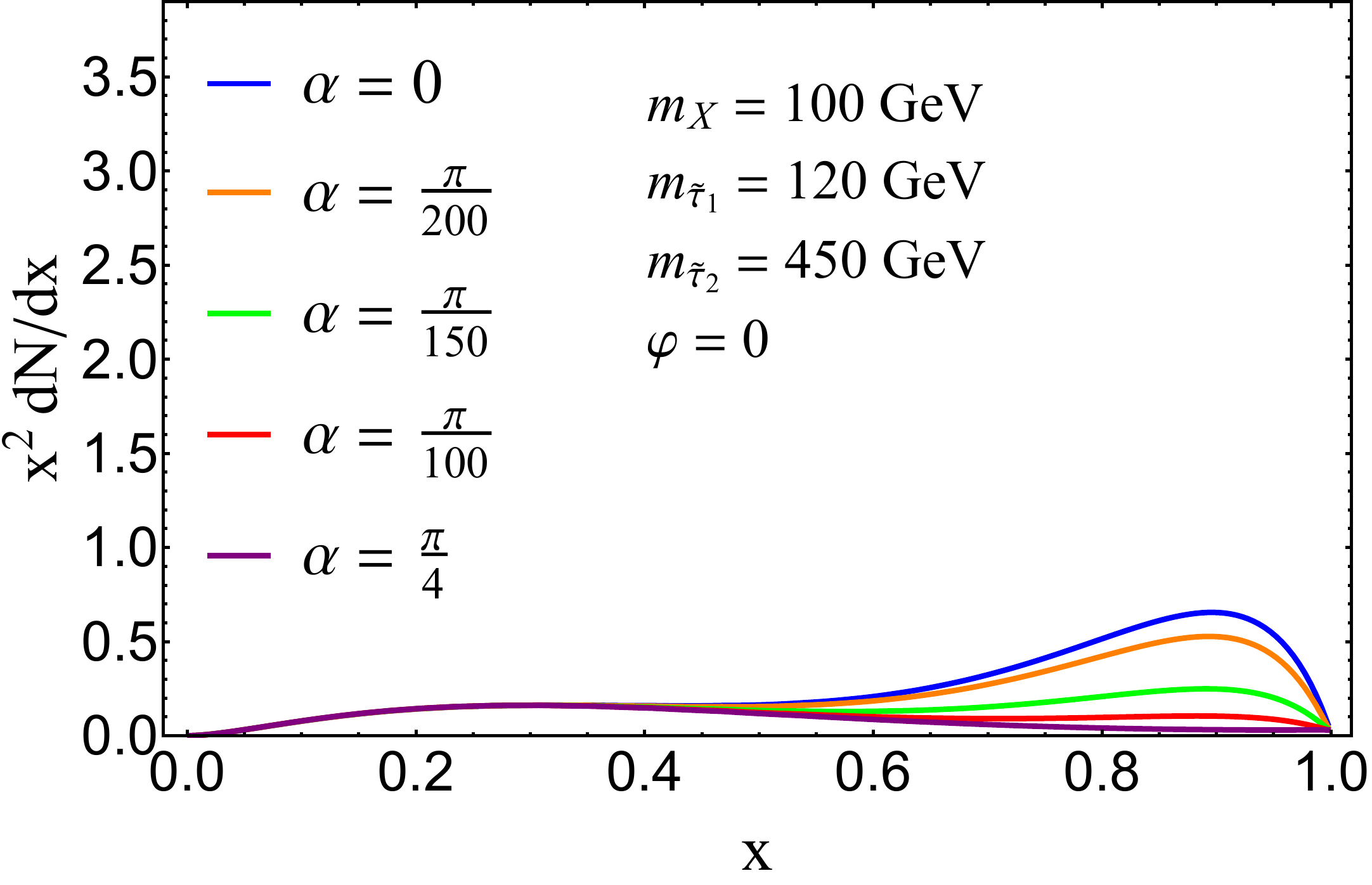}}
\caption{{Dependence of the continuum photon spectrum on $\alpha$ for the process $XX \rightarrow \bar f f \gamma$. The left panels show the $\mu$ channel and the right show the $\tau$ channel. The three $\widetilde{f}_{1}$ masses correspond to $\mu_{1}=1.01$, $1.05$ and $1.44$.
We take $\lambda_L = (\sqrt{2}/2)g$, $\lambda_R =\sqrt{2}g$.} }
\label{fig:ampcompare}
\end{figure*}

In Fig.~\ref{fig:ampcompare}, we plot the continuum photon spectrum for both $\mu$ and $\tau$ final states with three different lightest scalar masses.  As expected,
the $\alpha =0$ case produces a hard spectrum which falls rapidly at low energies. The peak feature is more prominent for degenerate $X$ and $\widetilde{f}_{1}$. Moreover, the normalization
of the spectrum remains stable as $m_f / m_X \rightarrow 0$.
Once $\alpha$ is large enough, the spectrum flattens due to the enhancement in emission of soft photons.  
If the lightest scalar mediator ($\widetilde{f}_{1}$,
without loss of generality)
is much lighter than the heavier scalar, one expects the crossover
between these behaviors to occur roughly when
\begin{eqnarray}
\tan^2 \alpha \times |\lambda_R/\lambda_L|^2  (m_{\widetilde{f}_1} / m_X)^4 \log (m_X^2 / m_f^2) \sim {\cal O}(1) \,,
\end{eqnarray}
which corresponds to the point where the suppression of soft FSR due to the small mixing angle is roughly canceled by the
enhancement for collinear emission.
Note that one expects the hard IB signal to dominate over
$2 \rightarrow 2$ scattering provided
\begin{eqnarray}
\tan^2 \alpha \times |\lambda_R/\lambda_L|^2 (m_{\widetilde{f}_1} / m_X )^4 < \alpha_{\text{em}}\,.
\end{eqnarray}
We thus see, for example, that if $m_X \sim 100$ GeV and $f=\mu$, then for a choice of parameters such that the
photon spectrum will interpolate between the hard and FSR regimes, the cross section for $XX \rightarrow \bar f f \gamma$
with a hard photon will be ${\cal O}(10\%)$ of the $XX \rightarrow \bar f f$ cross section. {For $f=\tau$, the high energy spectrum behaves in a similar way, but there is an $\alpha$-independent bump at the low energy end due to the photons from the hadronic decay of $\tau^\pm$.}

\begin{figure}
    \centering
    \includegraphics[width=0.45\textwidth]{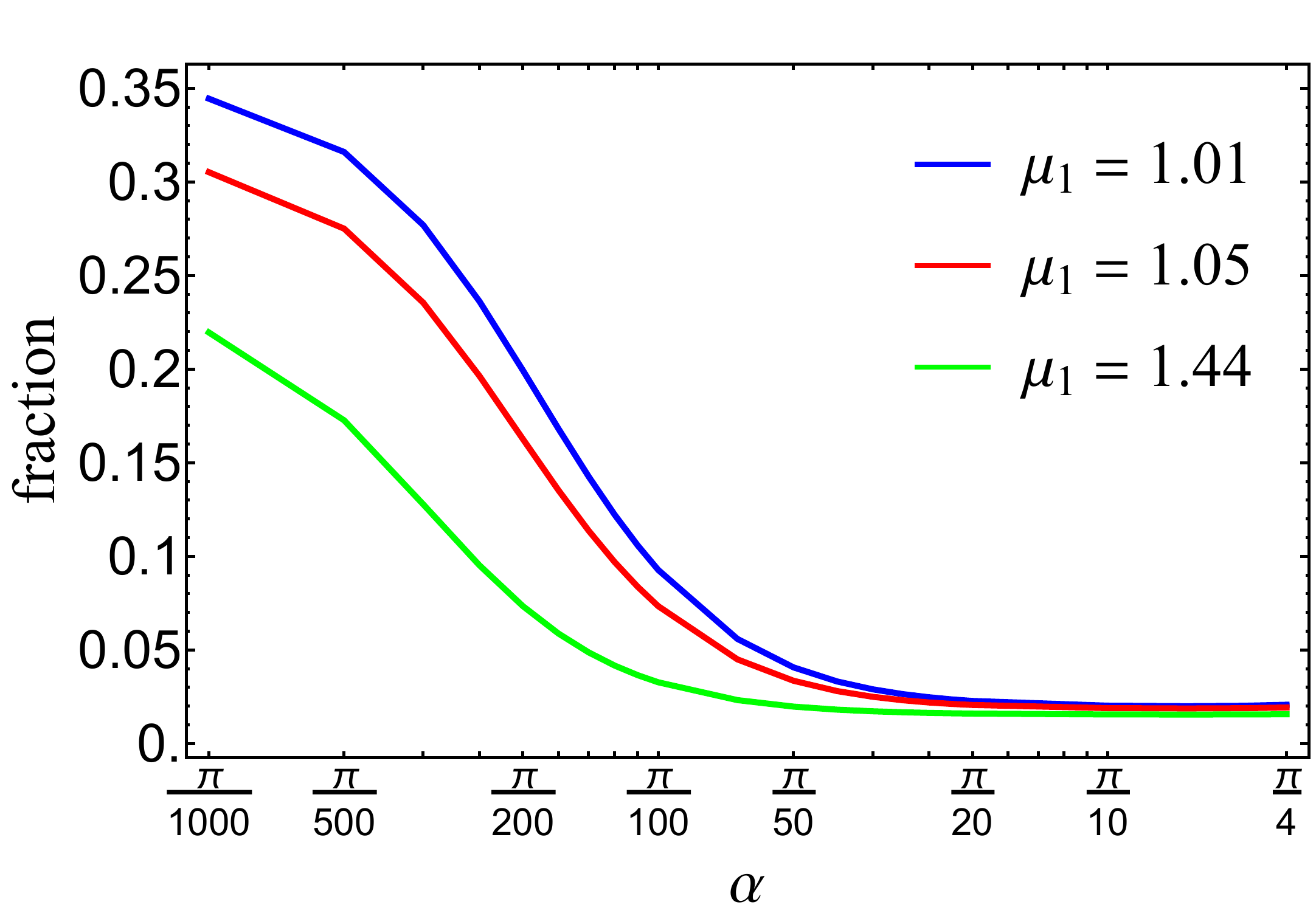}
    \caption{The photon number fraction in the peak of $\mu^{+}\mu^{-}\gamma$ final state for three different lightest scalar masses. The model parameters are the same as in the left panel of Fig.~\ref{fig:ampcompare}}
    \label{fig:fraction}
\end{figure}

To get a better idea of when the peak feature disappears, we plot in Fig.~\ref{fig:fraction} the ratio of the photon number in the peak to the total photon number (integrated from $x=0.01$ to the cutoff). To find out the peak for each $\alpha$, we integrate the photon number in a bin of which the width is $10\%$ of its central value and slide it from $x=0.6$ to the cutoff. The peak corresponds to the maximum photon number found in this process. For $f=\mu$, we may see that the transition happens around $\alpha\sim\pi/100$.

Finally we would like to clarify the procedure used to obtain the components of the spectra.  As mentioned above, prompt photon emission can also come from the decay of the charged SM particles produced by dark matter annihilation.
The prompt photon spectrum is usually simulated by event generators such as {\tt PYTHIA}, which first create the final state phase space for the decay of a hypothetical boson with various branching ratios into SM particles, then simulate the prompt evolution of the final state, and finally return the resultant photon spectrum.
In particular, the FSR of the decay, as the leading order contribution, is thus captured by the simulation.
But in our calculation of bremsstrahlung, FSR is necessarily included in order for the calculation to be gauge invariant.
Thus, as in Ref.~\cite{Bringmann08}, we must subtract the FSR from the {\tt PYTHIA} decay spectrum before adding the bremsstrahlung
spectrum.

Finally, in Table~\ref{tab:bmcs2}, we present, for reference a summary of the dark matter abundance and relevant annihilation cross sections for the benchmarks discussed here.

\begin{table*}[ht]
    \centering
    \renewcommand{\arraystretch}{1.2}
    \begin{tabular}{*{6}{|c}|}
         \hline
         \multirow{2}{*}{Model} & $(\sigma v)_{f\bar{f}}$ & $(\sigma v)_{\text{IB}}$ & $\Omega h^{2}$ & $(\sigma v)_{\gamma\gamma}$ & $(\sigma v)_{\gamma Z}$ \\
           & $\times 10^{-26}\,\text{cm}^{3}/\text{s}$ & $\times 10^{-27}\,\text{cm}^{3}/\text{s}$ & (thermal) & $\times 10^{-28}\,\text{cm}^{3}/\text{s}$ & $\times 10^{-32}\,\text{cm}^{3}/\text{s}$ \\ \hline
         $A$          & $2.127$ & $4.917$ & $0.1156$ & $4.256\times 10^{-3}$ & $2.89$  \\ \hline
         $B$        & $2.010$ & $2.872$ & $0.1212$ & $2.662\times 10^{-3}$ & $3.03$  \\ \hline
         $C$         & $2.128$ & $3.046$ & $0.1155$ & $4.513\times 10^{-3}$ & $2.88$ \\ \hline
         $D$ & $46.95$ & $108.5$ & Underabundant& $8.019\times 10^{-2}$ & $-$  \\ \hline
         $D'$ & $53.53$ & $124.7$ & Underabundant& $0.1355$ & $-$  \\ \hline
         $E$ & Forbidden & Forbidden & Overabundant & $2.9370$ & $-$ \\ \hline
    \end{tabular}
    \caption{Physical quantities derived from our benchmark models $A$, $B$ and $C$. $(\sigma v)_{\text{IB}}$ is integrated from $x=0.2$. Note that all models satisfy the constraints on the dipole moments of the SM leptons, with the exception of Benchmark $A$, which does not exacerbate the problem of the muon anomalous magnetic moment, but also does not produce the measured value.  If $\varphi$ is shifted sightly to $0.49\pi$, $a_{\mu}$ will fall into the $2\sigma$ range of current experimental measurement, while all the other quantities in the table above remain nearly unchanged. See Ref.~\cite{Fukushima14} for more details on the magnetic and electric dipole moments. }
    \label{tab:bmcs2}
\end{table*}

\subsection{Constraints from Fermi-LAT}

\begin{figure*}
\centering
\subfloat[$\mu_{1}=1.44$]{\includegraphics[width=0.48\textwidth]{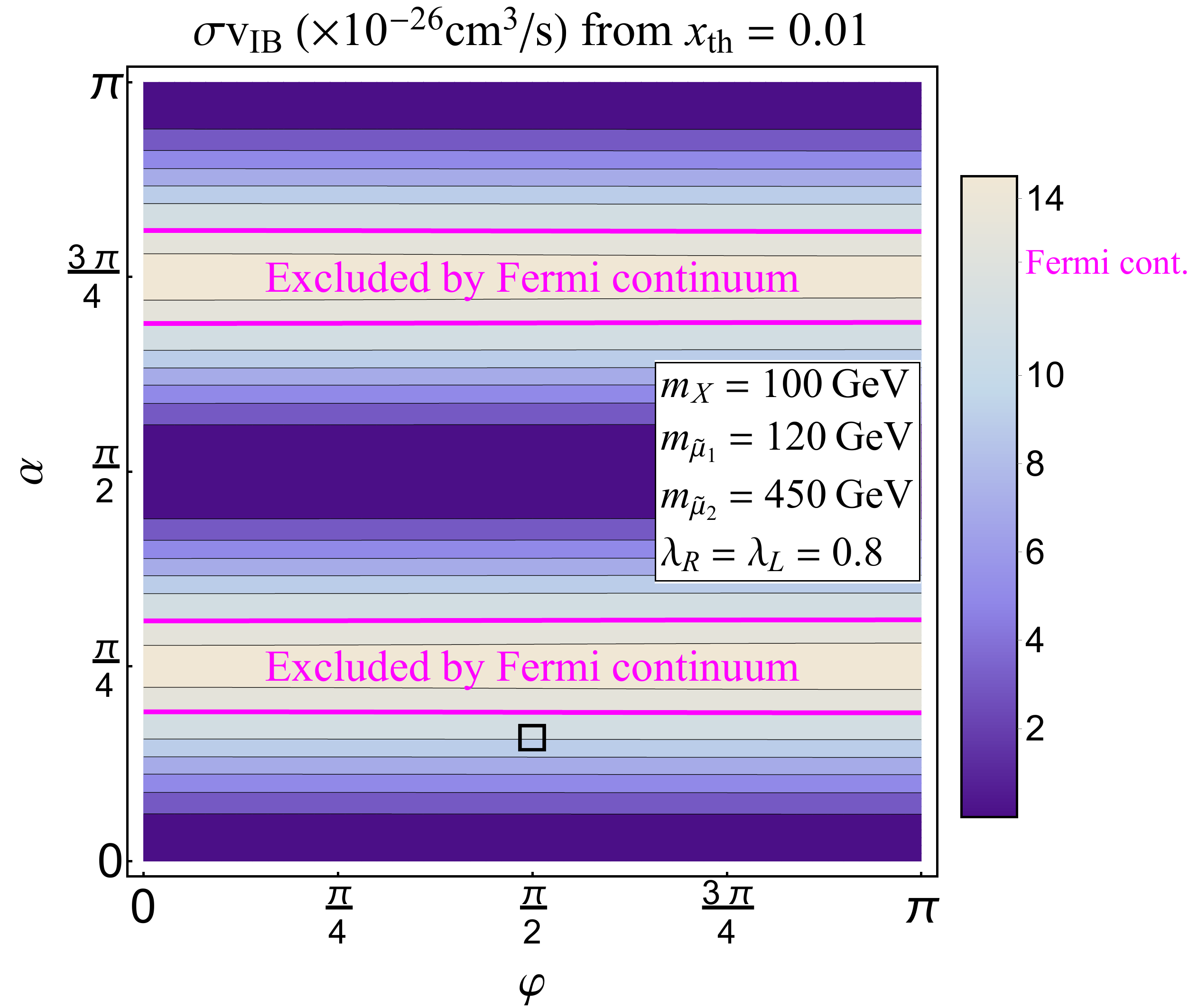}}\quad
\subfloat[$\mu_{1}=1.05$]{\includegraphics[width=0.48\textwidth]{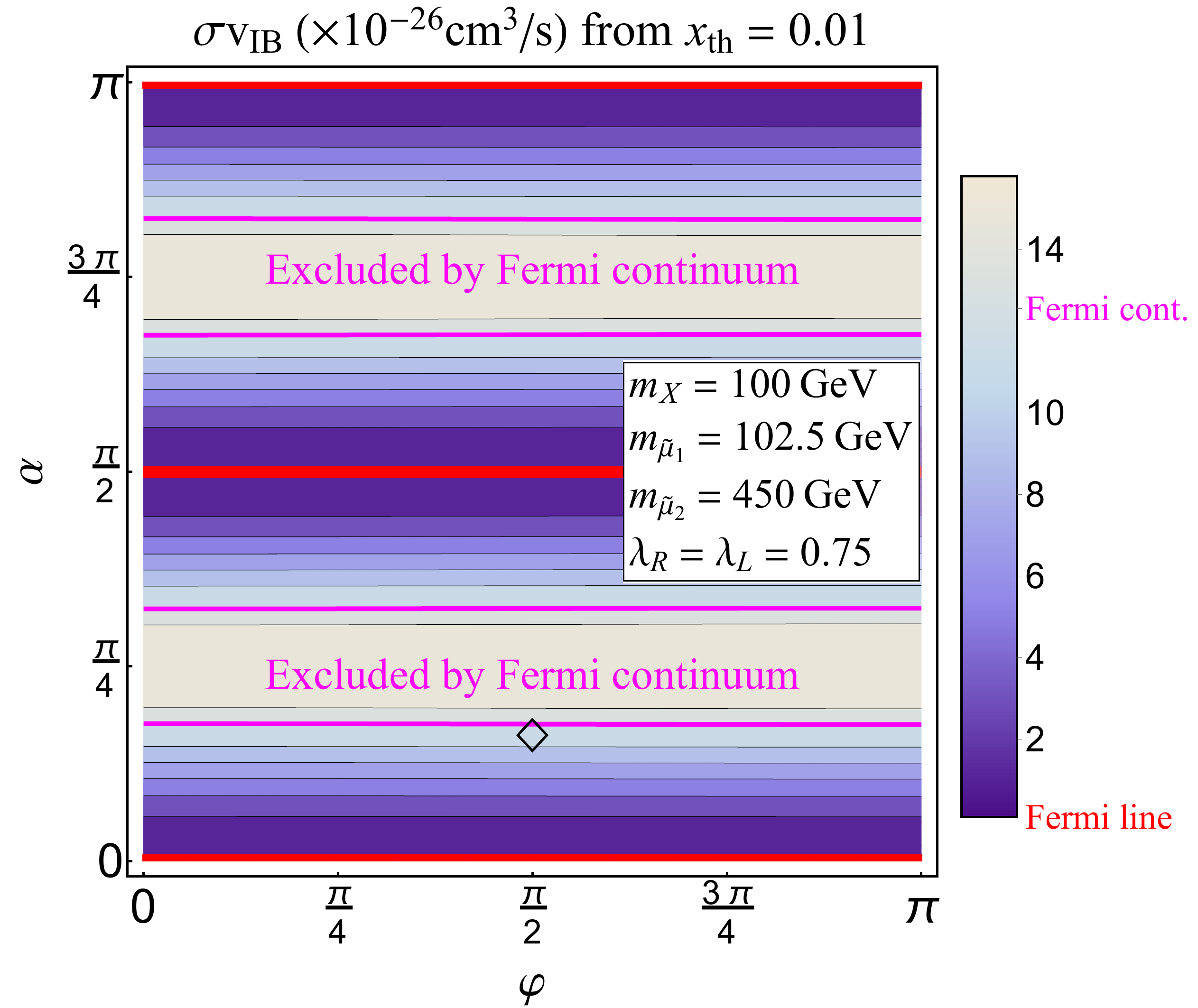}}
\caption{\label{fig:IBlimit} {The total IB cross section for (a) $\mu_{1}=1.44$ and $\lambda_L = \lambda_R = 0.8$ and (b) $\mu_{1}=1.05$ and $\lambda_L = \lambda_R = 0.75$. At large mixing, both models are constrained by the Fermi continuum limit; At $\alpha\sim 0$, $\pi/2$ and $\pi$, the spectrum of (b) is linelike so it is also constrained by the Fermi line limit. The benchmarks $D$ and $D'$ are labeled by the square and diamond. The Fermi continuum limit is taken from Ref.~\cite{GeringerSameth:2011iw}. }}
\end{figure*}

{In Fig.~\ref{fig:IBlimit}, we plot the Fermi-LAT exclusion contours for $f=\mu$ in the $(\alpha, \varphi)$ plane for Benchmark $D$ with $\lambda_L=\lambda_R=0.8$ and $\mu_{1}=1.44$ and Benchmark $D'$ with $\lambda_{L}=\lambda_{R}=0.75$ and $\mu_{1}=1.05$.}
Since the Fermi-LAT analysis searches for photons, and muon decay produces few photons, this is
essentially a search for the $XX \rightarrow \bar f f \gamma$ (for these parameters, the monoenergetic
photon final states are subdominant).  For the parameter range displayed, the maximum cross section for $XX \rightarrow
\mu^+ \mu^-\gamma$ in the $(\alpha, \varphi)$-plane is $\sim 1.4\times 10^{-25}\text{cm}^{3}/\text{s}$.
For $\mu_{1}=1.44$, the cross section for process $XX \rightarrow \mu^+ \mu^-$ is $(20.9~\pb) \times \sin^2 2\alpha$. The continuum
limit arises from a stacked search of dwarf spheroidals for photons with $E > 1~\gev$ and follows the
analysis of Ref.~\cite{GeringerSameth:2011iw}.  Although this is not the most recent analysis and does not provide the most
stringent limit from dwarf spheroidals, it is applicable here because it makes no assumption about the photon
spectrum. Constraints are phrased in terms of a particle physics factor, $\Phi_{\text{PP}}$,
\begin{equation}
 \Phi_{\text{PP}}=\frac{(\sigma v)_{\text{ann.}}}{8\pi m_{X}^{2}}\int_{x_{\text{th}}}^{1}dx\left(\frac{dN}{dx}\right)_{\text{cont.}}.
\end{equation}
{We take the constraint on $\Phi_{\text{PP}}$ from Ref.~\cite{GeringerSameth:2011iw},
\begin{equation*}
    \Phi_{\text{PP}}=5.0^{+4.3}_{-4.5}\times 10^{-30}\text{cm}^{3}\,\text{s}^{-1}\,\text{GeV}^{-2}\,.
\end{equation*}
For the $\mu$ final state, with small mixing angle, $\Phi_{\text{PP}}$ can be approximated by
\begin{equation}
    \Phi_{\text{PP}}\approx\frac{(\sigma v)_{\text{IB}}}{8\pi m_{X}^{2}}\,,
\end{equation}
such that it can be directly translated into an upper limit for $(\sigma v)_{\text{IB}}$. For small $\alpha$, the IB spectrum for $\mu_{1}=1.44$ (left panel of Fig.~\ref{fig:IBlimit}) might just marginally display a linelike feature, while for $\mu_{1}=1.05$ (right panel of Fig.~\ref{fig:IBlimit}), the spectrum is hard enough that it can be constrained by the Fermi-LAT line search \cite{Ackermann:2015lka}. The spectral features of both cases can be understood in light of Fig.~\ref{fig:ampcompare}.}

\section{\label{sec:Con}Conclusion}

In this paper, we have investigated possible gamma-ray signatures from dark matter annihilation in a class of  simplified models in which
the dark matter couples to light fermions via a pair of new charged scalars.
In particular, we have studied the effect of chiral mixing and a $CP$-violating phase on the gamma-ray signals from
dark matter annihilation into $\gamma\gamma$ and $\gamma Z$ as well as the internal bremsstrahlung spectrum associated with dark matter annihilation into a fermion pair.

We have found that varying the mixing angle results in a prompt photon spectrum for the process $XX \rightarrow \bar f f \gamma$
which interpolates between the standard regimes which are
dominated by either virtual internal bremsstrahlung or soft/collinear final state radiation.  In some regions of parameter space,
this deviation from the standard spectra will be observable, and can provide a clue as to the relative strength of deviations from
minimal flavor violation in the underlying theory.  For the $2 \rightarrow 3$ annihilation process, although the mixing angle is
very important in determining the spectrum, the $CP$-violating phase is less so.

On the other hand, the mixing angle and $CP$-violating phase are both important for monoenergetic annihilation signals.  In particular, varying the mixing angle will change the relative branching fractions to the final states
$\gamma \gamma$ and $\gamma Z$.  For sufficiently large couplings in the simplified model, this effect could be observed in future
experiments with excellent energy resolution.

Interestingly, a nonvanishing $CP$-violating phase can result in an asymmetry in the left- vs.~right-circularly polarized
photons arising from dark matter annihilation.  Current gamma-ray instruments are not capable of detecting such polarization
for the energy range of interest.  Experimental study of this scenario would require a new strategy.
Monoenergetic
photon signals from dark matter annihilation are sometimes considered the ultimate ``smoking gun" for indirect detection;
it would be interesting to further study the feasibility of observing the polarization asymmetry in this channel, which is a
hallmark of $CP$ violation.

\begin{acknowledgments}
We are grateful to Philip von Doetinchem, Peter Gorham, Danny Marfatia, and Kuver Sinha for useful discussions.
J.K. is supported in part by NSF CAREER Grant
No.~PHY-1250573.  P.S. is supported in part by NSF Grant No.~PHY-1417367.
\end{acknowledgments}

\bibliography{references}

\begin{thebibliography}{48}%
\makeatletter
\providecommand \@ifxundefined [1]{%
 \@ifx{#1\undefined}
}%
\providecommand \@ifnum [1]{%
 \ifnum #1\expandafter \@firstoftwo
 \else \expandafter \@secondoftwo
 \fi
}%
\providecommand \@ifx [1]{%
 \ifx #1\expandafter \@firstoftwo
 \else \expandafter \@secondoftwo
 \fi
}%
\providecommand \natexlab [1]{#1}%
\providecommand \enquote  [1]{``#1''}%
\providecommand \bibnamefont  [1]{#1}%
\providecommand \bibfnamefont [1]{#1}%
\providecommand \citenamefont [1]{#1}%
\providecommand \href@noop [0]{\@secondoftwo}%
\providecommand \href [0]{\begingroup \@sanitize@url \@href}%
\providecommand \@href[1]{\@@startlink{#1}\@@href}%
\providecommand \@@href[1]{\endgroup#1\@@endlink}%
\providecommand \@sanitize@url [0]{\catcode `\\12\catcode `\$12\catcode
  `\&12\catcode `\#12\catcode `\^12\catcode `\_12\catcode `\%12\relax}%
\providecommand \@@startlink[1]{}%
\providecommand \@@endlink[0]{}%
\providecommand \url  [0]{\begingroup\@sanitize@url \@url }%
\providecommand \@url [1]{\endgroup\@href {#1}{\urlprefix }}%
\providecommand \urlprefix  [0]{URL }%
\providecommand \Eprint [0]{\href }%
\providecommand \doibase [0]{http://dx.doi.org/}%
\providecommand \selectlanguage [0]{\@gobble}%
\providecommand \bibinfo  [0]{\@secondoftwo}%
\providecommand \bibfield  [0]{\@secondoftwo}%
\providecommand \translation [1]{[#1]}%
\providecommand \BibitemOpen [0]{}%
\providecommand \bibitemStop [0]{}%
\providecommand \bibitemNoStop [0]{.\EOS\space}%
\providecommand \EOS [0]{\spacefactor3000\relax}%
\providecommand \BibitemShut  [1]{\csname bibitem#1\endcsname}%
\let\auto@bib@innerbib\@empty
%</preamble>
\bibitem [{\citenamefont {Ade}\ \emph {et~al.}(2014)\citenamefont {Ade} \emph
  {et~al.}}]{planck}%
  \BibitemOpen
  \bibfield  {author} {\bibinfo {author} {\bibfnamefont {P.~A.~R.}\
  \bibnamefont {Ade}} \emph {et~al.} (\bibinfo {collaboration} {Planck}),\
  }\href {\doibase 10.1051/0004-6361/201321591} {\bibfield  {journal} {\bibinfo
   {journal} {Astron. Astrophys.}\ }\textbf {\bibinfo {volume} {571}},\
  \bibinfo {pages} {A16} (\bibinfo {year} {2014})},\ \Eprint
  {http://arxiv.org/abs/1303.5076} {arXiv:1303.5076 [astro-ph.CO]} \BibitemShut
  {NoStop}%
%%CITATION = ARXIV:1303.5076;%%
\bibitem [{\citenamefont {Zeldovich}(1965)}]{Zeldovich}%
  \BibitemOpen
  \bibfield  {author} {\bibinfo {author} {\bibfnamefont {Y.~B.}\ \bibnamefont
  {Zeldovich}},\ }\href@noop {} {\bibfield  {journal} {\bibinfo  {journal}
  {Adv. Astron. Astrophys.}\ }\textbf {\bibinfo {volume} {3}},\ \bibinfo
  {pages} {241} (\bibinfo {year} {1965})}\BibitemShut {NoStop}%
\bibitem [{\citenamefont {Chiu}(1966)}]{Chiu:1966kg}%
  \BibitemOpen
  \bibfield  {author} {\bibinfo {author} {\bibfnamefont {H.-Y.}\ \bibnamefont
  {Chiu}},\ }\href {\doibase 10.1103/PhysRevLett.17.712} {\bibfield  {journal}
  {\bibinfo  {journal} {Phys. Rev. Lett.}\ }\textbf {\bibinfo {volume} {17}},\
  \bibinfo {pages} {712} (\bibinfo {year} {1966})}\BibitemShut {NoStop}%
%%CITATION = PRLTA,17,712;%%
\bibitem [{\citenamefont {Lee}\ and\ \citenamefont
  {Weinberg}(1977)}]{Lee:1977ua}%
  \BibitemOpen
  \bibfield  {author} {\bibinfo {author} {\bibfnamefont {B.~W.}\ \bibnamefont
  {Lee}}\ and\ \bibinfo {author} {\bibfnamefont {S.}~\bibnamefont {Weinberg}},\
  }\href {\doibase 10.1103/PhysRevLett.39.165} {\bibfield  {journal} {\bibinfo
  {journal} {Phys. Rev. Lett.}\ }\textbf {\bibinfo {volume} {39}},\ \bibinfo
  {pages} {165} (\bibinfo {year} {1977})}\BibitemShut {NoStop}%
%%CITATION = PRLTA,39,165;%%
\bibitem [{\citenamefont {Steigman}(1979)}]{Steigman:1979kw}%
  \BibitemOpen
  \bibfield  {author} {\bibinfo {author} {\bibfnamefont {G.}~\bibnamefont
  {Steigman}},\ }\href {\doibase 10.1146/annurev.ns.29.120179.001525}
  {\bibfield  {journal} {\bibinfo  {journal} {Ann. Rev. Nucl. Part. Sci.}\
  }\textbf {\bibinfo {volume} {29}},\ \bibinfo {pages} {313} (\bibinfo {year}
  {1979})}\BibitemShut {NoStop}%
%%CITATION = ARNUA,29,313;%%
\bibitem [{\citenamefont {Scherrer}\ and\ \citenamefont
  {Turner}(1986)}]{Scherrer:1985zt}%
  \BibitemOpen
  \bibfield  {author} {\bibinfo {author} {\bibfnamefont {R.~J.}\ \bibnamefont
  {Scherrer}}\ and\ \bibinfo {author} {\bibfnamefont {M.~S.}\ \bibnamefont
  {Turner}},\ }\href {\doibase 10.1103/PhysRevD.33.1585,
  10.1103/PhysRevD.34.3263} {\bibfield  {journal} {\bibinfo  {journal} {Phys.
  Rev.}\ }\textbf {\bibinfo {volume} {D33}},\ \bibinfo {pages} {1585} (\bibinfo
  {year} {1986})},\ \bibinfo {note} {[Erratum: Phys.
  Rev.D34,3263(1986)]}\BibitemShut {NoStop}%
%%CITATION = PHRVA,D33,1585;%%
\bibitem [{\citenamefont {Jungman}\ \emph {et~al.}(1996)\citenamefont
  {Jungman}, \citenamefont {Kamionkowski},\ and\ \citenamefont
  {Griest}}]{Jungman96}%
  \BibitemOpen
  \bibfield  {author} {\bibinfo {author} {\bibfnamefont {G.}~\bibnamefont
  {Jungman}}, \bibinfo {author} {\bibfnamefont {M.}~\bibnamefont
  {Kamionkowski}}, \ and\ \bibinfo {author} {\bibfnamefont {K.}~\bibnamefont
  {Griest}},\ }\href {\doibase 10.1016/0370-1573(95)00058-5} {\bibfield
  {journal} {\bibinfo  {journal} {Phys. Rept.}\ }\textbf {\bibinfo {volume}
  {267}},\ \bibinfo {pages} {195} (\bibinfo {year} {1996})},\ \Eprint
  {http://arxiv.org/abs/hep-ph/9506380} {arXiv:hep-ph/9506380 [hep-ph]}
  \BibitemShut {NoStop}%
%%CITATION = HEP-PH/9506380;%%
\bibitem [{\citenamefont {Bergstrom}\ and\ \citenamefont
  {Ullio}(1997)}]{Bergstrom97}%
  \BibitemOpen
  \bibfield  {author} {\bibinfo {author} {\bibfnamefont {L.}~\bibnamefont
  {Bergstrom}}\ and\ \bibinfo {author} {\bibfnamefont {P.}~\bibnamefont
  {Ullio}},\ }\href {\doibase 10.1016/S0550-3213(97)00530-0} {\bibfield
  {journal} {\bibinfo  {journal} {Nucl. Phys.}\ }\textbf {\bibinfo {volume}
  {B504}},\ \bibinfo {pages} {27} (\bibinfo {year} {1997})},\ \Eprint
  {http://arxiv.org/abs/hep-ph/9706232} {arXiv:hep-ph/9706232 [hep-ph]}
  \BibitemShut {NoStop}%
%%CITATION = HEP-PH/9706232;%%
\bibitem [{\citenamefont {Bern}\ \emph {et~al.}(1997)\citenamefont {Bern},
  \citenamefont {Gondolo},\ and\ \citenamefont {Perelstein}}]{Bern97}%
  \BibitemOpen
  \bibfield  {author} {\bibinfo {author} {\bibfnamefont {Z.}~\bibnamefont
  {Bern}}, \bibinfo {author} {\bibfnamefont {P.}~\bibnamefont {Gondolo}}, \
  and\ \bibinfo {author} {\bibfnamefont {M.}~\bibnamefont {Perelstein}},\
  }\href {\doibase 10.1016/S0370-2693(97)00990-8} {\bibfield  {journal}
  {\bibinfo  {journal} {Phys. Lett.}\ }\textbf {\bibinfo {volume} {B411}},\
  \bibinfo {pages} {86} (\bibinfo {year} {1997})},\ \Eprint
  {http://arxiv.org/abs/hep-ph/9706538} {arXiv:hep-ph/9706538 [hep-ph]}
  \BibitemShut {NoStop}%
%%CITATION = HEP-PH/9706538;%%
\bibitem [{\citenamefont {Ullio}\ and\ \citenamefont
  {Bergstrom}(1998)}]{Ullio98}%
  \BibitemOpen
  \bibfield  {author} {\bibinfo {author} {\bibfnamefont {P.}~\bibnamefont
  {Ullio}}\ and\ \bibinfo {author} {\bibfnamefont {L.}~\bibnamefont
  {Bergstrom}},\ }\href {\doibase 10.1103/PhysRevD.57.1962} {\bibfield
  {journal} {\bibinfo  {journal} {Phys. Rev.}\ }\textbf {\bibinfo {volume}
  {D57}},\ \bibinfo {pages} {1962} (\bibinfo {year} {1998})},\ \Eprint
  {http://arxiv.org/abs/hep-ph/9707333} {arXiv:hep-ph/9707333 [hep-ph]}
  \BibitemShut {NoStop}%
%%CITATION = HEP-PH/9707333;%%
\bibitem [{\citenamefont {Bringmann}\ \emph {et~al.}(2008)\citenamefont
  {Bringmann}, \citenamefont {Bergstrom},\ and\ \citenamefont
  {Edsjo}}]{Bringmann08}%
  \BibitemOpen
  \bibfield  {author} {\bibinfo {author} {\bibfnamefont {T.}~\bibnamefont
  {Bringmann}}, \bibinfo {author} {\bibfnamefont {L.}~\bibnamefont
  {Bergstrom}}, \ and\ \bibinfo {author} {\bibfnamefont {J.}~\bibnamefont
  {Edsjo}},\ }\href {\doibase 10.1088/1126-6708/2008/01/049} {\bibfield
  {journal} {\bibinfo  {journal} {JHEP}\ }\textbf {\bibinfo {volume} {01}},\
  \bibinfo {pages} {049} (\bibinfo {year} {2008})},\ \Eprint
  {http://arxiv.org/abs/0710.3169} {arXiv:0710.3169 [hep-ph]} \BibitemShut
  {NoStop}%
%%CITATION = ARXIV:0710.3169;%%
\bibitem [{\citenamefont {Fukushima}\ \emph {et~al.}(2014)\citenamefont
  {Fukushima}, \citenamefont {Kelso}, \citenamefont {Kumar}, \citenamefont
  {Sandick},\ and\ \citenamefont {Yamamoto}}]{Fukushima14}%
  \BibitemOpen
  \bibfield  {author} {\bibinfo {author} {\bibfnamefont {K.}~\bibnamefont
  {Fukushima}}, \bibinfo {author} {\bibfnamefont {C.}~\bibnamefont {Kelso}},
  \bibinfo {author} {\bibfnamefont {J.}~\bibnamefont {Kumar}}, \bibinfo
  {author} {\bibfnamefont {P.}~\bibnamefont {Sandick}}, \ and\ \bibinfo
  {author} {\bibfnamefont {T.}~\bibnamefont {Yamamoto}},\ }\href {\doibase
  10.1103/PhysRevD.90.095007} {\bibfield  {journal} {\bibinfo  {journal} {Phys.
  Rev.}\ }\textbf {\bibinfo {volume} {D90}},\ \bibinfo {pages} {095007}
  (\bibinfo {year} {2014})},\ \Eprint {http://arxiv.org/abs/1406.4903}
  {arXiv:1406.4903 [hep-ph]} \BibitemShut {NoStop}%
%%CITATION = ARXIV:1406.4903;%%
\bibitem [{\citenamefont {Buckley}\ \emph {et~al.}(2013)\citenamefont
  {Buckley}, \citenamefont {Hooper},\ and\ \citenamefont
  {Kumar}}]{Buckley:2013sca}%
  \BibitemOpen
  \bibfield  {author} {\bibinfo {author} {\bibfnamefont {M.~R.}\ \bibnamefont
  {Buckley}}, \bibinfo {author} {\bibfnamefont {D.}~\bibnamefont {Hooper}}, \
  and\ \bibinfo {author} {\bibfnamefont {J.}~\bibnamefont {Kumar}},\ }\href
  {\doibase 10.1103/PhysRevD.88.063532} {\bibfield  {journal} {\bibinfo
  {journal} {Phys. Rev.}\ }\textbf {\bibinfo {volume} {D88}},\ \bibinfo {pages}
  {063532} (\bibinfo {year} {2013})},\ \Eprint {http://arxiv.org/abs/1307.3561}
  {arXiv:1307.3561} \BibitemShut {NoStop}%
%%CITATION = ARXIV:1307.3561;%%
\bibitem [{\citenamefont {Pierce}\ \emph {et~al.}(2013)\citenamefont {Pierce},
  \citenamefont {Shah},\ and\ \citenamefont {Freese}}]{Pierce:2013rda}%
  \BibitemOpen
  \bibfield  {author} {\bibinfo {author} {\bibfnamefont {A.}~\bibnamefont
  {Pierce}}, \bibinfo {author} {\bibfnamefont {N.~R.}\ \bibnamefont {Shah}}, \
  and\ \bibinfo {author} {\bibfnamefont {K.}~\bibnamefont {Freese}},\
  }\href@noop {} {\  (\bibinfo {year} {2013})},\ \Eprint
  {http://arxiv.org/abs/1309.7351} {arXiv:1309.7351 [hep-ph]} \BibitemShut
  {NoStop}%
%%CITATION = ARXIV:1309.7351;%%
\bibitem [{\citenamefont {Peskin}\ and\ \citenamefont
  {Schroeder}(1995)}]{Peskin}%
  \BibitemOpen
  \bibfield  {author} {\bibinfo {author} {\bibfnamefont {M.}~\bibnamefont
  {Peskin}}\ and\ \bibinfo {author} {\bibfnamefont {D.}~\bibnamefont
  {Schroeder}},\ }\href@noop {} {\emph {\bibinfo {title} {An Introduction to
  Quantum Field Theory}}}\ (\bibinfo  {publisher} {Westview Press},\ \bibinfo
  {year} {1995})\BibitemShut {NoStop}%
\bibitem [{\citenamefont {Chatrchyan}\ \emph {et~al.}(2012)\citenamefont
  {Chatrchyan} \emph {et~al.}}]{CMShiggs}%
  \BibitemOpen
  \bibfield  {author} {\bibinfo {author} {\bibfnamefont {S.}~\bibnamefont
  {Chatrchyan}} \emph {et~al.} (\bibinfo {collaboration} {CMS}),\ }\href
  {\doibase 10.1016/j.physletb.2012.02.064} {\bibfield  {journal} {\bibinfo
  {journal} {Phys. Lett.}\ }\textbf {\bibinfo {volume} {B710}},\ \bibinfo
  {pages} {26} (\bibinfo {year} {2012})},\ \Eprint
  {http://arxiv.org/abs/1202.1488} {arXiv:1202.1488 [hep-ex]} \BibitemShut
  {NoStop}%
%%CITATION = ARXIV:1202.1488;%%
\bibitem [{\citenamefont {Aad}\ \emph {et~al.}(2012)\citenamefont {Aad} \emph
  {et~al.}}]{ATLAShiggs}%
  \BibitemOpen
  \bibfield  {author} {\bibinfo {author} {\bibfnamefont {G.}~\bibnamefont
  {Aad}} \emph {et~al.} (\bibinfo {collaboration} {ATLAS}),\ }\href {\doibase
  10.1016/j.physletb.2012.02.044} {\bibfield  {journal} {\bibinfo  {journal}
  {Phys. Lett.}\ }\textbf {\bibinfo {volume} {B710}},\ \bibinfo {pages} {49}
  (\bibinfo {year} {2012})},\ \Eprint {http://arxiv.org/abs/1202.1408}
  {arXiv:1202.1408 [hep-ex]} \BibitemShut {NoStop}%
%%CITATION = ARXIV:1202.1408;%%
\bibitem [{\citenamefont {Aad}\ \emph {et~al.}(2014{\natexlab{a}})\citenamefont
  {Aad} \emph {et~al.}}]{ATLASsquark}%
  \BibitemOpen
  \bibfield  {author} {\bibinfo {author} {\bibfnamefont {G.}~\bibnamefont
  {Aad}} \emph {et~al.} (\bibinfo {collaboration} {ATLAS}),\ }\href {\doibase
  10.1007/JHEP09(2014)176} {\bibfield  {journal} {\bibinfo  {journal} {JHEP}\
  }\textbf {\bibinfo {volume} {09}},\ \bibinfo {pages} {176} (\bibinfo {year}
  {2014}{\natexlab{a}})},\ \Eprint {http://arxiv.org/abs/1405.7875}
  {arXiv:1405.7875 [hep-ex]} \BibitemShut {NoStop}%
%%CITATION = ARXIV:1405.7875;%%
\bibitem [{\citenamefont {Chatrchyan}\ \emph {et~al.}(2014)\citenamefont
  {Chatrchyan} \emph {et~al.}}]{CMSsquark}%
  \BibitemOpen
  \bibfield  {author} {\bibinfo {author} {\bibfnamefont {S.}~\bibnamefont
  {Chatrchyan}} \emph {et~al.} (\bibinfo {collaboration} {CMS}),\ }\href
  {\doibase 10.1007/JHEP06(2014)055} {\bibfield  {journal} {\bibinfo  {journal}
  {JHEP}\ }\textbf {\bibinfo {volume} {06}},\ \bibinfo {pages} {055} (\bibinfo
  {year} {2014})},\ \Eprint {http://arxiv.org/abs/1402.4770} {arXiv:1402.4770
  [hep-ex]} \BibitemShut {NoStop}%
%%CITATION = ARXIV:1402.4770;%%
\bibitem [{\citenamefont {Cabrera}\ \emph {et~al.}(2014)\citenamefont
  {Cabrera}, \citenamefont {Casas}, \citenamefont {de~Austri},\ and\
  \citenamefont {Bertone}}]{cabrera1}%
  \BibitemOpen
  \bibfield  {author} {\bibinfo {author} {\bibfnamefont {M.~E.}\ \bibnamefont
  {Cabrera}}, \bibinfo {author} {\bibfnamefont {A.}~\bibnamefont {Casas}},
  \bibinfo {author} {\bibfnamefont {R.~R.}\ \bibnamefont {de~Austri}}, \ and\
  \bibinfo {author} {\bibfnamefont {G.}~\bibnamefont {Bertone}},\ }\href
  {\doibase 10.1007/JHEP12(2014)114} {\bibfield  {journal} {\bibinfo  {journal}
  {JHEP}\ }\textbf {\bibinfo {volume} {12}},\ \bibinfo {pages} {114} (\bibinfo
  {year} {2014})},\ \Eprint {http://arxiv.org/abs/1311.7152} {arXiv:1311.7152
  [hep-ph]} \BibitemShut {NoStop}%
%%CITATION = ARXIV:1311.7152;%%
\bibitem [{\citenamefont {Cabrera-Catalan}\ \emph {et~al.}(2015)\citenamefont
  {Cabrera-Catalan}, \citenamefont {Ando}, \citenamefont {Weniger},\ and\
  \citenamefont {Zandanel}}]{cabrera2}%
  \BibitemOpen
  \bibfield  {author} {\bibinfo {author} {\bibfnamefont {M.~E.}\ \bibnamefont
  {Cabrera-Catalan}}, \bibinfo {author} {\bibfnamefont {S.}~\bibnamefont
  {Ando}}, \bibinfo {author} {\bibfnamefont {C.}~\bibnamefont {Weniger}}, \
  and\ \bibinfo {author} {\bibfnamefont {F.}~\bibnamefont {Zandanel}},\ }\href
  {\doibase 10.1103/PhysRevD.92.035018} {\bibfield  {journal} {\bibinfo
  {journal} {Phys. Rev.}\ }\textbf {\bibinfo {volume} {D92}},\ \bibinfo {pages}
  {035018} (\bibinfo {year} {2015})},\ \Eprint
  {http://arxiv.org/abs/1503.00599} {arXiv:1503.00599 [hep-ph]} \BibitemShut
  {NoStop}%
%%CITATION = ARXIV:1503.00599;%%
\bibitem [{\citenamefont {Heister}\ \emph {et~al.}(2002)\citenamefont {Heister}
  \emph {et~al.}}]{ALEPH}%
  \BibitemOpen
  \bibfield  {author} {\bibinfo {author} {\bibfnamefont {A.}~\bibnamefont
  {Heister}} \emph {et~al.} (\bibinfo {collaboration} {ALEPH}),\ }\href
  {\doibase 10.1016/S0370-2693(02)02471-1} {\bibfield  {journal} {\bibinfo
  {journal} {Phys. Lett.}\ }\textbf {\bibinfo {volume} {B544}},\ \bibinfo
  {pages} {73} (\bibinfo {year} {2002})},\ \Eprint
  {http://arxiv.org/abs/hep-ex/0207056} {arXiv:hep-ex/0207056 [hep-ex]}
  \BibitemShut {NoStop}%
%%CITATION = HEP-EX/0207056;%%
\bibitem [{\citenamefont {Achard}\ \emph {et~al.}(2004)\citenamefont {Achard}
  \emph {et~al.}}]{L3}%
  \BibitemOpen
  \bibfield  {author} {\bibinfo {author} {\bibfnamefont {P.}~\bibnamefont
  {Achard}} \emph {et~al.} (\bibinfo {collaboration} {L3}),\ }\href {\doibase
  10.1016/j.physletb.2003.10.010} {\bibfield  {journal} {\bibinfo  {journal}
  {Phys. Lett.}\ }\textbf {\bibinfo {volume} {B580}},\ \bibinfo {pages} {37}
  (\bibinfo {year} {2004})},\ \Eprint {http://arxiv.org/abs/hep-ex/0310007}
  {arXiv:hep-ex/0310007 [hep-ex]} \BibitemShut {NoStop}%
%%CITATION = HEP-EX/0310007;%%
\bibitem [{\citenamefont {Abdallah}\ \emph {et~al.}(2003)\citenamefont
  {Abdallah} \emph {et~al.}}]{DELPHI}%
  \BibitemOpen
  \bibfield  {author} {\bibinfo {author} {\bibfnamefont {J.}~\bibnamefont
  {Abdallah}} \emph {et~al.} (\bibinfo {collaboration} {DELPHI}),\ }\href
  {\doibase 10.1140/epjc/s2003-01355-5} {\bibfield  {journal} {\bibinfo
  {journal} {Eur. Phys. J.}\ }\textbf {\bibinfo {volume} {C31}},\ \bibinfo
  {pages} {421} (\bibinfo {year} {2003})},\ \Eprint
  {http://arxiv.org/abs/hep-ex/0311019} {arXiv:hep-ex/0311019 [hep-ex]}
  \BibitemShut {NoStop}%
%%CITATION = HEP-EX/0311019;%%
\bibitem [{\citenamefont {Abbiendi}\ \emph {et~al.}(2004)\citenamefont
  {Abbiendi} \emph {et~al.}}]{OPAL}%
  \BibitemOpen
  \bibfield  {author} {\bibinfo {author} {\bibfnamefont {G.}~\bibnamefont
  {Abbiendi}} \emph {et~al.} (\bibinfo {collaboration} {OPAL}),\ }\href
  {\doibase 10.1140/epjc/s2003-01466-y} {\bibfield  {journal} {\bibinfo
  {journal} {Eur. Phys. J.}\ }\textbf {\bibinfo {volume} {C32}},\ \bibinfo
  {pages} {453} (\bibinfo {year} {2004})},\ \Eprint
  {http://arxiv.org/abs/hep-ex/0309014} {arXiv:hep-ex/0309014 [hep-ex]}
  \BibitemShut {NoStop}%
%%CITATION = HEP-EX/0309014;%%
\bibitem [{\citenamefont {Aad}\ \emph {et~al.}(2014{\natexlab{b}})\citenamefont
  {Aad} \emph {et~al.}}]{ATLASslepton}%
  \BibitemOpen
  \bibfield  {author} {\bibinfo {author} {\bibfnamefont {G.}~\bibnamefont
  {Aad}} \emph {et~al.} (\bibinfo {collaboration} {ATLAS}),\ }\href {\doibase
  10.1007/JHEP05(2014)071} {\bibfield  {journal} {\bibinfo  {journal} {JHEP}\
  }\textbf {\bibinfo {volume} {05}},\ \bibinfo {pages} {071} (\bibinfo {year}
  {2014}{\natexlab{b}})},\ \Eprint {http://arxiv.org/abs/1403.5294}
  {arXiv:1403.5294 [hep-ex]} \BibitemShut {NoStop}%
%%CITATION = ARXIV:1403.5294;%%
\bibitem [{\citenamefont {Khachatryan}\ \emph {et~al.}(2014)\citenamefont
  {Khachatryan} \emph {et~al.}}]{CMSslepton}%
  \BibitemOpen
  \bibfield  {author} {\bibinfo {author} {\bibfnamefont {V.}~\bibnamefont
  {Khachatryan}} \emph {et~al.} (\bibinfo {collaboration} {CMS}),\ }\href
  {\doibase 10.1140/epjc/s10052-014-3036-7} {\bibfield  {journal} {\bibinfo
  {journal} {Eur. Phys. J.}\ }\textbf {\bibinfo {volume} {C74}},\ \bibinfo
  {pages} {3036} (\bibinfo {year} {2014})},\ \Eprint
  {http://arxiv.org/abs/1405.7570} {arXiv:1405.7570 [hep-ex]} \BibitemShut
  {NoStop}%
%%CITATION = ARXIV:1405.7570;%%
\bibitem [{\citenamefont {Eckel}\ \emph {et~al.}(2014)\citenamefont {Eckel},
  \citenamefont {Ramsey-Musolf}, \citenamefont {Shepherd},\ and\ \citenamefont
  {Su}}]{Eckel1}%
  \BibitemOpen
  \bibfield  {author} {\bibinfo {author} {\bibfnamefont {J.}~\bibnamefont
  {Eckel}}, \bibinfo {author} {\bibfnamefont {M.~J.}\ \bibnamefont
  {Ramsey-Musolf}}, \bibinfo {author} {\bibfnamefont {W.}~\bibnamefont
  {Shepherd}}, \ and\ \bibinfo {author} {\bibfnamefont {S.}~\bibnamefont
  {Su}},\ }\href {\doibase 10.1007/JHEP11(2014)117} {\bibfield  {journal}
  {\bibinfo  {journal} {JHEP}\ }\textbf {\bibinfo {volume} {11}},\ \bibinfo
  {pages} {117} (\bibinfo {year} {2014})},\ \Eprint
  {http://arxiv.org/abs/1408.2841} {arXiv:1408.2841 [hep-ph]} \BibitemShut
  {NoStop}%
%%CITATION = ARXIV:1408.2841;%%
\bibitem [{\citenamefont {Fukushima}\ and\ \citenamefont
  {Kumar}(2013)}]{Fukushima:2013efa}%
  \BibitemOpen
  \bibfield  {author} {\bibinfo {author} {\bibfnamefont {K.}~\bibnamefont
  {Fukushima}}\ and\ \bibinfo {author} {\bibfnamefont {J.}~\bibnamefont
  {Kumar}},\ }\href {\doibase 10.1103/PhysRevD.88.056017} {\bibfield  {journal}
  {\bibinfo  {journal} {Phys. Rev.}\ }\textbf {\bibinfo {volume} {D88}},\
  \bibinfo {pages} {056017} (\bibinfo {year} {2013})},\ \Eprint
  {http://arxiv.org/abs/1307.7120} {arXiv:1307.7120} \BibitemShut {NoStop}%
%%CITATION = ARXIV:1307.7120;%%
\bibitem [{\citenamefont {Lefranc}\ and\ \citenamefont {Moulin}(2015)}]{HESS}%
  \BibitemOpen
  \bibfield  {author} {\bibinfo {author} {\bibfnamefont {V.}~\bibnamefont
  {Lefranc}}\ and\ \bibinfo {author} {\bibfnamefont {E.}~\bibnamefont {Moulin}}
  (\bibinfo {collaboration} {HESS}),\ }in\ \href
  {http://inspirehep.net/record/1393231/files/arXiv:1509.04123.pdf} {\emph
  {\bibinfo {booktitle} {{Proceedings, 34th International Cosmic Ray Conference
  (ICRC 2015)}}}}\ (\bibinfo {year} {2015})\ \Eprint
  {http://arxiv.org/abs/1509.04123} {arXiv:1509.04123 [astro-ph.HE]}
  \BibitemShut {NoStop}%
%%CITATION = ARXIV:1509.04123;%%
\bibitem [{\citenamefont {Actis}\ \emph {et~al.}(2011)\citenamefont {Actis}
  \emph {et~al.}}]{CTA}%
  \BibitemOpen
  \bibfield  {author} {\bibinfo {author} {\bibfnamefont {M.}~\bibnamefont
  {Actis}} \emph {et~al.} (\bibinfo {collaboration} {CTA Consortium}),\ }\href
  {\doibase 10.1007/s10686-011-9247-0} {\bibfield  {journal} {\bibinfo
  {journal} {Exper. Astron.}\ }\textbf {\bibinfo {volume} {32}},\ \bibinfo
  {pages} {193} (\bibinfo {year} {2011})},\ \Eprint
  {http://arxiv.org/abs/1008.3703} {arXiv:1008.3703 [astro-ph.IM]} \BibitemShut
  {NoStop}%
%%CITATION = ARXIV:1008.3703;%%
\bibitem [{\citenamefont {Wood}\ \emph {et~al.}(2013)\citenamefont {Wood},
  \citenamefont {Buckley}, \citenamefont {Digel}, \citenamefont {Funk},
  \citenamefont {Nieto},\ and\ \citenamefont {Sanchez-Conde}}]{Wood:2013taa}%
  \BibitemOpen
  \bibfield  {author} {\bibinfo {author} {\bibfnamefont {M.}~\bibnamefont
  {Wood}}, \bibinfo {author} {\bibfnamefont {J.}~\bibnamefont {Buckley}},
  \bibinfo {author} {\bibfnamefont {S.}~\bibnamefont {Digel}}, \bibinfo
  {author} {\bibfnamefont {S.}~\bibnamefont {Funk}}, \bibinfo {author}
  {\bibfnamefont {D.}~\bibnamefont {Nieto}}, \ and\ \bibinfo {author}
  {\bibfnamefont {M.~A.}\ \bibnamefont {Sanchez-Conde}},\ }in\ \href
  {http://www.slac.stanford.edu/econf/C1307292/docs/submittedArxivFiles/1305.0302.pdf}
  {\emph {\bibinfo {booktitle} {{Community Summer Study 2013: Snowmass on the
  Mississippi (CSS2013) Minneapolis, MN, USA, July 29-August 6, 2013}}}}\
  (\bibinfo {year} {2013})\ \Eprint {http://arxiv.org/abs/1305.0302}
  {arXiv:1305.0302 [astro-ph.HE]} \BibitemShut {NoStop}%
%%CITATION = ARXIV:1305.0302;%%
\bibitem [{\citenamefont {Ackermann}\ \emph
  {et~al.}(2015{\natexlab{a}})\citenamefont {Ackermann} \emph
  {et~al.}}]{Ackermann:2015lka}%
  \BibitemOpen
  \bibfield  {author} {\bibinfo {author} {\bibfnamefont {M.}~\bibnamefont
  {Ackermann}} \emph {et~al.} (\bibinfo {collaboration} {Fermi-LAT}),\ }\href
  {\doibase 10.1103/PhysRevD.91.122002} {\bibfield  {journal} {\bibinfo
  {journal} {Phys. Rev.}\ }\textbf {\bibinfo {volume} {D91}},\ \bibinfo {pages}
  {122002} (\bibinfo {year} {2015}{\natexlab{a}})},\ \Eprint
  {http://arxiv.org/abs/1506.00013} {arXiv:1506.00013 [astro-ph.HE]}
  \BibitemShut {NoStop}%
%%CITATION = ARXIV:1506.00013;%%
\bibitem [{\citenamefont {Topchiev}\ \emph {et~al.}(2015)\citenamefont
  {Topchiev} \emph {et~al.}}]{G400}%
  \BibitemOpen
  \bibfield  {author} {\bibinfo {author} {\bibfnamefont {N.~P.}\ \bibnamefont
  {Topchiev}} \emph {et~al.},\ }in\ \href
  {http://inspirehep.net/record/1384340/files/arXiv:1507.06246.pdf} {\emph
  {\bibinfo {booktitle} {{Proceedings, 34th International Cosmic Ray Conference
  (ICRC 2015)}}}}\ (\bibinfo {year} {2015})\ \Eprint
  {http://arxiv.org/abs/1507.06246} {arXiv:1507.06246 [astro-ph.IM]}
  \BibitemShut {NoStop}%
%%CITATION = ARXIV:1507.06246;%%
\bibitem [{\citenamefont {Huang}\ \emph {et~al.}(2016)\citenamefont {Huang}
  \emph {et~al.}}]{HERD}%
  \BibitemOpen
  \bibfield  {author} {\bibinfo {author} {\bibfnamefont {X.}~\bibnamefont
  {Huang}} \emph {et~al.},\ }\href {\doibase
  10.1016/j.astropartphys.2016.02.003} {\bibfield  {journal} {\bibinfo
  {journal} {Astropart. Phys.}\ } (\bibinfo {year} {2016}),\
  10.1016/j.astropartphys.2016.02.003},\ \Eprint
  {http://arxiv.org/abs/1509.02672} {arXiv:1509.02672 [astro-ph.HE]}
  \BibitemShut {NoStop}%
%%CITATION = ARXIV:1509.02672;%%
\bibitem [{\citenamefont {Bringmann}\ and\ \citenamefont
  {Calore}(2014)}]{Bringmann14}%
  \BibitemOpen
  \bibfield  {author} {\bibinfo {author} {\bibfnamefont {T.}~\bibnamefont
  {Bringmann}}\ and\ \bibinfo {author} {\bibfnamefont {F.}~\bibnamefont
  {Calore}},\ }\href {\doibase 10.1103/PhysRevLett.112.071301} {\bibfield
  {journal} {\bibinfo  {journal} {Phys. Rev. Lett.}\ }\textbf {\bibinfo
  {volume} {112}},\ \bibinfo {pages} {071301} (\bibinfo {year} {2014})},\
  \Eprint {http://arxiv.org/abs/1308.1089} {arXiv:1308.1089 [hep-ph]}
  \BibitemShut {NoStop}%
%%CITATION = ARXIV:1308.1089;%%
\bibitem [{\citenamefont {Hahn}(2001)}]{Hahn01}%
  \BibitemOpen
  \bibfield  {author} {\bibinfo {author} {\bibfnamefont {T.}~\bibnamefont
  {Hahn}},\ }\href {\doibase 10.1016/S0010-4655(01)00290-9} {\bibfield
  {journal} {\bibinfo  {journal} {Comput. Phys. Commun.}\ }\textbf {\bibinfo
  {volume} {140}},\ \bibinfo {pages} {418} (\bibinfo {year} {2001})},\ \Eprint
  {http://arxiv.org/abs/hep-ph/0012260} {arXiv:hep-ph/0012260 [hep-ph]}
  \BibitemShut {NoStop}%
%%CITATION = HEP-PH/0012260;%%
\bibitem [{\citenamefont {Hahn}\ and\ \citenamefont
  {Perez-Victoria}(1999)}]{Hahn98}%
  \BibitemOpen
  \bibfield  {author} {\bibinfo {author} {\bibfnamefont {T.}~\bibnamefont
  {Hahn}}\ and\ \bibinfo {author} {\bibfnamefont {M.}~\bibnamefont
  {Perez-Victoria}},\ }\href {\doibase 10.1016/S0010-4655(98)00173-8}
  {\bibfield  {journal} {\bibinfo  {journal} {Comput. Phys. Commun.}\ }\textbf
  {\bibinfo {volume} {118}},\ \bibinfo {pages} {153} (\bibinfo {year}
  {1999})},\ \Eprint {http://arxiv.org/abs/hep-ph/9807565}
  {arXiv:hep-ph/9807565 [hep-ph]} \BibitemShut {NoStop}%
%%CITATION = HEP-PH/9807565;%%
\bibitem [{\citenamefont {Ellis}\ \emph {et~al.}(2012)\citenamefont {Ellis},
  \citenamefont {Kunszt}, \citenamefont {Melnikov},\ and\ \citenamefont
  {Zanderighi}}]{Ellis12}%
  \BibitemOpen
  \bibfield  {author} {\bibinfo {author} {\bibfnamefont {R.~K.}\ \bibnamefont
  {Ellis}}, \bibinfo {author} {\bibfnamefont {Z.}~\bibnamefont {Kunszt}},
  \bibinfo {author} {\bibfnamefont {K.}~\bibnamefont {Melnikov}}, \ and\
  \bibinfo {author} {\bibfnamefont {G.}~\bibnamefont {Zanderighi}},\ }\href
  {\doibase 10.1016/j.physrep.2012.01.008} {\bibfield  {journal} {\bibinfo
  {journal} {Phys. Rept.}\ }\textbf {\bibinfo {volume} {518}},\ \bibinfo
  {pages} {141} (\bibinfo {year} {2012})},\ \Eprint
  {http://arxiv.org/abs/1105.4319} {arXiv:1105.4319 [hep-ph]} \BibitemShut
  {NoStop}%
%%CITATION = ARXIV:1105.4319;%%
\bibitem [{\citenamefont {Yaguna}(2009)}]{Yaguna09}%
  \BibitemOpen
  \bibfield  {author} {\bibinfo {author} {\bibfnamefont {C.~E.}\ \bibnamefont
  {Yaguna}},\ }\href {\doibase 10.1103/PhysRevD.80.115002} {\bibfield
  {journal} {\bibinfo  {journal} {Phys. Rev.}\ }\textbf {\bibinfo {volume}
  {D80}},\ \bibinfo {pages} {115002} (\bibinfo {year} {2009})},\ \Eprint
  {http://arxiv.org/abs/0909.4181} {arXiv:0909.4181 [hep-ph]} \BibitemShut
  {NoStop}%
%%CITATION = ARXIV:0909.4181;%%
\bibitem [{\citenamefont {Bergstrom}\ \emph {et~al.}(2013)\citenamefont
  {Bergstrom}, \citenamefont {Bringmann}, \citenamefont {Cholis}, \citenamefont
  {Hooper},\ and\ \citenamefont {Weniger}}]{Bergstrom:2013jra}%
  \BibitemOpen
  \bibfield  {author} {\bibinfo {author} {\bibfnamefont {L.}~\bibnamefont
  {Bergstrom}}, \bibinfo {author} {\bibfnamefont {T.}~\bibnamefont
  {Bringmann}}, \bibinfo {author} {\bibfnamefont {I.}~\bibnamefont {Cholis}},
  \bibinfo {author} {\bibfnamefont {D.}~\bibnamefont {Hooper}}, \ and\ \bibinfo
  {author} {\bibfnamefont {C.}~\bibnamefont {Weniger}},\ }\href {\doibase
  10.1103/PhysRevLett.111.171101} {\bibfield  {journal} {\bibinfo  {journal}
  {Phys. Rev. Lett.}\ }\textbf {\bibinfo {volume} {111}},\ \bibinfo {pages}
  {171101} (\bibinfo {year} {2013})},\ \Eprint {http://arxiv.org/abs/1306.3983}
  {arXiv:1306.3983 [astro-ph.HE]} \BibitemShut {NoStop}%
%%CITATION = ARXIV:1306.3983;%%
\bibitem [{\citenamefont {Di~Mauro}\ \emph {et~al.}(2016)\citenamefont
  {Di~Mauro}, \citenamefont {Donato}, \citenamefont {Fornengo},\ and\
  \citenamefont {Vittino}}]{DiMauro:2015jxa}%
  \BibitemOpen
  \bibfield  {author} {\bibinfo {author} {\bibfnamefont {M.}~\bibnamefont
  {Di~Mauro}}, \bibinfo {author} {\bibfnamefont {F.}~\bibnamefont {Donato}},
  \bibinfo {author} {\bibfnamefont {N.}~\bibnamefont {Fornengo}}, \ and\
  \bibinfo {author} {\bibfnamefont {A.}~\bibnamefont {Vittino}},\ }\href
  {\doibase 10.1088/1475-7516/2016/05/031} {\bibfield  {journal} {\bibinfo
  {journal} {JCAP}\ }\textbf {\bibinfo {volume} {1605}},\ \bibinfo {pages}
  {031} (\bibinfo {year} {2016})},\ \Eprint {http://arxiv.org/abs/1507.07001}
  {arXiv:1507.07001 [astro-ph.HE]} \BibitemShut {NoStop}%
%%CITATION = ARXIV:1507.07001;%%
\bibitem [{\citenamefont {Ackermann}\ \emph
  {et~al.}(2015{\natexlab{b}})\citenamefont {Ackermann} \emph
  {et~al.}}]{Ackermann:2015zua}%
  \BibitemOpen
  \bibfield  {author} {\bibinfo {author} {\bibfnamefont {M.}~\bibnamefont
  {Ackermann}} \emph {et~al.} (\bibinfo {collaboration} {Fermi-LAT}),\ }\href
  {\doibase 10.1103/PhysRevLett.115.231301} {\bibfield  {journal} {\bibinfo
  {journal} {Phys. Rev. Lett.}\ }\textbf {\bibinfo {volume} {115}},\ \bibinfo
  {pages} {231301} (\bibinfo {year} {2015}{\natexlab{b}})},\ \Eprint
  {http://arxiv.org/abs/1503.02641} {arXiv:1503.02641 [astro-ph.HE]}
  \BibitemShut {NoStop}%
%%CITATION = ARXIV:1503.02641;%%
\bibitem [{\citenamefont {Rico}\ \emph {et~al.}(2015)\citenamefont {Rico},
  \citenamefont {Wood}, \citenamefont {Drlica-Wagner},\ and\ \citenamefont
  {Aleksić}}]{Rico:2015nya}%
  \BibitemOpen
  \bibfield  {author} {\bibinfo {author} {\bibfnamefont {J.}~\bibnamefont
  {Rico}}, \bibinfo {author} {\bibfnamefont {M.}~\bibnamefont {Wood}}, \bibinfo
  {author} {\bibfnamefont {A.}~\bibnamefont {Drlica-Wagner}}, \ and\ \bibinfo
  {author} {\bibfnamefont {J.}~\bibnamefont {Aleksić}} (\bibinfo
  {collaboration} {Fermi-LAT, MAGIC}),\ }in\ \href
  {https://inspirehep.net/record/1389141/files/arXiv:1508.05827.pdf} {\emph
  {\bibinfo {booktitle} {{Proceedings, 34th International Cosmic Ray Conference
  (ICRC 2015)}}}}\ (\bibinfo {year} {2015})\ \Eprint
  {http://arxiv.org/abs/1508.05827} {arXiv:1508.05827 [astro-ph.HE]}
  \BibitemShut {NoStop}%
%%CITATION = ARXIV:1508.05827;%%
\bibitem [{\citenamefont {Bergstrom}(1989)}]{Bergstrom89}%
  \BibitemOpen
  \bibfield  {author} {\bibinfo {author} {\bibfnamefont {L.}~\bibnamefont
  {Bergstrom}},\ }\href {\doibase 10.1016/0370-2693(89)90585-6} {\bibfield
  {journal} {\bibinfo  {journal} {Phys. Lett.}\ }\textbf {\bibinfo {volume}
  {B225}},\ \bibinfo {pages} {372} (\bibinfo {year} {1989})}\BibitemShut
  {NoStop}%
%%CITATION = PHLTA,B225,372;%%
\bibitem [{\citenamefont {Kumar}\ and\ \citenamefont
  {Marfatia}(2013)}]{Kumar:2013iva}%
  \BibitemOpen
  \bibfield  {author} {\bibinfo {author} {\bibfnamefont {J.}~\bibnamefont
  {Kumar}}\ and\ \bibinfo {author} {\bibfnamefont {D.}~\bibnamefont
  {Marfatia}},\ }\href {\doibase 10.1103/PhysRevD.88.014035} {\bibfield
  {journal} {\bibinfo  {journal} {Phys. Rev.}\ }\textbf {\bibinfo {volume}
  {D88}},\ \bibinfo {pages} {014035} (\bibinfo {year} {2013})},\ \Eprint
  {http://arxiv.org/abs/1305.1611} {arXiv:1305.1611 [hep-ph]} \BibitemShut
  {NoStop}%
%%CITATION = ARXIV:1305.1611;%%
\bibitem [{\citenamefont {Geringer-Sameth}\ and\ \citenamefont
  {Koushiappas}(2011)}]{GeringerSameth:2011iw}%
  \BibitemOpen
  \bibfield  {author} {\bibinfo {author} {\bibfnamefont {A.}~\bibnamefont
  {Geringer-Sameth}}\ and\ \bibinfo {author} {\bibfnamefont {S.~M.}\
  \bibnamefont {Koushiappas}},\ }\href {\doibase
  10.1103/PhysRevLett.107.241303} {\bibfield  {journal} {\bibinfo  {journal}
  {Phys. Rev. Lett.}\ }\textbf {\bibinfo {volume} {107}},\ \bibinfo {pages}
  {241303} (\bibinfo {year} {2011})},\ \Eprint {http://arxiv.org/abs/1108.2914}
  {arXiv:1108.2914 [astro-ph.CO]} \BibitemShut {NoStop}%
%%CITATION = ARXIV:1108.2914;%%
\bibitem [{\citenamefont {Elvang}\ and\ \citenamefont
  {Huang}(2013)}]{Elvang13}%
  \BibitemOpen
  \bibfield  {author} {\bibinfo {author} {\bibfnamefont {H.}~\bibnamefont
  {Elvang}}\ and\ \bibinfo {author} {\bibfnamefont {Y.-t.}\ \bibnamefont
  {Huang}},\ }\href@noop {} {\  (\bibinfo {year} {2013})},\ \Eprint
  {http://arxiv.org/abs/1308.1697} {arXiv:1308.1697 [hep-th]} \BibitemShut
  {NoStop}%
%%CITATION = ARXIV:1308.1697;%%
\end{thebibliography}%

\appendix
\section{\label{sec:photoncs}Analytic Two-photon Cross Section}
\begin{figure*}
\centering
\begin{minipage}[t]{0.24\textwidth}
\includegraphics[width=\textwidth]{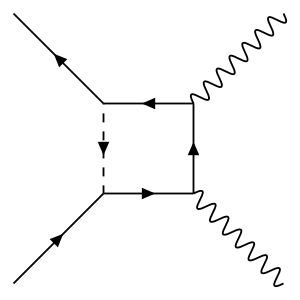}
\end{minipage}%
\begin{minipage}[t]{0.24\textwidth}
\includegraphics[width=\textwidth]{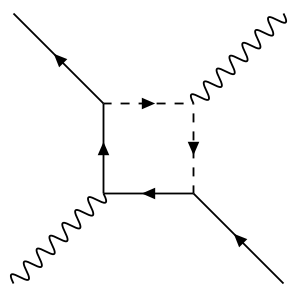}
\end{minipage}
\begin{minipage}[t]{0.24\textwidth}
\includegraphics[width=\textwidth]{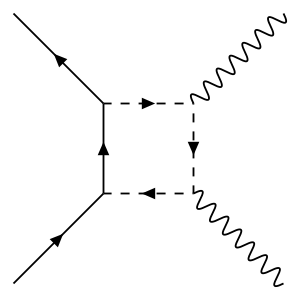}
\end{minipage}
\begin{minipage}[t]{0.24\textwidth}
\includegraphics[width=\textwidth]{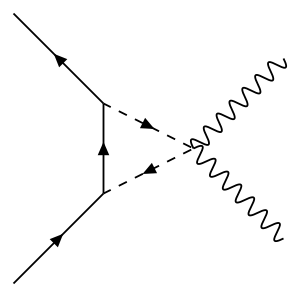}
\end{minipage}
%\captionsetup{justification=raggedright,singlelinecheck=false}
\caption{Feynman diagrams that contribute to the one-loop process $XX\rightarrow\gamma\gamma$. Diagrams with initial and final state particles exchanged are not listed.}
\label{fig:feyndiag}
\end{figure*}

The interaction between the fermion, scalar, and photon is given by
\begin{align}
\mathcal{L}_{\text{qed}}&=ie\left(\widetilde{f}_{1}^{\ast}A^{\mu}\partial_{\mu}\widetilde{f}^{}_{1}+\widetilde{f}_{2}^{\ast}A^{\mu}\partial_{\mu}\widetilde{f}^{}_{2}-\text{c.c}\right)\nonumber\\
&\quad +e^{2}A^{\mu}A_{\mu}\left(\widetilde{f}_{1}^{\ast}\widetilde{f}^{}_{1}+\widetilde{f}_{2}^{\ast}\widetilde{f}^{}_{2}\right)+e\,\overline{f}\gamma^{\mu}A_{\mu}f\,.
\end{align}
We denote the momenta of the two initial state dark matter particles as $k_{1}$ and $k_{2}$, and the momenta of the two final state photons as $k_{3}$ and $k_{4}$. Since the annihilation takes place between two dark matter particles at rest, the momentum configuration is
\begin{align}
&k_{1}=k_{2}=k=\left(m_{{X}},0,0,0\right)\,,\nonumber\\
&k_{3}=\left(m_{{X}},m_{{X}}\,\hat{n}\right)\,,\\
 &k_{4}=\left(m_{{X}},-m_{{X}}\,\hat{n}\right),\nonumber
\end{align}
where the unit vector $\hat{n}$ gives the direction of the photon momentum. Using the spinor helicity formalism, we choose the polarization vectors as
\begin{align}
&\epsilon^{\mu}_{3}(+)\equiv\epsilon^{\mu}_{+}\left(k_{3};k_{4}\right)=\frac{1}{\sqrt{2}}\frac{\left[k_{3}\left|\gamma^{\mu}\right|k_{4}\right\rangle}{\left\langle k_{4}k_{3}\right\rangle}\,,\nonumber\\
&\epsilon^{\mu}_{3}(-)\equiv\epsilon^{\mu}_{-}\left(k_{3};k_{4}\right)=\frac{1}{\sqrt{2}}\frac{\left\langle k_{3}\left|\gamma^{\mu}\right|k_{4}\right]}{\left[k_{4}k_{3}\right]}\,,\nonumber\\
&\epsilon_{4}^{\mu}(+)\equiv\epsilon^{\mu}_{+}\left(k_{4};k_{3}\right)=\frac{1}{\sqrt{2}}\frac{\left[k_{4}\left|\gamma^{\mu}\right|k_{3}\right\rangle}{\left\langle k_{3}k_{4}\right\rangle}\,,\nonumber\\
&\epsilon^{\mu}_{4}(-)\equiv\epsilon^{\mu}_{-}\left(k_{4};k_{3}\right)=\frac{1}{\sqrt{2}}\frac{\left\langle k_{4}\left|\gamma^{\mu}\right|k_{3}\right]}{\left[k_{3}k_{4}\right]}\,,
\end{align}
where the notation follows Ref.~\cite{Elvang13}. The benefit of this choice is that the inner products between opposite helicity vectors are always zero. Feynman diagrams that contribute to the amplitude are displayed in Fig.~\eqref{fig:feyndiag}.

We write the total amplitude $\pmb{\mathcal{A}}$ as
\begin{equation}
\bm{\mathcal{A}}=\frac{i\alpha_{\text{em}}}{2\pi}\left(\epsilon_{3}\cdot\epsilon_{4}\right)\left[\frac{\overline{u}(k_{1})\gamma^{5}v(k_{2})}{2m_{{X}}}\right]\pmb{\mathscr{A}}(h)\,,
\end{equation}
where $\alpha_{\text{em}}$ is the fine structure constant and $h\equiv(h_{3}+h_{4})/2$ such that $h=1$ for the $(++)$ final and $h=-1$ for the $(--)$ final state. The structure $\left(\epsilon_{3}\cdot\epsilon_{4}\right)\left(\overline{u}_{1}\gamma^{5}v_{2}\right)$ reflects the $s$-wave nature of this amplitude, since this factor is nonzero only for the $L=0$ initial state configuration. Then all the contributions from the loop integrals in Fig.~\eqref{fig:feyndiag} are collected in $\pmb{\mathscr{A}}(h)$, which we write as
\begin{widetext}
\begin{align}
\pmb{\mathscr{A}}(h)&=\mathcal{I}_{1}\left(|\lambda_{L}^{2}|\cos^{2}\alpha+|\lambda^{2}_{R}|\sin^{2}\alpha\right)+\mathcal{I}_{2}\left(|\lambda_{L}^{2}|\sin^{2}\alpha+|\lambda_{R}^{2}|\cos^{2}\alpha\right)\nonumber\\
&\quad+2|\lambda_{L}\lambda_{R}|\sin\alpha\cos\alpha\left(\frac{m_{f}}{m_{{X}}}\right)\left[(\mathcal{J}_{1}-\mathcal{J}_{2})\cos\varphi+ih(\mathcal{K}_{1}-\mathcal{K}_{2})\sin\varphi\right].
\end{align}
Because of the term $ih(\mathcal{K}_{1}-\mathcal{K}_{2})\sin\varphi$, the probabilities of having $(++)$ and $(--)$ photon final states are unequal (note that $\mathcal{I}_i$, $\mathcal{J}_i$ and $\mathcal{K}_i$ are, in general, complex functions), which is a potentially measurable effect of $CP$ violation. The coefficients $\mathcal{I}_{i}$, $\mathcal{J}_{i}$ and $\mathcal{K}_{i}$ are given by
\begin{align}
\label{eq:Ii}
\mathcal{I}_{i}&=\frac{m_{i}^{2}\,I_{2}(m_{i},m_{f})}{m_{i}^{2}-m_{l}^{2}}-\frac{2m_{f}^{2}\,I_{1}(m_{f})}{m_{i}^{2}+m_{{X}}^{2}-m_{f}^{2}}+\frac{m_{f}^{2}(m_{i}^{2}-m_{{X}}^{2}-m_{f}^{2})I_{3}(m_{i},m_{f})}{(m_{i}^{2}-m_{f}^{2})(m_{i}^{2}+m_{{X}}^{2}-m_{f}^{2})}\,,\\
\mathcal{J}_{i}&=\frac{2m_{{X}}^{2}\,\left[I_{1}(m_{f})-I_{3}(m_{i},m_{f})\right]}{m_{i}^{2}+m_{{X}}^{2}-m_{f}^{2}}\,,\\
\mathcal{K}_{i}&=\frac{2(m_{{X}}^{2}-m_{f}^{2})\,I_{1}(m_{f})}{m_{i}^{2}+m_{{X}}^{2}-m_{f}^{2}}+\frac{2m_{i}^{2}\,I_{2}(m_{i},m_{f})}{m_{i}^{2}-m_{f}^{2}}-\frac{2m_{i}^{2}m_{{X}}^{2}\,I_{3}(m_{i},m_{f})}{(m_{i}^{2}-m_{f}^{2})(m_{i}^{2}+m_{{X}}^{2}-m_{f}^{2})}\,,\nonumber\\*
&\quad -\frac{2m_{i}^{2}}{m_{i}^{2}-m_{{X}}^{2}-m_{f}^{2}}\left[I_{2}(m_{i},m_{f})-I_{1}(m_{i})\right]\,,
\end{align}
where $m_{i}$ is the mass for the internal scalars (here we have adopted a simplified notation, $m_{i}\equiv m_{\widetilde{f}_{i}}$ in the main text). We observe that if $m_{1}=m_{2}$, we have $(\mathcal{I}_{1},\mathcal{J}_{1},\mathcal{K}_{1})=(\mathcal{I}_{2},\mathcal{J}_{2},\mathcal{K}_{2})$ such that the amplitude will not vanish but it will depend neither on the mixing angle $\alpha$ nor on the $CP$-violation phase $\varphi$. However, as seen from \eqref{eq:AnniCS}, the $s$-wave $2 \rightarrow 2$ annihilation cross section is identically zero in this case. {If the integral $\mathcal{K}_{1}-\mathcal{K}_{2}$ is complex (which is the case for both the $\mu$ and $\tau$ channels), the amplitudes of the $(++)$ and $(--)$ final states do not have the same magnitude, which leads to an asymmetry ratio $R$ as discussed in the main text.} Here $I_{1}$ and $I_{2}$ are the same as $2m^{2}_{X}I_{3}^{[1]}$ and $2m^{2}_{X}I_{3}^{[2]}$ in Ref.~\cite{Bern97}. They are related to the standard three-point scalar loop integrals through
\begin{align}
\frac{I_{1}(m_{a})}{2m^{2}_{{X}}}&=C_{0}\left[0,0,4m^{2}_{{X}},m_{a}^{2},m_{a}^{2},m_{a}^{2}\right]\\
\frac{I_{2}(m_{a},m_{b})}{2m^{2}_{{X}}}&=C_{0}\left[0,m^{2}_{{X}},-m^{2}_{{X}},m_{a}^{2},m_{a}^{2},m_{b}^{2}\right],
\end{align}
in which we follow the convention of {\tt LoopTools} \cite{Hahn98}. The analytic expressions for $I_{1,2,3}$ are
\begin{align}
I_{1}(m_{a})&=\left\{\begin{array}{ll}
\frac{1}{4}\left[\log\left(\frac{1+\sqrt{1-m_{a}^{2}/m_{{X}}^{2}}}{1-\sqrt{1-m^{2}_{a}/m_{{X}}^{2}}}\right)+i\pi\right]^{2} & m_{a}\leq m_{{X}} \\
-\left[\arctan\sqrt{\frac{1}{m_{a}^{2}/m_{{X}}^{2}-1}}\,\right]^{2} & m_{a}>m_{{X}} \\
\end{array}\right.\,,\\
I_{2}(m_{a},m_{b})&=\left[-\Li_{2}\left(\frac{m_{a}^{2}-m_{b}^{2}+m_{{X}}^{2}-\sqrt{\Delta_{1}}}{2m_{a}^{2}}\right)-\Li_{2}\left(\frac{m_{a}^{2}-m_{b}^{2}+m_{{X}}^{2}+\sqrt{\Delta_{1}}}{2m_{a}^{2}}\right)\right.\nonumber\\*
&\qquad\left.+\Li_{2}\left(\frac{m_{a}^{2}-m_{b}^{2}-m_{{X}}^{2}-\sqrt{\Delta_{2}}}{2m_{a}^{2}}\right)+\Li_{2}\left(\frac{m_{a}^{2}-m_{b}^{2}-m_{{X}}^{2}+\sqrt{\Delta_{2}}}{2m_{a}^{2}}\right)\right],\\*
I_{3}(m_{a},m_{b})&\equiv I_{2}(m_{b},m_{a})\,,
\end{align}
where
\begin{align}
&\Delta_{1}=(m_{a}^{2}-m_{b}^{2}-m_{{X}}^{2})^{2}-4m_{{X}}^{2}m_{b}^{2}\,, \nonumber\\
&\Delta_{2}=(m_{a}^{2}-m_{b}^{2}+m_{{X}}^{2})^{2}+4m_{{X}}^{2}m_{b}^{2}\,.
\end{align}
We note that both the $\mathcal{I}_{i}$ and $\mathcal{J}_{i}$ terms are contained in the analytic expression in Refs.\cite{Bergstrom97,Bern97}, but the $\mathcal{K}_{i}$ term is missing. Finally, we define the square of the total unpolarized amplitude as
\begin{equation}
\left|\pmb{\mathcal{M}}\right|^{2}=\frac{1}{4}\sum_{s_{1},s_{2}}\sum_{h_{3},h_{4}}\left|\pmb{\mathcal{A}}\right|^{2}=\frac{\alpha_{\text{em}}^{2}}{8\pi^{2}}\sum_{h=\pm 1}\left|\pmb{\mathscr{A}}(h)\right|^{2},
\end{equation}
and the total cross section is
\begin{equation}
\label{eq:sigmavAA}
(\sigma v)_{\gamma\gamma}=\frac{1}{2}\times\frac{\left|\pmb{\mathcal{M}}\right|^{2}}{32\pi m_{{X}}^{2}}=\frac{\alpha_{\text{em}}^{2}}{512\pi^{3}m^{2}_{{X}}}\sum_{h=\pm 1}\left|\pmb{\mathscr{A}}(h)\right|^{2},
\end{equation}
where the factor $1/2$ accounts for the fact that the final state consists of identical particles.
\end{widetext}

\end{document}